\documentclass[12pt]{iopart}
 \expandafter\let\csname equation*\endcsname\relax
  \expandafter\let\csname endequation*\endcsname\relax
\usepackage{amsmath,amsfonts,amssymb,bm,graphicx,mathrsfs,units,cite}

\usepackage{color,colordvi}
\definecolor{blue}{rgb}{0,0,1}
\definecolor{grey}{rgb}{0.6,0.6,0.6}
\usepackage{bbold}
\definecolor{myurlcolor}{rgb}{0,0,0.7}
\definecolor{myrefcolor}{rgb}{0.8,0,0}
\usepackage[unicode=true,pdfusetitle, bookmarks=false,bookmarksnumbered=false,
bookmarksopen=false, breaklinks=false,pdfborder={0 0 0},backref=false,
colorlinks=true, linkcolor=myrefcolor,citecolor=myurlcolor,urlcolor=myurlcolor]{hyperref}

\newcount\colveccount
\newcommand*\colvec[1]{
        \global\colveccount#1
        \begin{pmatrix}
        \colvecnext
}
\def\colvecnext#1{
        #1
        \global\advance\colveccount-1
        \ifnum\colveccount>0
                \\
                \expandafter\colvecnext
        \else
                \end{pmatrix}
        \fi
}

\begin{document}


\title[Thermodynamically consistent master equation]{A thermodynamically consistent Markovian master equation beyond the secular approximation}

\today

\author{Patrick P. Potts,$^{1,2}$ Alex Arash Sand Kalaee,$^{1}$ Andreas Wacker$^{1}$}

\address{$^1$Physics Department and NanoLund, Lund University, Box 118, 22100 Lund, Sweden.}
\address{$^2$Department of Physics, University of Basel, Klingelbergstrasse 82, 4056 Basel, Switzerland.}

\ead{patrick.potts@unibas.ch}

\date{\today}

\begin{abstract}
Markovian master equations provide a versatile tool for describing open quantum systems
when memory effects of the environment may be neglected. As these equations are of an approximate
nature, they often do not respect the laws of thermodynamics when no secular approximation is performed
in their derivation. Here we introduce a Markovian master equation that is 
thermodynamically consistent and provides an accurate description whenever memory effects can be 
neglected. The thermodynamic consistency is obtained through a rescaled Hamiltonian for the thermodynamic bookkeeping,
exploiting the fact that a Markovian description implies a limited resolution for heat. Our results enable a thermodynamically consistent description of a variety of systems where the
secular approximation breaks down.
\end{abstract}


\vspace{2pc}
\maketitle
\section{Introduction}
Theoretical descriptions of open quantum systems are crucial for understanding various scenarios, as a complete shielding from the environment is usually not feasible or not even desirable. The latter is the case in the field of quantum thermodynamics, where heat flows in out-of-equilibrium situations are the key issue of concern. As a microscopic description of the environmental degrees of freedom becomes quickly intractable, numerous approaches exist to approximate the behavior of the degrees of freedom of the system alone \cite{breuer:book,weiss:book,kamenev:book,schaller:book}. 
Of particular interest are Markovian master equations in so-called GKLS form, after Gorini, Kosakowski, Sudarshan \cite{gorini:1976}, and Linblad \cite{lindblad:1976}. By neglecting any memory effects induced by the environment, they provide a particularly tractable description that gives access to the full density matrix of the system alone (i.e., the reduced system). These equations usually rely on Born-Markov approximations which do not ensure GKLS form. For this reason additional approximations are usually employed.

The most prominent approximation is the so-called secular approximation \cite{breuer:book}, where oscillating terms are dropped from the master equation (for a recent generalization to time-dependent systems, see Ref.~\cite{dann:2018}). This approximation has the desirable feature that it ensures consistency with the laws of thermodynamics. The secular approximation has also been termed the \textit{global} approach, because it uses the (delocalized) eigenstates of the Hamiltonian. This is in contrast to the \textit{local} approach which is widely used in, e.g., quantum optics \cite{carmichael:book,restrepo:2014} (for a comparison between the global and the local approach, see Refs.~\cite{cresser:1992,scala:2007,rivas:2010,hofer:2017njp,gonzalez:2017,seah:2018,naseem:2018,cattaneo:2019,farina:2020,elouard:2020,scali:2021}). The difference between the local and the global approach becomes particularly apparent when considering a system of weakly coupled components (e.g., two or three qubits or harmonic oscillators weakly coupled to each other). In this case, the local approach can be obtained by deriving a master equation for the individual, uncoupled components and simply adding the coupling term. We stress however that the local approach is by no means phenomenological, as a microscopic derivation exists \cite{hofer:2017njp,scali:2021}. This approach is appealing mainly for two reasons: First, it does not require diagonalization of the Hamiltonian and can thus be applied to problems where a secular approximation is difficult to implement analytically. Second, it holds for systems that consist of weakly coupled degenerate units, where the secular approximation breaks down due to the near-degeneracies. Such systems are widely studied in the field of quantum thermodynamics as they are promising for manipulating and exploiting heat flows \cite{brunner:2012,mitchison:2019}. However, the local approach was criticized for not being thermodynamically consistent as it may result in violations of the second law of thermodynamics \cite{levy:2014}. These violations where shown to be small as long as the local approach is justified \cite{novotny:2002,trushechkin:2016}. One may thus argue that sizable violations of the second law provide a useful \textit{red flag}, indicating that the approach is applied outside its regime of validity. Nevertheless, thermodynamic consistency is desirable in any Markovian master equation as it allows for falling back on well established laws \cite{dann:2020}. It was shown that the local approach can be rendered thermodynamically consistent by re-defining heat. This was motivated from a collisional model, where work is required to maintain the collisions \cite{chiara:2018}, as well as directly from the master equation itself \cite{hegwill:2021}. Below, we show how thermodynamic consistency of the local approach can be obtained starting from the standard microscopic system-bath picture by exploiting a crucial insight: \textit{the approximations that result in a Markovian master equation impose limitations on the energy-resolution for heat}. Within this limited resolution, we may re-define heat without compromising the accuracy of the approach.

Recently, a novel approach termed PERLind (Position and Energy Resolved LINDblad equation) was introduced \cite{kirsanskas:2018} in an attempt to interpolate between the local and global approach. Since then, multiple microscopic derivations were given \cite{ptaszynski:2019,kleinherbers:2020,nathan:2020,davidovic:2020}, showing that the PERLind approach does not require any additional assumptions going beyond the ones already present when performing Born-Markov approximations. Compared to the global and local approaches, the PERLind approach thus has an increased regime of validity, making it a very promising approach. However, it also comes with disadvantages: it requires diagonalization of the Hamiltonian and it is not thermodynamically consistent. Furthermore, an analytical treatment is often complicated by tedious expressions as shown below.
Another approach that goes beyond the secular approximation is based on introducing a coarse-graining time \cite{schaller:2008,majenz:2013,cresser:2017,seah:2018,farina:2019,mozgunov:2020}. While taking this time to infinity recovers the global master equation, a GKLS master equation can be obtained by choosing a finite coarse-graining time. In general, coarse-grained master equations are not thermodynamically consistent. However, just as for the local approach, thermodynamic consistency can be recovered by re-defining heat as motivated by a collisional model \cite{schaller:2020}. Yet an alternative approach for obtaining a GKLS master equation beyond the secular approximation is provided by truncating the Redfield equation \cite{becker:2021}.

Here we introduce a novel Markovian master equation in GKLS form that goes beyond the secular approximation. Its main merit compared to previous approaches is that it is thermodynamically consistent. Our approach is based on the same principle that is at the heart of all Markovian master equations: the environment properties are slowly changing in energy. This allows us to not only neglect the broadening of energy levels but to also treat transition energies that are close as having the same value. Employing the same approximation in the definition of heat then naturally results in a thermodynamically consistent treatment. Our approach may reduce to the global approach (when no transition energies are close) or to a thermodynamically consistent version of the local approach (when all transition energies are close). We note that for a time-independent Hamiltonian, our results are in agreement with the \textit{unified GKLS master equation} which was very recently derived in an independent work \cite{trushechkin:2021}.

The rest of this paper is structured as follows. In Sec.~\ref{sec:compare}, we compare different Markovian master equations and illustrate the main results without going into technical details. In particular, we introduce a thermodynamically consistent local master equation as an example of our general master equation, which is derived in Sec.~\ref{sec:derivation}. Sections \ref{sec:fermions} and \ref{sec:bosons} illustrate our master equation with the examples of a fermionic and a bosonic heat engine, covering both the time-independent as well as the time-dependent case. In Sec.~\ref{sec:beyond}, we apply our master equation to an interacting double quantum dot, a problem where we do not have an exact solution and where our master equation may provide a description that differs from both the global and the local approach. We conclude in Sec.~\ref{sec:conclusions}.

\section{Comparing different master equations}
\label{sec:compare}
\begin{figure}
	\centering
	\includegraphics[width=.75\textwidth]{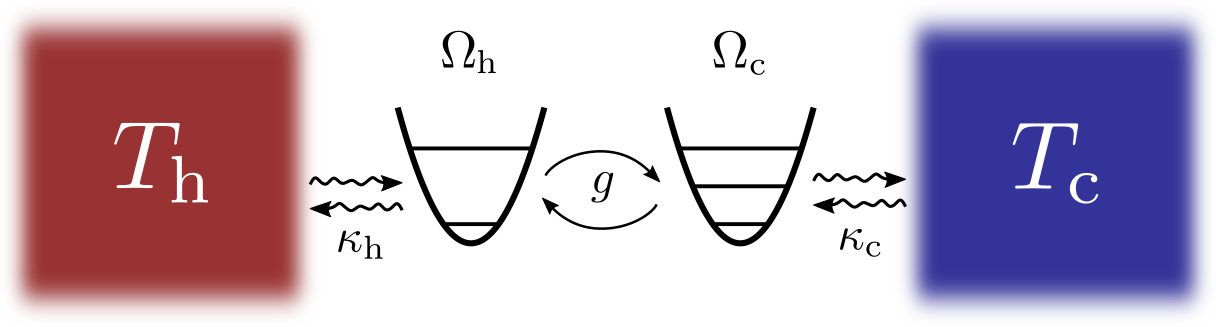}
	\caption{\label{fig:oscillators} System of coupled harmonic oscillators to illustrate the validity of different Markovian master equations. Two harmonic oscillators with frequencies $\Omega_{\rm h}$ and $\Omega_{\rm c}$ are coupled to each other with coupling strength $g$ and to a thermal reservoir of temperature $T_{\rm h}$ and $T_{\rm c}$ (with coupling strength $\kappa_{\rm h}$ and $\kappa_{\rm c}$) respectively.}
\end{figure}

In this section, we compare different master equations, in order to illustrate some of our main results without going into mathematical details. To shed light on the discussion about the thermodynamic consistency, we focus on the system that was considered in Ref.~\cite{levy:2014}, where it was shown that a local approach (here referred to as \textit{conventional local} approach) may violate the second law of thermodynamics. The system (sketched in Fig.~\ref{fig:oscillators}) consists of two coupled harmonic oscillators and is described by the Hamiltonian
\begin{equation}
\label{eq:hamtwoosc}
\hat{H}_{\rm S} = \Omega_{\rm c}\hat{a}_{\rm c}^\dagger\hat{a}_{\rm c} + \Omega_{\rm h}\hat{a}_{\rm h}^\dagger\hat{a}_{\rm h}+g(\hat{a}_{\rm h}^\dagger\hat{a}_{\rm c}+\hat{a}_{\rm c}^\dagger\hat{a}_{\rm h}) = \Omega_+\hat{a}^\dagger_+\hat{a}_+ +\Omega_-\hat{a}^\dagger_-\hat{a}_-,
\end{equation}
with the standard commutation relations
\begin{equation}
\label{eq:commrels}
[\hat{a}_\alpha,\hat{a}^\dagger_\beta] = \delta_{\alpha,\beta},\hspace{1.5cm}[\hat{a}_\alpha,\hat{a}_\beta]=0.
\end{equation}
The two oscillators are labeled by c and h, as they are coupled to a cold and a hot bath respectively. Diagonalizing the Hamiltonian results in the eigenfrequencies
\begin{equation}
\label{eq:eigenfreq}
\Omega_\pm = \bar{\Omega}\pm\sqrt{\Delta^2+g^2},\hspace{.75cm}\bar{\Omega} = \frac{\Omega_{\rm c}+\Omega_{\rm h}}{2},\hspace{.75cm}\Delta = \frac{\Omega_{\rm h}-\Omega_{\rm c}}{2},
\end{equation}
and the corresponding ladder operators
\begin{equation}
\label{eq:ladderpm}
\hat{a}_+ = \cos(\theta/2)\hat{a}_{\rm h}+\sin(\theta/2)\hat{a}_{\rm c},\hspace{.75cm}\hat{a}_- = -\sin(\theta/2)\hat{a}_{\rm h}+\cos(\theta/2)\hat{a}_{\rm c},
\end{equation}
where the angle $\theta$ is defined through
\begin{equation}
\label{eq:theta}
\cos(\theta) = \frac{\Delta}{\sqrt{\Delta^2+g^2}},
\end{equation}
where we use the branch $0\le \theta\le \pi$, which corresponds to $g>0$, which we tacitly assume througout this work.

We will now compare five different approaches to describe the heat transport through this system:
\begin{enumerate}
	\item The transmission function approach (valid for arbitrary system-bath coupling).
	\item The conventional local approach (may violate the second law of thermodynamics).
	\item Our local approach (ensures the laws of thermodynamics).
	\item The global approach (relies on the secular approximation).
	\item The PERLind approach (interpolates between local and global).
\end{enumerate}
As we show below, the master equation we introduce in this work reduces either to the local, or to the global approach depending on the parameters. We stress that the approach we refer to as \textit{local} is thermodynamically consistent, in contrast to the \textit{conventional local} approach investigated, e.g., in Ref.~\cite{levy:2014}. The transmission function approach is used as a benchmark, because it is not perturbative in the system-bath coupling. Furthermore, by resolving the energy dependence of heat transport, it indicates specific shortcomings in the other approaches. We include the local approach used in Ref.~\cite{levy:2014} to illustrate when and why the second law of thermodynamics may be violated.

In all five approaches, we take the so-called wide-band limit, assuming an energy-independent coupling between system and bath and neglecting any Lamb shift of the Hamiltonian. While this is the only assumption for the transmission approach, all master equation approaches further rely on the Born-Markov approximations. For the current scenario, these approximations are valid whenever
\begin{equation}
\label{eq:skap}
\kappa_\alpha \ll\Omega_\alpha,
\end{equation}
where $\kappa_\alpha$ denotes the transition rate between system and reservoir $\alpha= c,h$. All master equation approaches are of the form
\begin{equation}
\label{eq:masterequation}
\partial_t \hat{\rho} = -i[\hat{H}_{\rm S},\hat{\rho}] +\mathcal{L}^{j}_{\rm c} \hat{\rho}+\mathcal{L}^{j}_{\rm h} \hat{\rho},
\end{equation}
with different dissipators $\mathcal{L}^j_\alpha$, where $j$ labels the approach.

\subsection{The transmission function}
For non-interacting particles, the heat current can be written using a Landauer-like formula \cite{wang:2014}. For bosons (with vanishing chemical potential) it takes on the form
\begin{equation}
\label{eq:landauer}
J^{\rm t} = \int_0^{\infty} \frac{d\omega}{2\pi} \mathcal{T}(\omega)\omega[n_{\rm B}^{\rm h}(\omega)-n_{\rm B}^{\rm c}(\omega)],
\end{equation}
where we introduced the Bose-Einstein distribution
\begin{equation}
\label{eq:bed}
n_{\rm B}^\alpha(\omega) = \frac{1}{e^{\frac{\omega}{k_{\rm B} T_\alpha}}-1},
\end{equation} 
and the transmission function $\mathcal{T}(\omega)$ can be obtained using non-equilibrium Green's functions \cite{meir:1992,zhang:2007}.
For the present system, it reads (see for instance Ref.~\cite{agarwalla:2018} for the analogous case of Fermions)
\begin{equation}
\label{eq:transmission}
\mathcal{T}(\omega) = \frac{g^2\kappa_{\rm c}\kappa_{\rm h}}{|(\omega-\Omega_{\rm h}+i\frac{\kappa_{\rm h}}{2})(\omega-\Omega_{\rm c}+i\frac{\kappa_{\rm c}}{2})-g^2|^2}.
\end{equation}
The Landauer-like formula has an intuitive interpretation. At each energy $\omega$, bosons traverse the system with the rate $\mathcal{T}(\omega) d\omega$. The transmission function has two peaks located at $\Omega_+$ and $\Omega_-$. For small $g$ and $\Delta$, these peaks merge as illustrated in the insets of Fig.~\ref{fig:compar1}\,(b).

\subsection{The conventional local approach}
In this approach, the dissipators act locally on the two oscillators
\begin{equation}
\label{eq:disslocal}
\mathcal{L}^{\rm cl}_\alpha = \kappa_\alpha\left\{n_{\rm B}^\alpha(\Omega_\alpha)\mathcal{D}[\hat{a}_\alpha^\dagger]+[n_{\rm B}^\alpha(\Omega_\alpha)+1]\mathcal{D}[\hat{a}_\alpha]\right\},
\end{equation}
where cl stands for conventional local. Here we used the GKLS superoperators
\begin{equation}
\label{eq:supop}
\mathcal{D}[\hat{A}]\hat{\rho} = \hat{A}\hat{\rho}\hat{A}^\dagger-\frac{1}{2}\{\hat{A}^\dagger\hat{A},\hat{\rho}\},
\end{equation}
with $\{\cdot,\cdot\}$ denoting the anti-commutator.
The heat current in this approach is defined as
\begin{equation}
\label{eq:heatcurrl}
J^{\rm cl} = {\rm Tr}\left\{\hat{H}_{\rm S}\mathcal{L}^{\rm cl}_{\rm h}\hat{\rho}\right\}.
\end{equation}
In steady state, this reduces to
\begin{equation}
\label{eq:heatcurrlss}
J^{\rm cl} = \frac{\kappa_{\rm c}\Omega_{\rm h}+\kappa_{\rm h}\Omega_{\rm c}}{(\kappa_{\rm c}+\kappa_{\rm h})^2}\frac{4g^2\kappa_{\rm c}\kappa_{\rm h}[n_{\rm B}^{\rm h}(\Omega_{\rm h})-n_{\rm B}^{\rm c}(\Omega_{\rm c})]}{4g^2+\kappa_{\rm c}\kappa_{\rm h}+16\Delta^2\kappa_{\rm c}\kappa_{\rm h}/(\kappa_{\rm c}+\kappa_{\rm h})^2}.
\end{equation}

\begin{figure}
	\includegraphics[width=\textwidth]{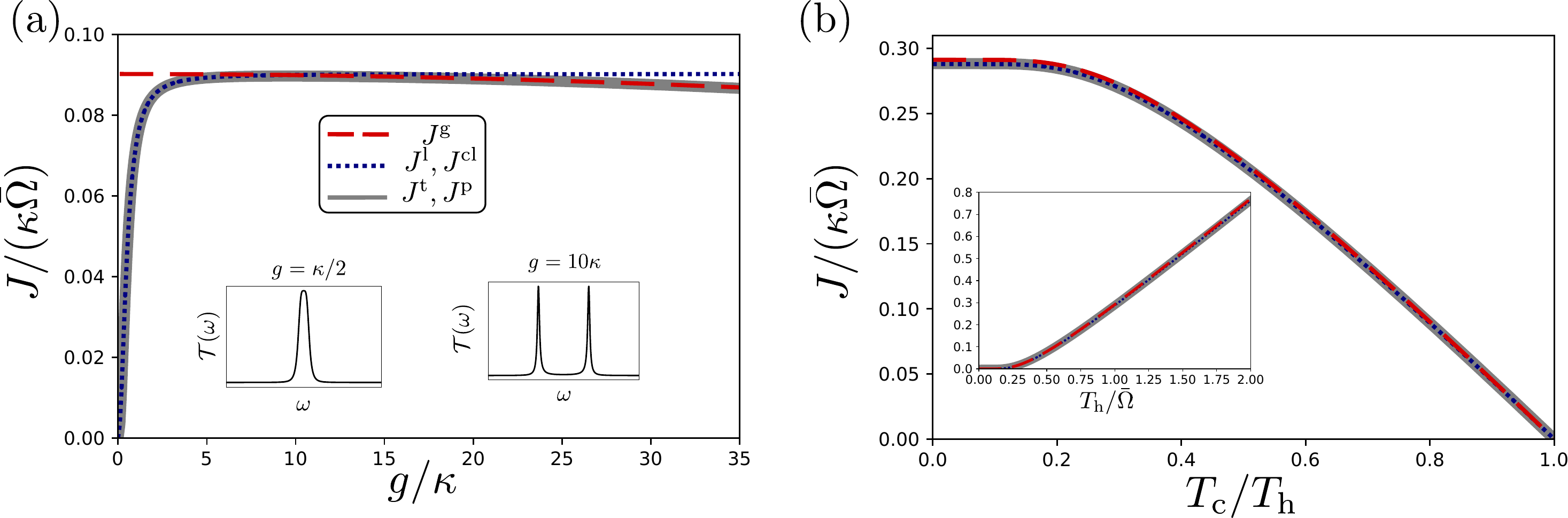}
	\caption{\label{fig:compar1} Steady state heat current for degenerate harmonic oscillators ($\Delta=0$). Grey: Transmission approach ($J^{\rm t}$) which serves as a benchmark. The PERLind approach ($J^{\rm p}$) agrees perfectly with the transmission approach for these figures. Blue: local approach ($J^{\rm l}$). In steady state, this approach reduces to the conventional local approach ($J^{\rm cl}$) for $\Delta=0$. Red: Global approach ($J^{\rm g}$) which relies on the secular approximation. (a) As expected, the transmission approach interpolates between the local and global approaches as a function of the coupling strength $g$. In agreement with Eq.~\eqref{eq:regvallocal}, the local approach may be justified even for $g>\kappa$. The insets illustrate the transmission function as a function of energy for two different values of $g$, where the local or the global approach is well justified. (b) The master equation approaches agree well with the transmission function approach for a wide range of temperatures. The inset (where $T_{\rm c}=0$) illustrates the regime where both temperatures become small. We note that in this regime, the relative error associated to the master equation approaches may become large but the absolute error stays small. Parameters: $\kappa\equiv\kappa_{\rm c}=\kappa_{\rm h}=0.02\,\bar{\Omega}$, $\Delta = 0$, $k_{\rm B}T_{\rm h} = \bar{\Omega}$, (a) $k_{\rm B}T_{\rm c} = 0.8\,\bar{\Omega}$,  (b) $g = 5\,\kappa$, inset: $T_{\rm c} = 0$.}
\end{figure}

It is important to note that there is a microscopic derivation that results in this local approach, which is valid for the current system as long as
$n_{\rm B}^\alpha(\Omega_{\rm h})\simeq n_{\rm B}^\alpha(\Omega_{\rm c})$ \cite{hofer:2017njp}, which is the case for
\begin{equation}
\label{eq:regvallocal}
g,\Delta \ll\bar{\Omega}.
\end{equation}
When this inequality is satisfied, we may expect that the laws of thermodynamics hold, see Fig.~\ref{fig:compar1}. When this approximation is not justified, the second law of thermodynamics is not guaranteed as illustrated in Fig.~\ref{fig:compar2}. The reason for this is that the Bose-Einstein distributions are evaluated at $\Omega_{\rm h}$ and $\Omega_{\rm c}$ respectively in Eq.~\eqref{eq:heatcurrlss}. This implies that the bosons change their energy when traversing the system. From the Landauer-like formula [cf.~Eq.~\eqref{eq:landauer}], it is apparent that this cannot happen. We note that while Eq.~\eqref{eq:regvallocal} ensures $|n_{\rm B}^\alpha(\Omega_{\rm h})-n_{\rm B}^\alpha(\Omega_{\rm c})|\ll 1$, the relative error in the heat current can still be sizable when the occupation numbers (and thus the heat current) become small. This is the case because Eq.~\eqref{eq:regvallocal} does not ensure $|n_{\rm B}^\alpha(\Omega_{\rm h})-n_{\rm B}^\alpha(\Omega_{\rm c})|\ll n_{\rm B}^\alpha(\Omega_{\rm \alpha})$

\subsection{Our local approach}
\begin{figure}
	\includegraphics[width=\textwidth]{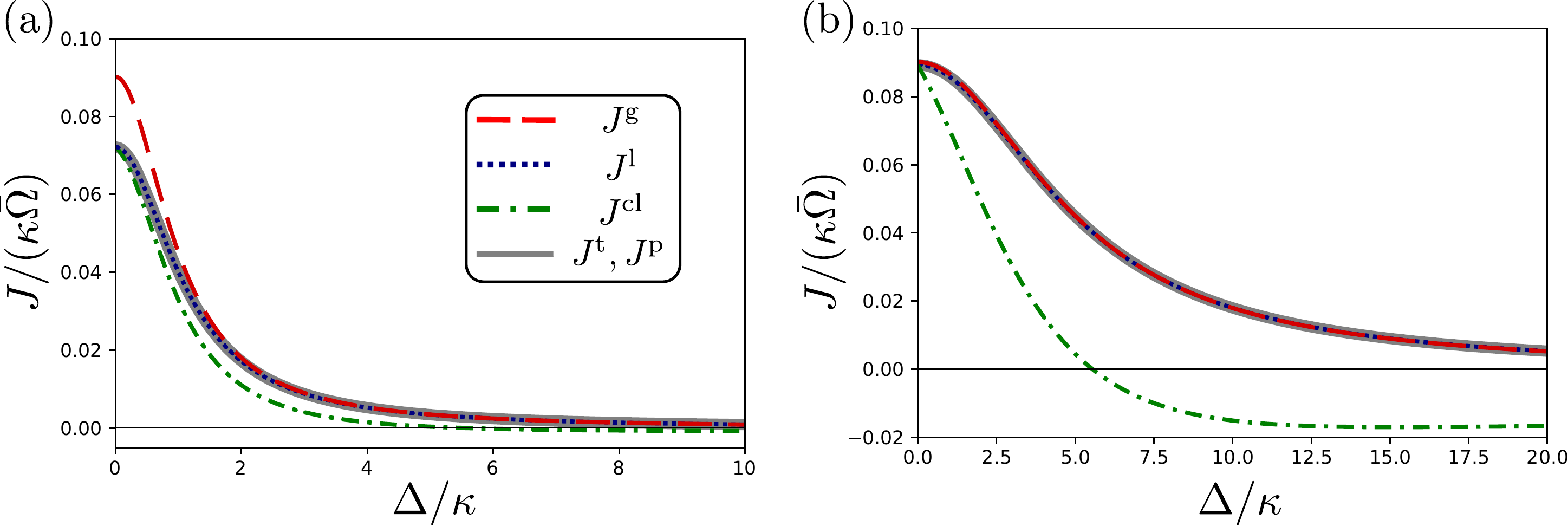}
	\caption{\label{fig:compar2} Steady state heat current as a function of the detuning between the oscillator frequencies $\Delta=\Omega_{\rm h}-\Omega_{\rm c}$. Grey: Transmission approach ($J^{\rm t}$) which serves as a benchmark. The PERLind approach ($J^{\rm p}$) agrees perfectly with the transmission approach for these figures. Blue: local approach ($J^{\rm l}$). Green: Conventional local approach ($J^{\rm cl}$). Red: Global approach ($J^{\rm g}$). (a) For small coupling $g$, only a very small violation of the second law (negative heat current) is observed in the conventional local approach. (b) For larger $g$, the violation of the second law becomes larger. Interestingly, the local approach provides an accurate heat current, even though it is no longer microscopically justified for these parameters. Parameters: $\kappa\equiv\kappa_{\rm c}=\kappa_{\rm h}=0.02\,\bar{\Omega}$, $k_{\rm B}T_{\rm c} = 0.8\,\bar{\Omega}$, $k_{\rm B}T_{\rm h} = \bar{\Omega}$ (a) $g=0.02\,\bar{\Omega}$ (b) $g=0.1\,\bar{\Omega}$.}
\end{figure}

Here we present a thermodynamically consistent local approach that follows from our general master equation introduced below in Sec.~\ref{sec:derivation}, where we motivate this approach microscopically and show how it can be generalized to more complicated systems. As in the previous approach, the dissipators act locally in this approach
\begin{equation}
\label{eq:dissreflocal}
\mathcal{L}^{\rm l}_\alpha = \kappa_\alpha\left\{n_{\rm B}^\alpha(\bar{\Omega})\mathcal{D}[\hat{a}_\alpha^\dagger]+[n_{\rm B}^\alpha(\bar{\Omega})+1]\mathcal{D}[\hat{a}_\alpha]\right\},
\end{equation}
the only difference being that the Bose-Einstein distributions are now all evaluated at the frequency $\bar{\Omega}$. This approach has the same regime of validity as the previous local approach. Indeed, Eq.~\eqref{eq:regvallocal} implies
\begin{equation}
\label{eq:nbs}
|n^\alpha_{\rm B}(\bar{\Omega})-n_{\rm B}^\alpha(\Omega_\alpha)|\ll 1,
\end{equation}
and we expect the dissipators in Eqs.~\eqref{eq:disslocal} and \eqref{eq:dissreflocal} to result in approximately the same dynamics as illustrated in Fig.~\ref{fig:compar1}. If Eq.~\eqref{eq:regvallocal} is not satisfied, the two local approaches may result in different results, see Fig.~\ref{fig:compar2}. 

To obtain a thermodynamically consistent description, the approximation in Eq.~\eqref{eq:regvallocal} is exploited in the definition of the heat current which reads
\begin{equation}
\label{eq:heatcurrrl}
J^{\rm l} = {\rm Tr}\left\{\hat{H}_{\rm TD}\mathcal{L}^{\rm l}_{\rm h}\hat{\rho}\right\},
\end{equation}
where we introduced a separate Hamiltonian for the thermodynamic bookkeeping (TD)
\begin{equation}
\label{eq:hamtd}
\hat{H}_{\rm TD} = \bar{\Omega}(\hat{a}_{\rm c}^\dagger\hat{a}_{\rm c} + \hat{a}_{\rm h}^\dagger\hat{a}_{\rm h}).
\end{equation}
Whenever Eq.~\eqref{eq:regvallocal} holds, this Hamiltonian provides the (approximately) correct energy flows. Consider the case where $\Delta = 0$. Then, $\hat{H}_{\rm TD}$ is simply obtained by dropping the coupling between the oscillators. The thermodynamic bookkeeping thus neglects both the system-bath couplings, as well as the coupling between the oscillators. Of course, all these couplings are crucial for the dynamics, where they appear either in the Hamiltonian or in the dissipators. In Sec.~\ref{sec:derivation}, we illustrate how such a thermodynamic Hamiltonian can be constructed in general, providing thermodynamic consistency of the corresponding master equation at the price of a slightly reduced energy resolution of thermodynamic quantities.

In steady state, the heat current reads
\begin{equation}
\label{eq:heatcurrrlss}
J^{\rm l} = \frac{\bar{\Omega}}{\kappa_{\rm c}+\kappa_{\rm h}}\frac{4g^2\kappa_{\rm c}\kappa_{\rm h}[n_{\rm B}^{\rm h}(\bar{\Omega})-n_{\rm B}^{\rm c}(\bar{\Omega})]}{4g^2+\kappa_{\rm c}\kappa_{\rm h}+16\Delta^2\kappa_{\rm c}\kappa_{\rm h}/(\kappa_{\rm c}+\kappa_{\rm h})^2}.
\end{equation}
We note that for $\Delta=0$, we find $J^{\rm l}=J^{\rm cl}$ in the steady state. However, in the transient regime the two local approaches may result in different heat currents even for $\Delta=0$. The difference between the two approaches is in this case the energy associated to the coupling term (proportional to $g$), which is dropped in the local approach. We note that the heat current associated to this coupling term has been interpreted as a quantum contribution before \cite{elouard:2020}.

Comparing the local master equation with the Landauer-like formula [cf.~Eq.~\eqref{eq:landauer}], one can see that transmission is approximated to happen at a single energy $\bar{\Omega}$. This is a good approximation when Eq.~\eqref{eq:regvallocal} holds. In this case, the transmission function is only non-zero around $\bar{\Omega}$ and the Bose-Einstein distributions may be assumed constant across all energies where transmission is non-zero. Interestingly, the local approach agrees well with the transmission approach for the present system, even when Eq.~\eqref{eq:regvallocal} is not satisfied, see Fig.~\ref{fig:compar2}\,(b). This is in stark contrast to the conventional local approach.

\subsection{The global approach}
In the global approach, the dissipators are introducing jumps between the eigenstates of the Hamiltonian
\begin{equation}
\label{eq:dissglobal}
\mathcal{L}^{\rm g}_\alpha =\sum_{\sigma=\pm} \kappa^\sigma_\alpha\left\{n_{\rm B}^\alpha(\Omega_\sigma)\mathcal{D}[\hat{a}_\sigma^\dagger]+[n_{\rm B}^\alpha(\Omega_\sigma)+1]\mathcal{D}[\hat{a}_\sigma]\right\},
\end{equation}
where the coupling strengths reflect the spatial distribution of the eigenstates
\begin{equation}
\label{eq:kappags}
\kappa_{\rm h}^+ = \kappa_{\rm h} \cos^2(\theta/2),\hspace{.25cm}\kappa_{\rm h}^- = \kappa_{\rm h} \sin^2(\theta/2),\hspace{.25cm}\kappa_{\rm c}^+ = \kappa_{\rm c} \sin^2(\theta/2),\hspace{.25cm}\kappa_{\rm c}^- = \kappa_{\rm c} \cos^2(\theta/2).
\end{equation}
The heat current in the global approach reads
\begin{equation}
\label{eq:heatcurrg}
J^{\rm g} = {\rm Tr}\left\{\hat{H}_{\rm S}\mathcal{L}^{\rm g}_{\rm h}\hat{\rho}\right\},
\end{equation}
which in steady state reduces to
\begin{equation}
\label{eq:heatcurrssg}
J^{\rm g} = \sum_{\sigma=\pm}\Omega_\sigma\frac{\kappa_{\rm c}^\sigma\kappa_{\rm h}^\sigma}{\kappa_{\rm c}^\sigma+\kappa_{\rm h}^\sigma}[n_{\rm B}^{\rm h}(\Omega_\sigma)-n_{\rm B}^{\rm c}(\Omega_\sigma)].
\end{equation}

This approach is thermodynamically consistent and we will always find heat flowing from hot to cold. As it relies on the secular approximation, it requires the condition
\begin{equation}
\label{eq:regvalglob}
\kappa_{\rm c},\kappa_{\rm h}\ll \sqrt{\Delta^2+g^2}.
\end{equation}

In the global master equation, the transmission is approximated to happen at the two energies $\Omega_\pm$. If Eq.~\eqref{eq:regvalglob} holds, the transmission function given in Eq.~\eqref{eq:transmission} consists of two narrow peaks located at $\Omega_\pm$, see the right inset in Fig.~\ref{fig:compar1}\,(b). Over the width of these peaks, the Bose-Einstein distributions can be assumed constant and the global master equation is valid. Note that if Eq.~\eqref{eq:regvalglob} does not hold, the peaks overlap and the secular approximation breaks down, see Fig.~\ref{fig:compar1}. We also note that Eqs.~\eqref{eq:regvallocal} and \eqref{eq:regvalglob} may both be valid. In this case, both the global as well as the local approach are expected to give accurate results.

\subsection{The PERLind approach}
In the PERLind approach, the dissipators read
\begin{equation}
\label{eq:dissperlind}
\mathcal{L}^{\rm p}_\alpha = \sum_{k=-1,1}\mathcal{D}[\hat{J}_{\alpha,k}],
\end{equation}
with the jump operators
\begin{equation}
\label{eq:jumperlind}
\begin{aligned}
\hat{J}_{\rm c,-1} &= \sqrt{\kappa_{\rm c}^+[n_{\rm B}^{\rm c}(\Omega_+)+1]}\hat{a}_++ \sqrt{\kappa_{\rm c}^-[n_{\rm B}^{\rm c}(\Omega_-)+1]}\hat{a}_-,\\
\hat{J}_{\rm c,1} &= \sqrt{\kappa_{\rm c}^+n_{\rm B}^{\rm c}(\Omega_+)}\hat{a}_+^\dagger+ \sqrt{\kappa_{\rm c}^-n_{\rm B}^{\rm c}(\Omega_-)}\hat{a}_-^\dagger,\\
\hat{J}_{\rm h,-1} &= \sqrt{\kappa_{\rm h}^+[n_{\rm B}^{\rm h}(\Omega_+)+1]}\hat{a}_+- \sqrt{\kappa_{\rm h}^-[n_{\rm B}^{\rm h}(\Omega_-)+1]}\hat{a}_-,\\
\hat{J}_{\rm h,1} &= \sqrt{\kappa_{\rm h}^+n_{\rm B}^{\rm h}(\Omega_+)}\hat{a}_+^\dagger- \sqrt{\kappa_{\rm h}^-n_{\rm B}^{\rm h}(\Omega_-)}\hat{a}_-^\dagger.
\end{aligned}
\end{equation}

In this approach, the heat current is defined as
\begin{equation}
\label{eq:heatcurrp}
J^{\rm p} = {\rm Tr}\left\{\hat{H}_{\rm S}\mathcal{L}^{\rm p}_{\rm h}\hat{\rho}\right\}.
\end{equation}
For the present system, the heat current in the PERLind approach agrees perfectly with the result obtained from the transmission approach for all considered parameter values. This is expected as we consider scenarios where Eq.~\eqref{eq:skap} is fulfilled and the Born-Markov approximations are justified. Clearly, the PERLind approach has strong advantages. However, it also has its disadvantages. First, the analytical expressions quickly become unwieldy, which is the case already for the simple system considered here. Second, the PERLind approach may violate the second law of thermodynamics \cite{kirsanskas:2018}. When heat is defined by Eq.~\eqref{eq:heatcurrp}, a necessary and sufficient condition for the second law to hold is given by \cite{spohn:1978,brandner:2016}
\begin{equation}
\label{eq:secondlawcond}
\mathcal{L}^{\rm p}_\alpha e^{-\beta_\alpha \hat{H}_{\rm S}} =0.
\end{equation}
This condition is not fulfilled as shown in Ref.~\cite{ptaszynski:2019}. However, as long as the Born-Markov approximations are justified, any second law violations should become vanishingly small. It is an open question if a different definition for heat, that is consistent with the Born-Markov approximation, could salvage the second law.

\subsection{Which approach to use?}
As we have shown above, all approaches have their advantages and disadvantages. The PERLind approach provides the most accurate description, while the conventional local approach is the most simple, not requiring diagonalization of the Hamiltonian. Both approaches may however result in violations of the second law. The global and the local approach both respect the laws of thermodynamics. Furthermore, together they provide an accurate description for all parameter values as they are justified in complementary, but overlapping, parameter regimes [cf.~Eqs.~\eqref{eq:regvallocal}, \eqref{eq:regvalglob}, and \eqref{eq:skap}, assuming the Born-Markov approximations are justified]. In the following, we illustrate how these thermodynamically consistent approaches follow from a unified framework that can be extended to scenarios where neither a global nor a local description is justified.

\section{A thermodynamically consistent master equation}
\label{sec:derivation}
We first revisit the general scenario under consideration and discuss the laws of thermodynamics. We then provide a detailed derivation and discuss how thermodynamic consistency is obtained by a consistent application of the approximations on the thermodynamic bookkeeping.

\subsection{The general scenario}
\label{sec:scenario}
We consider the general scenario described by the Hamiltonian
\begin{equation}
\label{eq:hamtot}
\hat{H}_{\rm tot}(t)=\hat{H}_{\rm S}(t)+\sum_{\alpha}\left(\hat{H}_\alpha+\hat{V}_{\alpha}\right)=\hat{H}_{\rm S}(t)+\hat{H}_{\rm B}+\hat{V},
\end{equation}
where the first term describes the Hamiltonian of the system (which may be time-dependent) and the second and third term describe the thermal reservoirs (labeled by $\alpha$) and their coupling to the system respectively. The system exchanges energy and particles with the reservoirs, such that energy changes can be divided into heat and work. The average heat that leaves bath $\alpha$ during the time interval $[0,t]$ is given by
\begin{equation}
\label{eq:heat}
\langle Q_\alpha\rangle =-{\rm Tr}\{(\hat{H}_\alpha-\mu_\alpha\hat{N}_\alpha)\hat{\rho}_{\rm tot}(t)\}+{\rm Tr}\{(\hat{H}_\alpha-\mu_\alpha\hat{N}_\alpha)\hat{\rho}_{\rm tot}(0)\},
\end{equation}
where $\hat{N}_\alpha$ denotes the particle number operator for reservoir $\alpha$ and $\mu_\alpha$ its chemical potential. The average work provided by reservoir $\alpha$ is
\begin{equation}
\label{eq:workbath}
\langle W_\alpha\rangle =-\mu_\alpha\left({\rm Tr}\{\hat{N}_\alpha\hat{\rho}_{\rm tot}(t)\}-{\rm Tr}\{\hat{N}_\alpha\hat{\rho}_{\rm tot}(0)\}\right).
\end{equation}
In addition, the time-dependence of the system Hamiltonian results in the external average power
\begin{equation}
\label{eq:powersys}
P_{\rm ext}(t)={\rm Tr}\{[\partial_t\hat{H}_{\rm S}(t)]\hat{\rho}_{\rm tot}(t)\}.
\end{equation}

\subsection{The laws of thermodynamics}
Before deriving a Markovian description, we discuss the laws of thermodynamics as they hold for the general scenario.

\subsubsection{The 0th law}
For a large environment in thermal equilibrium (i.e., described by a single inverse temperature $\beta$ and chemical potential $\mu$) and a time-independent system Hamiltonian, the reduced state of the system tends to \cite{geva:2000,subasi:2012}
\begin{equation}
\label{eq:0thtot}
\hat{\rho}_{\rm S} = {\rm Tr}_{\rm B}\left\{e^{-\beta(\hat{H}_{\rm tot}-\mu\hat{N}_{\rm{tot})}}/Z\right\},\hspace{1.5cm}Z={\rm Tr}\left\{e^{-\beta(\hat{H}_{\rm tot}-\mu\hat{N}_{\rm{tot}})}\right\},
\end{equation}
where ${\rm Tr_{B}}$ denotes the trace over the reservoir degrees of freedom.
In the weak system-bath coupling limit, Eq.~\eqref{eq:0thtot} reduces to the Gibbs state
\begin{equation}
\label{eq:0thgibbs}
\hat{\rho}_{\rm S} = \frac{e^{-\beta(\hat{H}_{\rm S}-\mu\hat{N}_{\rm S})}}{{\rm Tr}\left\{e^{-\beta(\hat{H}_{\rm S}-\mu\hat{N}_{\rm S})}\right\}}.
\end{equation}

\subsubsection{The 1st law}
For future reference, we first introduce the heat current and power provided by bath $\alpha$ as
\begin{equation}
\label{eq:heatcurrpowerbath}
J_\alpha(t) = \partial_t\langle Q_\alpha\rangle,\hspace{2cm}P_\alpha(t) = \partial_t\langle W_\alpha\rangle.
\end{equation}
The first law of thermodynamics is then given by
\begin{equation}
\label{eq:firstlaw}
\partial_t U(t) = P_{\rm ext}(t)+\sum_{\alpha}[J_\alpha(t)+P_\alpha(t)],\hspace{1.5cm}U={\rm Tr}\{[\hat{H}_{\rm S}(t)+\hat{V}]\hat{\rho}_{\rm tot}(t)\}.
\end{equation}
In the weak system-bath coupling limit, the coupling energy can be neglected and $U={\rm Tr}\{\hat{H}_{\rm S}(t)\hat{\rho}_{\rm tot}(t)\}$ reduces to the usual internal energy of the system.

\subsubsection{The 2nd law}
To express the second law of thermodynamics, we impose the initial condition
\begin{equation}
\label{eq:initcon}
\hat{\rho}_{\rm tot}(0) = \hat{\rho}_{\rm S}(0)\bigotimes_\alpha\hat{\tau}_\alpha,\hspace{1.5cm}\hat{\tau}_\alpha = \frac{e^{-\beta(\hat{H}_\alpha-\mu\hat{N}_\alpha)}}{{\rm Tr}\left\{e^{-\beta_\alpha(\hat{H}_{\alpha}-\mu_\alpha\hat{N}_{\alpha})}\right\}},
\end{equation}
i.e., all reservoirs are in local thermal equilibrium and uncorrelated with the system and the other reservoirs. With this initial condition, the second law of thermodynamics can be written as \cite{esposito:2010njp}
\begin{equation}
\label{eq:2ndlawgen}
\Sigma(t)\equiv\Delta S(t) - \sum_{\alpha}\frac{\langle Q_\alpha\rangle}{T_\alpha}=S\left[{\textstyle \hat{\rho}_{\rm tot}(t)||\hat{\rho}_{\rm S}(t)\bigotimes_\alpha\hat{\tau}_{\alpha}}\right]\geq 0,
\end{equation}
where $\hat{\rho}_{\rm S}={\rm Tr}_{\rm B}\{\hat{\rho}_{\rm tot}(t)\}$ denotes the reduced state of the system, $S[\hat{\rho}_1||\hat{\rho_2}]=\textrm{Tr}\{\hat{\rho}_1(\ln\hat{\rho}_1-\ln\hat{\rho}_2) \}$ is the quantum relative entropy (which is by definition positive for positive definite density matrices $\hat{\rho}_1,\hat{\rho}_2$ with unity trace), and $\Delta S$ denotes the change in the system's von Neumann entropy
\begin{equation}
\Delta S(t) = -k_{\rm B}{\rm Tr}\{\hat{\rho}_{\rm S}(t)\ln\hat{\rho}_{\rm S}(t)\}+k_{\rm B}{\rm Tr}\{\hat{\rho}_{\rm S}(0)\ln\hat{\rho}_{\rm S}(0)\}.
\end{equation}
 Equation \eqref{eq:2ndlawgen} has an intuitive interpretation: The entropy production $\Sigma$ denotes the information that is lost when describing system and reservoirs by the state $\hat{\rho}_{\rm S}(t)\bigotimes_\alpha\hat{\tau}_{\alpha}$. In this description, correlations between the system and the reservoirs, as well as any displacement from equilibrium of the reservoirs are neglected \cite{ptaszynski:2019ent}.
 
 While Eq.~\eqref{eq:2ndlawgen} is always non-negative, the same is not necessarily true for the entropy production rate
 \begin{equation}
 \label{eq:sigma}
 \dot{\Sigma}(t) \equiv  \partial_t{\Delta S}(t) -\sum_{\alpha}\frac{J_\alpha(t)}{T_\alpha}.
 \end{equation}
 A negative entropy production rate can be understood as \textit{information backflow} and is a hallmark of non-Markovian behavior \cite{strasberg:2019}. For systems amenable to a Markovian description, we expect $\dot{\Sigma}\geq 0$, as the initial condition in Eq.~\eqref{eq:initcon} can be translated in time without altering the dynamics.

\subsection{Distribution for heat and work exchanged with the reservoirs}
Throughout this work, we are mainly interested in average thermodynamic quantities. In order to obtain a Markovian description of these, it is nevertheless instructive to start with the full probability distribution for the heat and work exchanged with the reservoirs. With the initial condition given in Eq.~\eqref{eq:initcon}, this distribution can be written as
\begin{equation}
\label{eq:probaheatwork}
\begin{aligned}
P(\bm{Q},\bm{W}) = \sum_{\bm{E},\bm{E}',\bm{N},\bm{N}'}& P_t(\bm{E}',\bm{N}'|\bm{E},\bm{N})P_0(\bm{E},\bm{N})\\&\times\prod_{\alpha}\delta(W_\alpha-\mu_\alpha N_\alpha'+\mu_\alpha N_\alpha)\delta(Q_\alpha+W_\alpha-E_\alpha'+E_\alpha),
\end{aligned}
\end{equation}
where we grouped $Q_\alpha$, $W_\alpha$, $E_\alpha$, and $N_\alpha$ into vectors $\bm{Q}$, $\bm{W}$, $\bm{E}$, and $\bm{N}$ and similarly for $E_\alpha'$ and $N_\alpha'$. The joint  probability for each bath $\alpha$ having Energy $E_\alpha$ and particle number $N_\alpha$ at time $t=0$ is given by
\begin{equation}
\label{eq:prob0}
P_0(\bm{E},\bm{N}) = \prod_\alpha \frac{e^{-\beta_\alpha(E_\alpha-\mu_\alpha N_\alpha)}}{{\rm Tr}\left\{e^{-\beta_\alpha(\hat{H}_{\alpha}-\mu_\alpha\hat{N}_{\alpha})}\right\}}.
\end{equation}
The conditional probability that the reservoirs have energies $E_\alpha'$ and particle number $N_\alpha'$ at time $t$, given that their energies and particle numbers where $E_\alpha$ and $N_\alpha$ initially reads
\begin{equation}
\label{eq:condprob}
P_t(\bm{E}',\bm{N}'|\bm{E},\bm{N}) = {\rm Tr}\{\hat{M}\hat{\rho}_{\rm S}(0)\hat{M}^\dagger\},\hspace{1cm}\hat{M} = {\rm Tr}_{\rm B}\{|\bm{E},\bm{N}\rangle\langle \bm{E}',\bm{N}'|\hat{U}(t)\},
\end{equation}
where $\hat{H}_\alpha |\bm{E},\bm{N}\rangle = E_\alpha |\bm{E},\bm{N}\rangle$, $\hat{N}_\alpha |\bm{E},\bm{N}\rangle = N_\alpha |\bm{E},\bm{N}\rangle$, and we introduced the time-evolution operator
\begin{equation}
\label{eq:timev}
\hat{U}(t)=\mathcal{T}e^{-i\int_0^t dt'\hat{H}_{\rm tot}(t')},
\end{equation}
with $\mathcal{T}$ denoting the time-ordering operator. We note that the probability distribution given in \eqref{eq:probaheatwork} can in principle be measured by applying projective measurements on the reservoirs at the initial and final time. Importantly, while such measurements may not be experimentally feasible, they do not influence the dynamics of the system due to the chosen initial condition.

The moment generating function is provided by the Fourier transform of the probability distribution in \eqref{eq:probaheatwork}
\begin{equation}
\label{eq:momgenheatwork}
\Lambda(\bm{\lambda},\bm{\chi})\equiv\int d\bm{Q}d\bm{W} P(\bm{Q},\bm{W})e^{-i\bm{\lambda}\cdot\bm{Q}-i\bm{\chi}\cdot\bm{W}}={\rm Tr}\left\{\hat{\rho}_{\rm tot}(\bm{\lambda},\bm{\chi};t)\right\},
\end{equation}
where we introduced
\begin{equation}
\label{eq:denscountf}
\hat{\rho}_{\rm tot}(\bm{\lambda},\bm{\chi};t)=\hat{U}(\bm{\lambda},\bm{\chi};t)\hat{\rho}_{\rm tot}(0)\hat{U}^\dagger(-\bm{\lambda},-\bm{\chi};t)
\end{equation}
with the modified time-evolution operator
\begin{equation}
\label{eq:timeevcf}
\hat{U}(\bm{\lambda},\bm{\chi};t)=e^{-\frac{i}{2}\sum_{\alpha}[\lambda_{\alpha}(\hat{H}_{\alpha}-\mu_\alpha\hat{N}_\alpha)+\chi_\alpha\mu_\alpha\hat{N}_\alpha]}\hat{U}(t)e^{\frac{i}{2}\sum_{\alpha}[\lambda_{\alpha}(\hat{H}_{\alpha}-\mu_\alpha\hat{N}_\alpha)+\chi_\alpha\mu_\alpha\hat{N}_\alpha]}.
\end{equation}
The quantities $\lambda_{\alpha}$ and $\chi_\alpha$ are known as counting fields and allow us to keep track of work and heat exchanged with the reservoirs (for the use of counting fields in master equations, see Refs.~\cite{schaller:book,esposito:2009rmp}).
From the moment generating function, we recover the average values for heat and work given in Sec.~\ref{sec:scenario} as
\begin{equation}
\label{eq:heatworkmomgen}
\langle Q_\alpha \rangle = i\partial_{\lambda_{\alpha}}\Lambda(\bm{\lambda},0)|_{\bm{\lambda}=0},\hspace{1.5cm}\langle W_\alpha \rangle = i\partial_{\chi_{\alpha}}\Lambda(0,\bm{\chi})|_{\bm{\chi}=0}.
\end{equation}

We note that we do not include the external power in the probability distribution for heat and work as this goes beyond the scope of this paper. Indeed, fluctuations of the external power may not necessarily be described by a positive probability distribution without introducing a measurement scheme that potentially alters the dynamics of the system \cite{perarnau:2017,hofer:2017q,kerremans:2021}. 

\subsection{Born-Markov approximations}
We are now in a position to derive a Markovian master equation, keeping track of the thermodynamic bookkeeping associated to the heat and work exchanged with the bath. Together with a secular approximation, resulting in a \textit{global} master equation, such a procedure has been applied before \cite{gasparinetti:2014,silaev:2014,friedman:2018} (for a similar approach based on path-integrals, see \cite{kilgour:2019}). Here we will make an approximation that is different from the secular approximation, allowing for treating (near) degeneracies while still ensuring thermodynamic consistency. We follow standard procedure \cite{breuer:book,schaller:book} and write the system-bath coupling as
\begin{equation}
\label{eq:sysbath}
\hat{V} =\sum_{\alpha, k} \hat{S}_{\alpha,k}\hat{B}_{\alpha,k},
\end{equation}
where the operators $\hat{S}_{\alpha,k}$ ($\hat{B}_{\alpha,k}$) act only on the system (reservoir). We note that we do not assume that these operators are Hermitian. To determine the jump operators that will enter the Markovian master equation, we need the Fourier coefficients of the operators $\hat{S}_{\alpha,k}$ in the interaction picture, i.e.,
\begin{equation}
\label{eq:skfour}
\hat{U}^\dagger_{\rm S}(t)\hat{S}_{\alpha,k}\hat{U}_{\rm S}(t) = \sum_j e^{-i\omega_jt}\hat{S}_{\alpha,k;j},\hspace{1.5cm}\hat{U}_{\rm S}(t) = \mathcal{T}e^{-i\int_0^t dt'\hat{H}_{\rm S}(t')}.
\end{equation}
For a time-independent Hamiltonian, the frequencies $\omega_j$ denote the energy gaps in the Hamiltonian and the operators $\hat{S}_{\alpha,k;j}$ are ladder operators that induce transitions between the corresponding states. For a periodic Hamiltonian with period $t_{\rm p}=2\pi/\varpi$, the $\hat{S}_{\alpha,k;j}$ are ladder operators of an averaged Hamiltonian defined by
\begin{equation}
\label{eq:avham}
\hat{U}_{\rm S}(t_{\rm p}) = e^{-i\hat{H}_{\rm av}t_{\rm p}}.
\end{equation}
The frequencies then fulfill $\omega_j=\nu_j+l_j\varpi$, where $\nu_j$ denotes an energy gap of $\hat{H}_{\rm av}$ and $l_j$ denotes an integer, corresponding to the exchange of photons with the driving field \cite{kohler:1997,levy:2012pre,gasparinetti:2014}.

Standard application of Born-Markov approximations then results in the Redfield equation including counting fields (in the interaction picture)
\begin{equation}
\label{eq:redfield}
\begin{aligned}
\partial_t\hat{\rho}_{\rm S}(t)&=-\sum_{\alpha,k,k',j,j'}e^{i(\omega_j-\omega_{j'})t}\int_{0}^{\infty}ds\mathcal{I}(s,t),\\\mathcal{I}(s,t)&=e^{i\omega_{j'}s}C_{k,k'}^\alpha(s)\hat{S}^\dagger_{\alpha,k;j}\hat{S}_{\alpha,k';j'}\hat{\rho}_{\rm S}(t)+e^{-i\omega_{j}s}C_{k,k'}^\alpha(-s)\hat{\rho}_{\rm S}(t)\hat{S}^\dagger_{\alpha,k;j}\hat{S}_{\alpha,k';j'}\\&\hspace{-1 cm}-e^{i\mu_\alpha n_{\alpha,k}(\lambda_\alpha-\chi_\alpha)}\left[e^{i\omega_{j'}s}C_{k,k'}^\alpha(s+\lambda_\alpha)+e^{-i\omega_{j}s}C_{k,k'}^\alpha(-s+\lambda_\alpha)\right]\hat{S}_{\alpha,k';j'}\hat{\rho}_{\rm S}(t)\hat{S}^\dagger_{\alpha,k;j},
\end{aligned}
\end{equation}
where we suppressed the counting field dependence of the density matrix for ease of notation. We note that the term \textit{Redfield equation} sometimes refers to the non-Markovian equation that is obtained before taking the time-integration to infinity \cite{breuer:book} while we include this limit here.
Here we used global particle conservation, which ensures
\begin{equation}
\label{eq:partcons}
[\hat{S}_{\alpha,k;j},\hat{N}_{\rm S}]=n_{\alpha,k}\hat{S}_{\alpha,k;j},\hspace{1cm}[\hat{S}_{\alpha,k';j'},\hat{N}_{\rm S}]=n_{\alpha,k}\hat{S}_{\alpha,k';j'},
\end{equation}
for all pairs $k$ and $k'$ that appear together in Eq.~\eqref{eq:redfield}, i.e., $\hat{S}_{\alpha,k;j}$ and $\hat{S}_{\alpha,k';j'}$ change the particle number by the same amount, such that no superpositions of particle numbers are created. We further introduced the bath correlation functions
\begin{equation}
\label{eq:bathcorr}
C_{k,k'}^\alpha(s)={\rm Tr}\left\{e^{is\hat{H}_\alpha}\hat{B}^\dagger_{\alpha,k} e^{-is\hat{H}_\alpha}\hat{B}_{\alpha,k'}\hat{\tau}_{\alpha}\right\}.
\end{equation}
These correlation functions define a bath correlation time $\tau_{\rm B}$ by their characteristic decay time. The Markov approximation is generally justified when the bath correlation time is much shorter than the characteristic time-scale over which $\hat{\rho}_{\rm S}$ changes in the interaction picture, $\tau_{\rm S}$. The time $\tau_{\rm S}$ describes the relaxation time of the system and is determined by the inverse of the system-bath coupling. Note however that the counting fields $\lambda_{\alpha}$ enter the argument of the bath correlation times. For the counting-field dependent density matrix, the Markov approximation is only justified if $C_{k,k'}^\alpha(\pm\tau+\lambda_\alpha)\simeq 0$ for $\tau\gtrsim \tau_{\rm S}$. This implies the regime of validity
\begin{equation}
\label{eq:regimvalm1}
\tau_{\rm B}\ll\tau_{\rm S},\hspace{1.5cm}|\lambda_{\alpha}|\ll\tau_{\rm S}.
\end{equation}
The last inequality has a very important consequence. It implies that only the low frequency components of the heat distribution are to be trusted. Due to the uncertainty principle between Fourier conjugate variables \cite{folland:1997}, this implies that energy-differences in heat of the order of $1/\tau_{\rm S}$ cannot be resolved. Thus, whenever a Markovian description is employed, the heat exchanged with the reservoirs suffers from a limited energy-resolution. With this in mind, it is no surprise that Markovian descriptions may result in thermodynamic inconsistencies. There is however a straightforward solution to the problem: as our resolution of heat is finite, we may change the definition of heat such that it fulfills the laws of thermodynamics while remaining the same within our limited resolution.

\subsection{Frequency grouping for positivity}
\label{sec:grouping}
It is well known that the Redfield equation does not preserve positivity of the density matrix and can thus result in negative probabilities. Multiple schemes have been put forward to achieve positivity. Here we introduce a novel scheme that has the same regime of validity as the Born-Markov assumptions and thus goes beyond the regime of validity of the secular approximation. As we show in the next subsection, our approach allows for a thermodynamically consistent formulation.

Equation \eqref{eq:regimvalm1} ensures that for any two transition frequencies we either have $|\omega_j-\omega_{j'}|\ll 1/\tau_{\rm B}$ or $|\omega_j-\omega_{j'}|\gg 1/\tau_{\rm S}$. Indeed, both of these conditions may be fulfilled simultaneously as $\tau_{\rm B}\ll\tau_{\rm S}$. We may thus group the transition frequencies into sets $x_q$, such that the first (second) inequality holds if the transition frequencies are in the same (different) set, that is
\begin{equation}
\label{eq:setsind}
\begin{aligned}
&|\omega_j-\omega_{j'}|\ll1/\tau_{\rm B}\hspace{.5cm}{\rm for}\,\, \omega_j\in x_q,\, \omega_{j'}\in x_{q'} \,\,{\rm with}\,\, q=q',\\
&|\omega_j-\omega_{j'}|\gg1/\tau_{\rm S}\hspace{.5cm}{\rm for}\,\,\omega_j\in x_q,\, \omega_{j'}\in x_{q'} \,\,{\rm with}\,\, q\neq q'.
\end{aligned}
\end{equation}
It is important to note that this procedure may not always work, for instance if the $\omega_j$ form a continuum. However, in small quantum systems, where the number of transition frequencies is finite, this procedure is expected to work. It is particularly well suited for thermal machines that consist of weakly coupled sub-units. These naturally exhibit sets of near-degenerate transition frequencies  (see the examples below).

In the spirit of the secular approximation, we then drop terms in Eq.~\eqref{eq:redfield} where $\omega_j$ and $\omega_{j'}$ belong to different sets. This is justified as the corresponding terms exhibit fast oscillations that average to zero. For transition frequencies in the same set $x_q$, we follow the spirit of the Markov approximation and replace
\begin{equation}
\label{eq:2ndmarkov}
e^{i\omega_{j} s},e^{i\omega_{j'} s}\rightarrow e^{i\omega_{q} s},\hspace{1.5cm}|\omega_q-\omega_j|\ll 1/\tau_{\rm B}\,\,\forall\,\omega_j\in x_q,
\end{equation}
within $\mathcal{I}(s)$ in Eq.~(\ref{eq:redfield}), while keeping the frequency differences in the prefactor $e^{i(\omega_j-\omega_{j'})t}$ governing the coherent dynamics of the system.
For each set, we thus choose a \textit{set frequency} $\omega_q$. All transition frequencies $\omega_j$ in the set $x_q$ are then replaced by the set frequency, because they are virtually indistinguishable over the time-scale over which the integrand in Eq.~\eqref{eq:redfield} is finite.

This approximation results in the Lindblad master equation
\begin{equation}
\label{eq:mastereq}
\partial_t\hat{\rho}_{\rm S}(\bm{\lambda},\bm{\chi};t)=-i[\hat{H}_{\rm LS},\hat{\rho}_{\rm S}(\bm{\lambda},\bm{\chi};t)]+\sum_{\alpha}\tilde{\mathcal{L}}_\alpha^{\chi_{\alpha},\lambda_{\alpha}}\hat{\rho}_{\rm S}(\bm{\lambda},\bm{\chi};t),
\end{equation}
with
\begin{equation}
\label{eq:bathops}
\tilde{\mathcal{L}}_\alpha^{\chi_{\alpha},\lambda_{\alpha}}\hat{\rho}=\sum_{k;q}\Gamma_{k}^\alpha(\omega_q)\bigg[e^{-i\lambda_\alpha\omega_q-i(\chi_\alpha-\lambda_{\alpha}) \mu_\alpha n_{\alpha,k}}\hat{S}_{\alpha,k;q}(t)\hat{\rho}\hat{S}^\dagger_{\alpha,k;q}(t)-\frac{1}{2}\left\{\hat{S}^\dagger_{\alpha,k;q}(t)\hat{S}_{\alpha,k;q}(t),\hat{\rho}\right\}\bigg],
\end{equation}
where the tilde denotes the interaction picture and we introduced the jump operators
\begin{equation}
\label{eq:jumps}
 \hat{S}_{\alpha,k;q}(t)=\sum_{\{j |\omega_j\in x_q\}}e^{-i\omega_jt}\hat{S}_{\alpha,k;j},
\end{equation}
and the Lamb-shift Hamiltonian
\begin{equation}
\label{eq:lamb}
\hat{H}_{\rm LS}=\sum_{\alpha,k;q}\Delta_{k}^\alpha(\omega_q)\hat{S}^\dagger_{\alpha,k;q}(t)\hat{S}_{\alpha,k;q}(t),
\end{equation}
as well as the quantities
\begin{equation}
\label{eq:gamma}
\Gamma_{k}^\alpha(\omega)=\int_{-\infty}^{\infty}ds e^{i\omega s}C_{k,k}^\alpha(s)\hspace{1cm}\Delta_{k}^\alpha(\omega)=-\frac{i}{2}\int_{-\infty}^{\infty}ds e^{i\omega s}{\rm sign}(s)C_{k,k}^\alpha(s).
\end{equation}
For simplicity, we assumed $C_{k,k'}^\alpha\propto \delta_{k,k'}$ to derive Eq.~\eqref{eq:mastereq}. We note that relaxing this assumption is straightforward.

In the absence of counting fields, we may use the Kubo-Martin-Schwinger condition \cite{breuer:book} (which is slightly complicated by our definition of the bath-correlation functions) to write the master equation as
\begin{equation}
\label{eq:mastereqncfin}
\partial_t\hat{\rho}_{\rm S}=-i[\hat{H}_{\rm LS}(t),\hat{\rho}_{\rm S}]+\sum_{\alpha}\tilde{\mathcal{L}}_{\alpha}\hat{\rho}_{\rm S},
\end{equation}
with
\begin{equation}
\label{eq:bathdiagin}
\tilde{\mathcal{L}}_{\alpha}=\sum_{\{q|\omega_q>0\}}\sum_{k} \Gamma_k^\alpha(\omega_q)\left\{\mathcal{D}[\hat{S}_{\alpha,k;q}(t)]+e^{-\beta_\alpha(\omega_q-\mu_\alpha n_{\alpha,k})}\mathcal{D}[\hat{S}_{\alpha,k;q}^\dagger(t)]\right\}.
\end{equation}
For a time-independent Hamiltonian, the master equation in the Schrödinger picture is given by 
\begin{equation}
\label{eq:mastereqncf}
\partial_t\hat{\rho}_{\rm S}=-i[\hat{H}_{\rm S}+\hat{H}_{\rm LS},\hat{\rho}_{\rm S}]+\sum_{\alpha}\mathcal{L}_{\alpha}\hat{\rho}_{\rm S},
\end{equation}
with
\begin{equation}
\label{eq:bathdiag}
\mathcal{L}_{\alpha}=\sum_{\{q|\omega_q>0\}}\sum_{k} \Gamma_k^\alpha(\omega_q)\left\{\mathcal{D}[\hat{S}_{\alpha,k;q}]+e^{-\beta_\alpha(\omega_q-\mu_\alpha n_{\alpha,k})}\mathcal{D}[\hat{S}_{\alpha,k;q}^\dagger]\right\}.
\end{equation}	
where $\hat{S}_{\alpha,k;q}\equiv \hat{S}_{\alpha,k;q}(0)$ and $\hat{H}_{\rm LS}\equiv\hat{H}_{\rm LS}(0)$. For time-dependent Hamiltonians, one has to be more careful because in general $\hat{S}_{\alpha,k;q}(t)\neq\hat{U}^\dagger_{\rm S}(t) \hat{S}_{\alpha,k;q}\hat{U}_{\rm S}(t)$, see also \ref{app:tdtime}.

There are two simple limits for the master equation in Eq.~\eqref{eq:mastereqncf}. If all pairs of transition frequencies fulfill $|\omega_j-\omega_{j'}|\gg1/\tau_{\rm S}$, then each transition frequency may be associated to a separate set $x_j$. We then recover the secular approximation where $\mathcal{D}[\hat{S}_{\alpha,k;q}(t)]=\mathcal{D}[\hat{S}_{\alpha,k;j}]$ (the time-dependence of the jump operators drops out in the interaction picture). If all pairs of transition frequencies fulfill $|\omega_j-\omega_{j'}|\ll1/\tau_{\rm B}$, then all transition frequencies may be grouped into a single set $x_0$ such that $\hat{S}_{\alpha,k;q}(t)=\hat{U}_{\rm S}^\dagger(t)\hat{S}_{\alpha,k}\hat{U}_{\rm S}(t)$ (the time-dependence of the jump operators drops out upon returning to the Schrödinger picture). We then recover a local master equation, which has the appeal that the Hamiltonian does not need to be diagonalized in order to identify the jump operators. This local approach differs from the conventional local approach as only a single frequency enters the bath distribution function.

Importantly, the counting fields $\lambda_{\alpha}$ have an influence on the regime of validity of Eq.~\eqref{eq:mastereq}, just as for the Markov approximation. Indeed, for the approximation to be valid, $|\omega_q-\omega_{j}|$ does not only have to be much smaller than $1/\tau_{\rm B}$ but also has to be much smaller than $1/|\lambda_{\alpha}|$ [in complete analogy to Eq.~\eqref{eq:regimvalm1}]. Just like the Markov approximation, this frequency-grouping thus limits the energy-resolution for heat exchanged with the reservoirs. As a consequence, energy changes of the order of $|\omega_j-\omega_{j'}|$, where both frequencies are within the same set, can no longer be resolved. This finite resolution in energy may result in the false conclusion that particles change their energy when traversing the system. In the conventional local approach, this results in thermodynamic inconsistencies when using standard definitions for heat, see Sec.~\ref{sec:compare}.

We note that dropping terms that involve frequencies from different sets is reminiscent of the partial secular approximation developed in Refs.~\cite{schaller:2008,majenz:2013,tscherbul:2015,jeske:2015,seah:2018}. Here, as well as in Ref.~\cite{trushechkin:2021}, an additional coarse-graining in energy (replacing frequencies with set-frequencies) results in a master equation in GKLS form.

\subsection{Two Hamiltonians for thermodynamic consistency}
\label{sec:hamtd}
We now turn to the question of how to utilize the finite energy-resolution to ensure thermodynamic consistency. To do this, we introduce a second Hamiltonian, $\hat{H}_{\rm TD}$, that provides the correct thermodynamic bookkeeping under the approximations that resulted in our master equation. For simplicity, we consider here a time-independent system Hamiltonian. The time-dependent case is slightly more complicated and treated in \ref{app:tdtime}. Due to the frequency grouping outlined in the last subsection, all frequencies in the set $x_q$ are effectively replaced by the frequency $\omega_q$ from the point of view of the reservoirs. To ensure a consistent thermodynamic bookkeeping, the Hamiltonian $\hat{H}_{\rm TD}$ needs to fulfill
\begin{equation}
\label{eq:commhtd}
[\hat{S}_{\alpha,k;j},\hat{H}_{\rm TD}]=\omega_q\hat{S}_{\alpha,k;j},
\end{equation}
for all frequencies $\omega_j\in x_q$. To fulfill this, the thermodynamic Hamiltonian can be obtained from $\hat{H}_{\rm S}$ by changing its eigenvalues such that all frequencies $\omega_j\rightarrow\omega_q$ for $\omega_j\in x_q$. Such a rescaling is expected to always be possible if the frequency grouping is possible as discussed above.

With the thermodynamic Hamiltonian at hand, we define the internal energy of the system as
\begin{equation}
\label{eq:inten}
U = {\rm Tr}\{\hat{H}_{\rm TD} \hat{\rho}\}.
\end{equation}
Furthermore, from Eqs.~\eqref{eq:mastereq},\eqref{eq:heatcurrpowerbath}, and \eqref{eq:heatworkmomgen} we find that the heat current and power provided by bath $\alpha$ can be cast into
\begin{equation}
\label{eq:heatpowbathtd}
J_\alpha = {\rm Tr}\{(\hat{H}_{\rm TD}-\mu_\alpha\hat{N}_{\rm S})\mathcal{L}_\alpha \hat{\rho}_{\rm S}\},\hspace{1.5cm}P_\alpha = \mu_\alpha{\rm Tr}\{\hat{N}_{\rm S}\mathcal{L}_\alpha \hat{\rho}_{\rm S}\},
\end{equation}
where $\mathcal{L}_\alpha$ is defined in Eq.~\eqref{eq:bathdiag}. We note that while $\hat{H}_{\rm TD}$ determines the thermodynamic bookkeeping, it is still $\hat{H}_{\rm S}$ that determines the kinetics, i.e., enters the master equation in Eq.~\eqref{eq:mastereq}. For future reference, we also introduce the total power produced by the system as
\begin{equation}
\label{eq:powerout}
P_{\rm S}(t) = -\sum_{\alpha}P_\alpha(t) -P_{\rm ext}(t).
\end{equation}
This will be the quantity of interest when we consider heat engines.

\subsubsection{The 0th law}
Using Eq.~\eqref{eq:commhtd}, it is straightforward to show that
\begin{equation}
\label{eq:dissfix}
\mathcal{L}_{\alpha} e^{-\beta_\alpha(\hat{H}_{\rm TD}-\mu_\alpha\hat{N}_{\rm S})}=0.
\end{equation}
as well as $[\hat{H}_{\rm TD},\hat{H}_{\rm LS}]=0$. Furthermore, as $\hat{H}_{\rm TD}$ is obtained by changing only the eigenvalues of $\hat{H}_{\rm S}$, these two Hamiltonians also commute. When all reservoirs have the same inverse temperature $\beta$ and chemical potential $\mu$, it then follows that the Gibbs state with respect to the thermodynamic Hamiltonian,
\begin{equation}
\label{eq:gibbs}
\hat{\rho}_{\rm G}(\beta,\mu)=\frac{e^{-\beta(\hat{H}_{\rm TD}-\mu\hat{N}_{\rm S})}}{{\rm Tr}\{e^{-\beta(\hat{H}_{\rm TD}-\mu\hat{N}_{\rm S})}\}},
\end{equation}
is the steady state of Eq.~\eqref{eq:mastereqncf}. Compared to Eq.~\eqref{eq:0thgibbs}, this state neglects the system-bath coupling, as well as the differences between $\hat{H}_{\rm S}$ and $\hat{H}_{\rm TD}$. This is consistent with our approximations, which imply that these differences cannot be resolved within our Markovian treatment.

\subsubsection{The first law}
Using $[\hat{H}_{\rm TD},\hat{H}_{\rm S}+\hat{H}_{\rm LS}]=0$, the first law of thermodynamics follows directly from Eqs.~\eqref{eq:inten} and \eqref{eq:heatpowbathtd} and reads
\begin{equation}
\label{eq:firstlawmast}
\partial_t U = \sum_{\alpha}(P_\alpha+J_\alpha).
\end{equation}

\subsubsection{The second law}
The entropy production rate can be written as
\begin{equation}
\label{eq:entprodratemast}
\begin{aligned}
\dot{\Sigma}/k_{\rm B} &= -\partial_t{\rm Tr}\{\hat{\rho}_{\rm S}\ln\hat{\rho}_{\rm S}\}-\sum_{\alpha}\beta_\alpha J_\alpha\\&=-\sum_{\alpha}{\rm Tr}\{(\mathcal{L}_\alpha\hat{\rho}_{\rm S})[\ln\hat{\rho}_{\rm S}-\ln\hat{\rho}_{\rm G}(\beta_\alpha,\mu_\alpha)]\}\geq 0.
\end{aligned}
\end{equation}
Here we used Eqs.~\eqref{eq:mastereqncf}, \eqref{eq:heatpowbathtd}, and the last inequality is known as Spohns inequality \cite{spohn:1978,spohn:1978b} and relies on $\hat{\rho}_{\rm G}(\beta_\alpha,\mu_\alpha)$ being a fixed point of $\mathcal{L}_\alpha$, cf.~Eq.~\eqref{eq:dissfix}.

The master equation in Eq.~\eqref{eq:mastereqncf}, together with the thermodynamic bookkeeping introduced in Eq.~\eqref{eq:inten} and \eqref{eq:heatpowbathtd}, thus provide a thermodynamically consistent description. We stress that the only assumption that went into Eq.~\eqref{eq:mastereqncf} is the one that justifies the standard Born-Markov approximations, i.e., $\tau_{\rm B}\ll\tau_{\rm S}$. However, when considering the full probability distribution for heat and work exchanged with the reservoirs, cf.~\eqref{eq:mastereq}, then we further have restrictions on the counting fields $\lambda_{\alpha}$ which are the Fourier transform variables of heat. These restrictions imply that we lose energy-resolution on the scale of $1/\tau_{\rm S}$ as well as $|\omega_j-\omega_j'|$ for pairs of frequencies that are close to each other (and grouped into one set). Thermodynamic consistency is obtained by using an appropriate definition of heat that is consistent with our limited energy-resolution (i.e., we may freely add terms of order $|\omega_j-\omega_j'|$ to heat, as our results are not to be trusted to this order in the first place).

\section{Fermionic heat engine}
\label{sec:fermions}
\subsection{System}
\begin{figure}
	\centering
	\includegraphics[width=.75\textwidth]{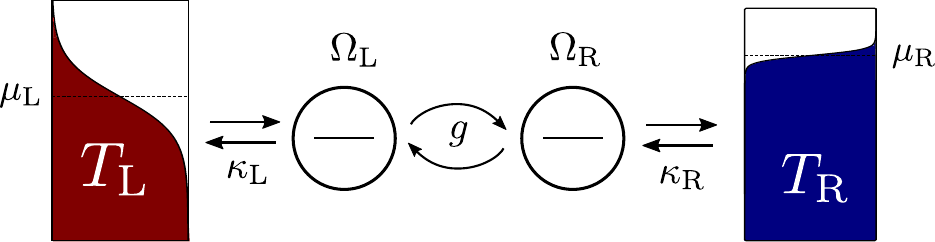}
	\caption{\label{fig:qdots} System of coupled quantum dots acting as a heat engine. Two quantum dots with on-site energies $\Omega_{\rm L}$ and $\Omega_{\rm R}$ are coupled to each other with coupling strength $g$ and to a thermal reservoir of temperatures $T_{\rm L}$, $T_{\rm R}$ and chemical potentials $\mu_{\rm L}$, $\mu_{\rm R}$ (with coupling strength $\kappa_{\rm h}$ and $\kappa_{\rm c}$) respectively.}
\end{figure}

In this section, we illustrate the derivation of our master equation for a simple but non-trivial example of non-interacting electrons. Here, as well as for all time-independent examples, we provide all equations in the Schrödinger picture. The system under consideration is sketched in Fig.~\ref{fig:qdots} and consists of two coupled single-level quantum dots
\begin{equation}
\label{eq:hamferms}
\hat{H}_{\rm S} = \Omega_{\rm L}\hat{d}_{\rm L}^\dagger\hat{d}_{\rm L} + \Omega_{\rm R}\hat{d}_{\rm R}^\dagger\hat{d}_{\rm R}
+ g(\hat{d}_{\rm L}^\dagger\hat{d}_{\rm R}+\hat{d}_{\rm R}^\dagger\hat{d}_{\rm L}),
\end{equation}
with the standard anti-commutation relations
\begin{equation}
\label{eq:anticommrels}
\{\hat{d}_\alpha,\hat{d}^\dagger_\beta\}=\delta_{\alpha,\beta},\hspace{1.5cm}\{\hat{d}_\alpha,\hat{d}_\beta\}=0.
\end{equation}
In Eq.~\eqref{eq:hamferms}, $\Omega_\alpha$ denote the on-site energies and $g$ the coupling strength. We note that this system is a fermionic version of the system considered in Sec.~\ref{sec:compare} and thus shares many of its properties. Due to the finite chemical potential of the reservoirs, it can however feature as a heat engine, leveraging a heat current to drive a particle current against a chemical potential bias. The reservoirs and their coupling to the system are described by
\begin{equation}
\label{eq:rescoupferm}
\hat{H}_\alpha = \sum_{l}\varepsilon_{\alpha,l}\hat{c}^\dagger_{\alpha,l}\hat{c}_{\alpha,l},\hspace{1.5cm}\hat{V}_{\alpha} = \sum_{l}t_{\alpha,l}(\hat{d}_\alpha^\dagger \hat{c}_{\alpha,l}+\hat{c}_{\alpha,l}^\dagger\hat{d}_\alpha),
\end{equation}
with $\alpha={\rm L, R}$.
From Eqs.~\eqref{eq:sysbath} and \eqref{eq:rescoupferm}, we identify the following system and bath operators
\begin{equation}
\label{eq:sysopferm}
\hat{S}_{\alpha,1}=\hat{d}_\alpha^\dagger, \hspace{2.5cm}\hat{S}_{\alpha,-1}=\hat{d}_\alpha,
\end{equation}
 and
\begin{equation}
\label{eq:bathopferm}
\hat{B}_{\alpha,1}=\sum_lt_{\alpha,l}\hat{c}_{\alpha,l},\hspace{1.5cm}\hat{B}_{\alpha,-1}=\sum_lt_{\alpha,l}\hat{c}^\dagger_{\alpha,l}.
\end{equation}
To derive the master equation and judge its validity, the bath correlation functions need to be inspected. We defer this discussion to \ref{app:bathcorrferm}, and simply state the conclusion that the Born Markov approximations are valid whenever
\begin{equation}
\label{eq:bornmarkvalferm}
\kappa_\alpha\ll\max\{k_{\rm B}T_\beta,\,|\omega_j-\mu_\beta|\},
\end{equation}
where the inequality should hold for all choices of $\alpha$, $\beta$, and $j$ and we assumed an energy-independent bath spectral density. Physically, Eq.~\eqref{eq:bornmarkvalferm} ensures that the rates
\begin{equation}
\label{eq:gammasferm}
\Gamma_{1}^\alpha(-\omega)=\kappa_\alpha n_{\rm F}^\alpha(\omega),\hspace{1cm}\Gamma_{-1}^\alpha(\omega) = \kappa_{\alpha} [1-n_{\rm F}^\alpha(\omega)],
\end{equation}
with the Fermi-Dirac distribution
\begin{equation}
\label{eq:fdd}
n_{\rm F}^\alpha(\omega) = \frac{1}{e^{\frac{\omega-\mu_\alpha}{k_{\rm B} T_\alpha}}+1},
\end{equation}
is flat over the energy scale $\kappa_\alpha$. At high temperatures, this is ensured because the Fermi-Dirac distribution becomes flat. For large $|\omega_j-\mu_\beta|$, the eigenenergies of the double quantum dot lie far away from the chemical potential, where the Fermi-Dirac distribution is flat and takes on the value zero or one.

To identify the transition frequencies and the jump operators, we require the Fourier coefficients of the system operators given in Eq.~\eqref{eq:sysopferm}
\begin{equation}
\label{eq:fermfour}
\begin{aligned}
&e^{i\hat{H}_{\rm S}t}\hat{d}_{\rm R}e^{-i\hat{H}_{\rm S}t} = e^{-i\Omega_- t}\cos(\theta/2)\hat{d}_- +e^{-i\Omega_+ t}\sin(\theta/2)\hat{d}_+,\\&e^{i\hat{H}_{\rm S}t}\hat{d}_{\rm L}e^{-i\hat{H}_{\rm S}t} = -e^{-i\Omega_- t}\sin(\theta/2)\hat{d}_- +e^{-i\Omega_+ t}\cos(\theta/2)\hat{d}_+,
\end{aligned}
\end{equation}
where we used a similar notation to Sec.~\ref{sec:compare}, i.e., the diagonlized Hamiltonian reads
\begin{equation}
\label{eq:hsfermd}
\hat{H}_{\rm S} = \Omega_+\hat{d}^\dagger_+\hat{d}_++\Omega_-\hat{d}^\dagger_-\hat{d}_-,
\end{equation}
where $\Omega_{\pm} = \bar{\Omega}\pm \sqrt{\Delta^2+g^2}$, with $\bar{\Omega}=(\Omega_{\rm R}+\Omega_{\rm L})/2$, $\Delta = (\Omega_{\rm R}-\Omega_{\rm L})/2$, and $\cos(\theta)= \Delta/\sqrt{\Delta^2+g^2}$. Comparing Eqs.~\eqref{eq:fermfour}, and their Hermitian conjugates, to Eq.~\eqref{eq:skfour}, we find the transition frequencies
\begin{equation}
\label{eq:transferm}
\{\omega_j\} = \{\Omega_+,\,\Omega_-,\,-\Omega_+,\,-\Omega_-\},
\end{equation}
and the corresponding jump operators
\begin{equation}
\label{eq:sjferm}
\begin{aligned}
&\{\hat{S}_{{\rm R},-1;j}\} = \{\sin(\theta/2)\hat{d}_+,\,\cos(\theta/2)\hat{d}_-,\,0,\,0\},\\
&\{\hat{S}_{{\rm L},-1;j}\} = \{\cos(\theta/2)\hat{d}_+,\,-\sin(\theta/2)\hat{d}_-,\,0,\,0\},\\
&\{\hat{S}_{{\rm R},1;j}\} = \{0,\,0,\,\sin(\theta/2)\hat{d}^\dagger_+,\,\cos(\theta/2)\hat{d}^\dagger_-\},\\
&\{\hat{S}_{{\rm L},1;j}\} = \{0,\,0,\,\cos(\theta/2)\hat{d}^\dagger_+,\,-\sin(\theta/2)\hat{d}^\dagger_-\}.
\end{aligned}
\end{equation}
We note that the frequencies $\Omega_{\pm}$ only feature in the jump operators with $k=-1$ (corresponding to electrons leaving the system), while the frequencies $-\Omega_{\pm}$ only feature in the jump operators with $k=1$ (corresponding to electrons entering the system). This implies that we may consider only the frequencies $\Omega_\pm$ when deciding how to group the transition frequencies into sets. We may group $\Omega_+$ and $\Omega_-$ in the same set, or we may assign them to different sets. For the frequencies $-\Omega_{\pm}$, we then choose an equivalent grouping.

\subsection{Global approach}
Grouping $\Omega_{\pm}$ into different sets is justified as long as their difference $2\sqrt{\Delta^2+g^2}$ is much larger than $\max\{\kappa_{\rm L},\kappa_{\rm R}\}$, which corresponds to $1/\tau_{\rm S}$. This grouping results in a different set for each frequency, such that $\{\omega_q\}=\{\omega_j\}$ and $\{\hat{S}_{\alpha,k;q}\}=\{\hat{S}_{\alpha,k;j}\}$. Having identified the set frequencies and jump operators, we may then use Eq.~\eqref{eq:bathdiag} to obtain the well-known dissipator in the secular approximation
\begin{equation}
\label{eq:dissglobalferm}
\mathcal{L}^{\rm g}_\alpha =\sum_{\sigma=\pm} \kappa^\sigma_\alpha\left\{n_{\rm F}^\alpha(\Omega_\sigma)\mathcal{D}[\hat{d}_\sigma^\dagger]+[1-n_{\rm F}^\alpha(\Omega_\sigma)]\mathcal{D}[\hat{d}_\sigma]\right\},
\end{equation}
with the coupling strengths
\begin{equation}
\label{eq:kappagsferm}
\kappa_{\rm L}^+ = \kappa_{\rm L} \cos^2(\theta/2),\hspace{.25cm}\kappa_{\rm L}^- = \kappa_{\rm L} \sin^2(\theta/2),\hspace{.25cm}\kappa_{\rm R}^+ = \kappa_{\rm R} \sin^2(\theta/2),\hspace{.25cm}\kappa_{\rm R}^- = \kappa_{\rm R} \cos^2(\theta/2).
\end{equation}
As the jump operators in Eq.~\eqref{eq:sjferm} are ladder operators of $\hat{H}_{\rm S}$, no rescaling is required for obtaining the thermodynamic Hamiltonian and we find $\hat{H}_{\rm TD}=\hat{H}_{\rm S}$, resulting in the standard definition for heat currents [cf.~Eq.~\eqref{eq:heatpowbathtd}].

\subsection{Local approach}
Alternatively, we may group $\Omega_+$ and $\Omega_-$ in the same set. In this case, we need to choose a set frequency. The average value $\bar{\Omega}$ is a natural choice. Using the same arguments that result in Eq.~\eqref{eq:bornmarkvalferm}, this grouping is justified as long as $2\sqrt{\Delta^2+g^2}\ll\max\{k_{\rm B}T_\beta,\,|\omega_j-\mu_\beta|\}$. In this case, we end up with two sets with the set frequencies
\begin{equation}
\label{eq:setfrlferm}
\{\omega_q\} = \{\bar{\Omega},\,-\bar{\Omega}\},
\end{equation}
and the jump operators
\begin{equation}
\label{eq:sqlferm}
\begin{aligned}
&\{\hat{S}_{{\rm R},-1;q}\} = \{\hat{d}_{\rm R},\,0\},\\
&\{\hat{S}_{{\rm L},-1;q}\} = \{\hat{d}_{\rm L},\,0\},\\
&\{\hat{S}_{{\rm R},1;q}\} = \{0,\,\hat{d}_{\rm R}^\dagger\},\\
&\{\hat{S}_{{\rm L},1;q}\} = \{0,\,\hat{d}_{\rm L}^\dagger\},
\end{aligned}
\end{equation}
which are simply obtained by adding the jump operators in Eq.~\eqref{eq:sjferm} which belong to the same set. Inserting these quantities into Eq.~\eqref{eq:bathops} results in 
the local approach with the dissipator
\begin{equation}
\label{eq:disslocalferm}
\mathcal{L}^{\rm l}_\alpha = \kappa_\alpha\left\{n_{\rm F}^\alpha(\bar{\Omega})\mathcal{D}[\hat{d}_\alpha^\dagger]+[1-n_{\rm F}^\alpha(\bar{\Omega})]\mathcal{D}[\hat{d}_\alpha]\right\},
\end{equation}
for $\alpha={\rm L,R}$.

The thermodynamic Hamiltonian is obtained by rescaling all transition frequencies to their set frequency. In the present case this rescaling is obtained by $\Omega_{\pm}\rightarrow\bar{\Omega}$ resulting in
\begin{equation}
\label{eq:hamtdferm}
\hat{H}_{\rm TD} =\bar{\Omega}(\hat{d}_{+}^\dagger\hat{d}_{+} + \hat{d}_{-}^\dagger\hat{d}_{-})= \bar{\Omega}(\hat{d}_{\rm L}^\dagger\hat{d}_{\rm L} + \hat{d}_{\rm R}^\dagger\hat{d}_{\rm R}).
\end{equation}

\subsection{Results}
\begin{figure}
	\includegraphics[width=\textwidth]{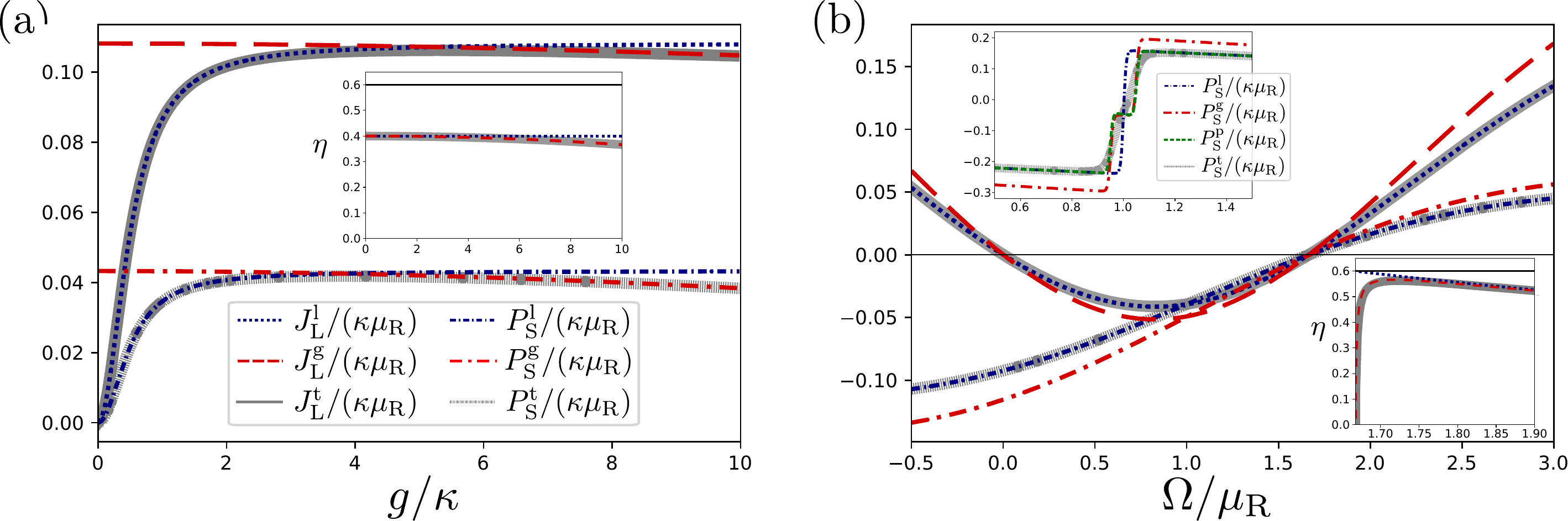}
	\caption{\label{fig:heatengineferm} Steady state heat current $J_{\rm L}^j$ from the hot bath and output power $P_{\rm S}^j$  as a function of (a) the coupling constant $g$, and (b) the on-site energy $\Omega\equiv \Omega_{\rm L}=\Omega_{\rm R}$. The superscripts l, g, p, t refer to the local, global, PERLind, and transmission approach respectively. When not shown, the PERLind approach agrees perfectly with the transmission approach. (a) The different approaches work well in their respective regime of validity. Inset: Efficiency $\eta=P_{\rm S}/J_{\rm L}$; the black line denotes the Carnot efficiency $1-T_{\rm R}/T_{\rm L}$.  (b) For the chosen parameters ($g/\kappa = 1$), the local approach agrees well with the transmission approach. Lower inset: Efficiency; the black line denotes the Carnot efficiency. Upper inset: For low temperatures ($k_{\rm B}T_{\rm R}=0.1\,\kappa$), the Born-Markov approximation breaks down for on-site energies close to the chemical potential $\mu_{\rm R}$. In this regime, the PERLind approach differs from the transmission approach. Parameters: $\mu_{\rm L}=0$, $\kappa\equiv\kappa_{\rm c}=\kappa_{\rm h}=0.05\,\mu_{\rm R}$, $k_{\rm B}T_{\rm L} = 2.5\,\mu_{\rm R}$, $k_{\rm B}T_{\rm R} = \mu_{\rm R}$ (a) $\Omega\equiv\Omega_{\rm L}=\Omega_{\rm R} =2.5\, \mu_{\rm R}$ (b) $g=\kappa$, upper inset: $k_{\rm B}T_{\rm R} =0.005\,\mu_{\rm R}$.}
\end{figure}

Depending on how the transition frequencies are grouped, we obtain either the global or the local approach. As long as the Born-Markov approximations are justified, i.e., as long as Eq.~\eqref{eq:bornmarkvalferm} holds, at least one of the two procedures is justified. As the parameters of the system are changed, we may thus need to change from the local to the global approach to maintain an adequate description of the system. This is a general drawback of our master equation: as parameters are varied, the grouping of transition frequencies into sets may need adaptation. This results in discontinuities which are expected to be small as long as the Born-Markov approximations are justified.

In Fig.~\ref{fig:heatengineferm}, we illustrate the global, the local, as well as the benchmark provided by the transmission approach for this system when operated as a heat engine (see \ref{app:currsferm} for analytical expressions). We find that our master equation reproduces the transmission approach well if the frequency grouping is chosen such that the local (global) approach is obtained for $g$ below (above) $5\kappa$. Interestingly, we find that the efficiency is generally better reproduced by the global approach (see insets in Fig.~\ref{fig:heatengineferm}). While the local approach predicts Carnot efficiency when $n_{\rm F}^{\rm L}(\bar{\Omega})=n_{\rm F}^{\rm R}(\bar{\Omega})$, the transmission and the global approach predict a drop in efficiency. The reason for this drop is that the heat current remains finite at vanishing power when transmission occurs at more than one energy \cite{brunner:2012}.

The upper inset in Fig.~\ref{fig:heatengineferm}\,(b) shows that at temperatures $k_{\rm B}T_\alpha\lesssim\kappa_\alpha$, the Born-Markov approximation breaks down if one of the transition frequencies $\Omega_{\pm}$ is close to the chemical potential $\mu_{\alpha}$, cf.~Eq.~\eqref{eq:bornmarkvalferm}. In this case, the step-like behavior of the Fermi-Dirac distribution renders the assumption of a Markovian bath unjustified. By approximating transmission to occur only at the transition frequencies of the system, the Markovian master equations predict a step-like behavior of heat currents and power. The transmission approach shows that the step-like feature is smeared out by the transmission function, which is a continuous function of energy.

\section{Bosonic heat engine - a time-dependent example}
\label{sec:bosons}
\subsection{System}
In this section, we provide an example of a time-dependent system. To this end, we consider the heat engine introduced by Kosloff in 1984 \cite{kosloff:1984} with the system Hamiltonian
\begin{equation}
\label{eq:hambos}
\hat{H}_{\rm S}(t) = \Omega_{\rm c}\hat{a}_{\rm c}^\dagger\hat{a}_{\rm c} + \Omega_{\rm h}\hat{a}_{\rm h}^\dagger\hat{a}_{\rm h}+g(\hat{a}_{\rm h}^\dagger\hat{a}_{\rm c}e^{-i\varpi t}+\hat{a}_{\rm c}^\dagger\hat{a}_{\rm h}e^{i\varpi t}),
\end{equation}
where the bosonic annihilation and creation operators fulfill the standard commutation relations given in Eq.~\eqref{eq:commrels}, and the frequency of the external drive reads
\begin{equation}
\label{eq:freqbos}
\varpi = \Omega_{\rm h}-\Omega_{\rm c}-2\Delta.
\end{equation}
This Hamiltonian describes two bosonic modes with different frequencies that are coupled via a time-dependent term with a detuning quantified by $\Delta$. For a physical implementation of this Hamiltonian based on a superconducting circuit, see Ref.~\cite{hofer:2016prb}. The reservoirs and their coupling to the system are described by
\begin{equation}
\label{eq:rescoupbos}
\hat{H}_\alpha = \sum_{l}\varepsilon_{\alpha,l}\hat{b}^\dagger_{\alpha,l}\hat{b}_{\alpha,l},\hspace{1.5cm}\hat{V}_{\alpha} = \sum_{l}t_{\alpha,l}(\hat{a}_\alpha^\dagger \hat{b}_{\alpha,l}+\hat{b}_{\alpha,l}^\dagger\hat{a}_\alpha),
\end{equation}
with $\alpha={\rm c, h}$ and we consider a vanishing chemical potential for the reservoirs, $\mu_{\alpha}=0$.

From Eqs.~\eqref{eq:sysbath} and \eqref{eq:rescoupbos}, we identify the following system and bath operators
\begin{equation}
\label{eq:sysopbos}
\hat{S}_{\alpha,1}=\hat{a}_\alpha^\dagger, \hspace{2.5cm}\hat{S}_{\alpha,-1}=\hat{a}_\alpha,
\end{equation}
 and
\begin{equation}
\label{eq:bathopbos}
\hat{B}_{\alpha,1}=\sum_lt_{\alpha,l}\hat{b}_{\alpha,l},\hspace{1.5cm}\hat{B}_{\alpha,-1}=\sum_lt_{\alpha,l}\hat{b}^\dagger_{\alpha,l}.
\end{equation}
We find that the Born-Markov approximations are justified when (see \ref{app:bathcorrbos} and Refs.~\cite{rivas:2010,hofer:2017njp})
\begin{equation}
\label{eq:bornmarkvalbos}
\kappa_{\alpha} \ll \omega_j,
\end{equation}
where the inequality should hold for all values of $\alpha$ and $j$. This ensures that the Bose-Einstein distribution is flat around the transition energies over the energy scale $\kappa_{\alpha}$. Assuming a flat bath spectral density, the same holds for the rates
\begin{equation}
\label{eq:gammasbos}
\Gamma^{\alpha}_{1}(-\omega)=\kappa_\alpha n_{\rm B}^\alpha(\omega),\hspace{1cm}\Gamma^{\alpha}_{-1}(\omega) = \kappa_{\alpha} [1-n_{\rm B}^\alpha(\omega)],
\end{equation}
where $\omega>0$ as we are dealing with Bosons with zero chemical potential.

The time-dependence in Eq.~\eqref{eq:hambos} can be removed by a suitable unitary transformation, see Ref.~\cite{hofer:2017njp}, where this model was used to compare local and global master equations. Here, we use the model to illustrate how a time-dependent Hamiltonian affects our master equation and we thus remain in the lab frame. As discussed in Sec.~\ref{sec:derivation}, the average Hamiltonian defined in Eq.~\eqref{eq:avham} is an important quantity. For our system, we find (see \ref{app:avham})
\begin{equation}
\label{eq:hamavbos}
\hat{H}_{\rm av} = \Omega_{\rm c}^+\hat{a}_+^\dagger\hat{a}_++\Omega_{\rm c}^-\hat{a}_-^\dagger\hat{a}_-,
\end{equation}
where the eigenmodes are given in Eqs.~(\ref{eq:ladderpm},\ref{eq:theta}) and we introduced the frequencies
\begin{equation}
\label{eq:bosfreq}
\Omega_{\rm c}^\pm = \Omega_{\rm c}+\Delta\pm\sqrt{\Delta^2+g^2},\hspace{2cm}\Omega_{\rm h}^\pm = \Omega_{\rm h}-\Delta\pm\sqrt{\Delta^2+g^2}.
\end{equation}
We note that this choice for the average Hamiltonian is not unique. Indeed, $\hat{H}'_{\rm av} = \Omega_{\rm h}^+\hat{a}_+^\dagger\hat{a}_++\Omega_{\rm h}^-\hat{a}_-^\dagger\hat{a}_-$ does also fulfill the defining relation given in Eq.~\eqref{eq:avham}.

The Fourier coefficients of the bath operators given in Eq.~\eqref{eq:bathopbos} are determined by
\begin{equation}
\label{eq:bosfour}
\begin{aligned}
&\hat{U}^\dagger_{\rm S}(t)\hat{a}_{\rm c}\hat{U}_{\rm S}(t) = e^{-it\Omega_{\rm c}^+}\sin(\theta/2)\hat{a}_+ +e^{-it\Omega_{\rm c}^-}\cos(\theta/2)\hat{a}_-,\\&\hat{U}^\dagger_{\rm S}(t)\hat{a}_{\rm h}\hat{U}_{\rm S}(t) = e^{-it\Omega_{\rm h}^+}\cos(\theta/2)\hat{a}_+ -e^{-it\Omega_{\rm h}^-}\sin(\theta/2)\hat{a}_-,
\end{aligned}
\end{equation}
where $\theta$ is defined in Eq.~\eqref{eq:theta} [where Eq.~\eqref{eq:freqbos} determines $\Delta$].
From these equations (and their Hermitian conjugates) we can identify the transition rates and jump operators. As we have seen for the fermionic heat engine, it is sufficient to consider only the frequencies and operators related to particles leaving the system (the others can be treated separately and equivalently). These transition frequencies read 
\begin{equation}
\label{eq:wjbos}
\{\omega_j\} = \{\Omega_{\rm c}^+,\,\Omega_{\rm c}^-,\,\Omega_{\rm h}^+,\,\Omega_{\rm h}^-\}.
\end{equation}
As discussed above, these can be written as $\nu_j+l\varpi$, where $\nu_j$ denotes a transition frequency of the average Hamiltonian [i.e. $\Omega_{\rm c}^\pm$ for Eq.~\eqref{eq:hamavbos}] and $l=0,1$. The corresponding jump operators read
\begin{equation}
\label{eq:sjbos}
\begin{aligned}
&\{\hat{S}_{{\rm c},-1;j}\} = \left\{\sin(\theta/2)\hat{a}_+,\,\cos(\theta/2)\hat{a}_-,\,0,\,0\right\},\\
&\{\hat{S}_{{\rm h},-1;j}\} = \left\{0,\,0,\,\cos(\theta/2)\hat{a}_+,\,-\sin(\theta/2)\hat{a}_-\right\}.
\end{aligned}
\end{equation}

\subsection{Global approach}
Again, we have two choices for grouping the frequencies which result in the local and global approach respectively. We may group $\Omega_{\alpha}^\pm$ into different sets or assign them to the same set. Grouping them into different sets results in a single frequency per set such that $\{\omega_q\}=\{\omega_j\}$ and  $\{\hat{S}_{\alpha,k;q}(t)\}=\{\hat{S}_{\alpha,k;j}\exp(-i\omega_jt)\}$ [cf.~Eq.~\eqref{eq:jumps}]. Inserting these quantities into Eq.~\eqref{eq:bathdiagin}, we find the dissipator (in the interaction picture)
\begin{equation}
\label{eq:dissglbos}
\tilde{\mathcal{L}}_\alpha^{\rm g} = \sum_{\sigma=\pm}\frac{\kappa_{\alpha}^\sigma}{2}\left\{n_{\rm B}^\alpha(\Omega_{\alpha}^\sigma)\mathcal{D}[\hat{a}^\dagger_{\sigma}]+[n_{\rm B}^\alpha(\Omega_{\alpha}^\sigma)+1]\mathcal{D}[\hat{a}_{\sigma}]\right\},
\end{equation}
whith $\kappa_{\alpha}^\sigma$ being defined in Eq.~\eqref{eq:kappags}.
As outlined in \ref{app:tdtime}, a single-thermodynamic Hamiltonian is usually not sufficient for the time-dependent scenario. Furthermore, even when a secular approximation is performed, a rescaling needs to be performed \cite{levy:2012pre}. In the present case, the rescaling is obtained starting from $\hat{H}_{\rm av}$ in Eq.~\eqref{eq:hamavbos} and rescaling the relevant transition frequencies to $\omega_q$. It turns out that it is sufficient to consider two thermodynamic Hamiltonians, one for each reservoir
\begin{equation}
\label{eq:htdsecbos}
\hat{H}_{\rm TD}^\alpha = \Omega_\alpha^+\hat{a}_+^\dagger\hat{a}_++\Omega_\alpha^-\hat{a}_-^\dagger\hat{a}_-.
\end{equation}
The expression for the heat current given in Eq.~\eqref{eq:heatbathtd} then reduces to
\begin{equation}
\label{eq:heatcurrglbos}
J_\alpha = {\rm Tr}\{\hat{H}_{\rm TD}^\alpha \tilde{\mathcal{L}}_\alpha^{\rm g}\hat{\rho}_{\rm S}\}.
\end{equation}
Note that the two thermodynamic Hamiltonians correspond to two possible choices for $\hat{H}_{\rm av}$.

In this scenario, the external power can no longer be accessed by the standard expression given in Eq.~\eqref{eq:powersys}. As a consequence of the secular approximation, this quantity evaluates to zero \cite{hofer:2017njp}. The time-averaged power can however be recovered by relying on the first law as in Eq.~\eqref{eq:firstlawav}.

\subsection{Local approach}
Alternatively, we may group $\Omega_{\alpha}^\pm$ into the same set. This grouping is justified as long as $g,\Delta\ll\Omega_{\rm c},\Omega_{\rm h}$ [in analogy to Eq.~\eqref{eq:bornmarkvalbos}]. A natural choice for the set frequencies then reads
\begin{equation}
\label{eq:wqbos}
\{\omega_q\} = \{\Omega_{\rm c}+\Delta,\,\Omega_{\rm h}-\Delta\},
\end{equation}
with the jump operators
\begin{equation}
\label{eq:sqbos}
\begin{aligned}
&\{\hat{S}_{{\rm c},-1;q}(t)\} = \left\{\hat{U}_{\rm S}^\dagger(t)\hat{a}_{\rm c}\hat{U}_{\rm S}(t),\,0\right\},\\
&\{\hat{S}_{{\rm h},-1;q}(t)\} = \left\{0,\,\hat{U}_{\rm S}^\dagger(t)\hat{a}_{\rm h}\hat{U}_{\rm S}(t)\right\}.
\end{aligned}
\end{equation}
Upon returning to the Schrödinger picture, these quantities result in the local dissipator
\begin{equation}
\label{eq:disslocbos}
\begin{aligned}
&\mathcal{L}^{\rm l}_{\rm c} = \kappa_{\rm c}\{n_{\rm B}(\Omega_{\rm c}+\Delta)\mathcal{D}[\hat{a}^\dagger_{\rm c}]+[n_{\rm B}(\Omega_{\rm c}+\Delta)+1]\mathcal{D}[\hat{a}_{\rm c}]\},\\
&\mathcal{L}^{\rm l}_{\rm h} = \kappa_{\rm h}\{n_{\rm B}(\Omega_{\rm h}-\Delta)\mathcal{D}[\hat{a}^\dagger_{\rm h}]+[n_{\rm B}(\Omega_{\rm h}-\Delta)+1]\mathcal{D}[\hat{a}_{\rm h}]\}.
\end{aligned}
\end{equation}

Two thermodynamic Hamiltonians can again be obtained by re-scaling the transition frequencies of the average Hamiltonian in Eq.~\eqref{eq:hamavbos}. However, due to the local structure of the jump operators, we can find a single thermodynamic Hamiltonian in the Schrödinger picture [cf.~\ref{app:tdtime}]
\begin{equation}
\label{eq:hamtdlocbos}
\hat{H}_{\rm TD} = (\Omega_{\rm c}+\Delta) \hat{a}_{\rm c}^\dagger\hat{a}_{\rm c}+(\Omega_{\rm h}-\Delta) \hat{a}_{\rm h}^\dagger\hat{a}_{\rm h}.
\end{equation}
From Eq.~\eqref{eq:firstlawtimeinq}, we recover the usual expression for the power produced by the system
\begin{equation}
\label{eq:powerbos}
P_{\rm S} = -P_{\rm ext} =i{\rm Tr}\{[\hat{H}_{\rm TD},\hat{H}_{\rm S}(t)]\hat{\rho}_{\rm S}\} = -{\rm Tr}\{[\partial_t\hat{H}_{\rm S}(t)]\hat{\rho}_{\rm S}\}.
\end{equation}
For $\Delta=0$, the thermodynamic Hamiltonian in Eq.~\eqref{eq:hamtdlocbos} was used before in Ref.~\cite{kerremans:2021} and has an intuitive explanation: For the thermodynamic bookkeeping, we neglect the coupling energy between the bosonic modes, just as we neglect the coupling energy between system and bath.

\subsection{Results}
\begin{figure}
	\includegraphics[width=\textwidth]{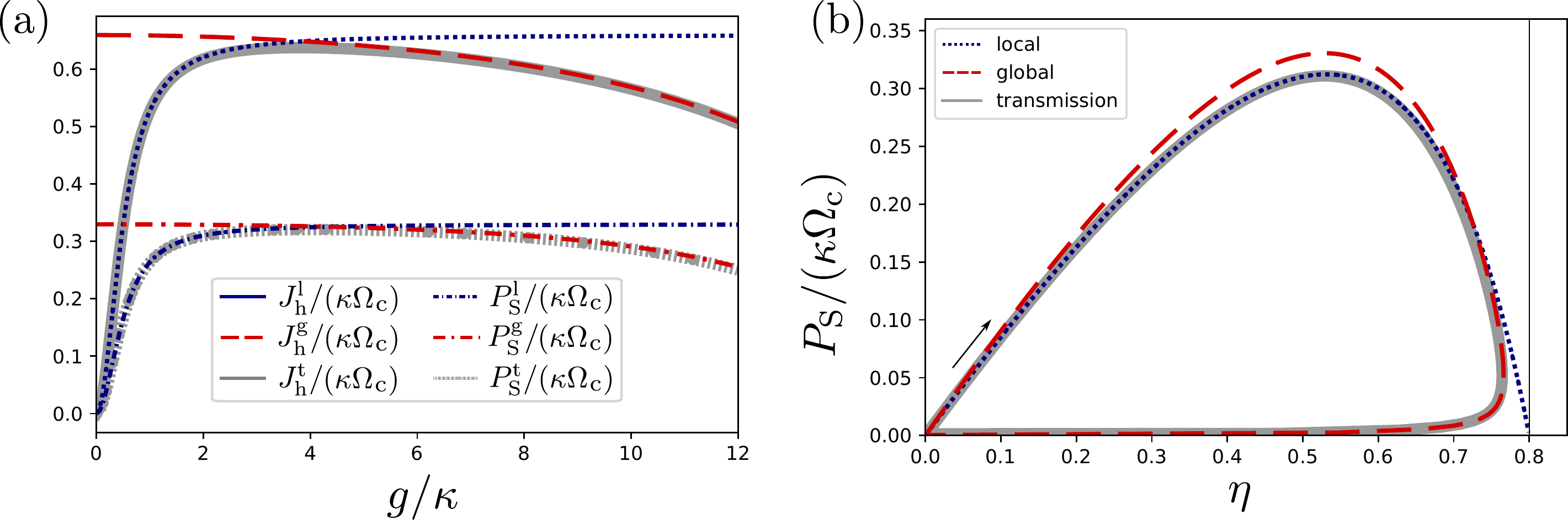}
	\caption{\label{fig:heatenginebos} Bosonic heat engine for a resonant drive ($\Delta=0$). (a)  Steady state heat current from the hot bath $J^j_{\rm h}$ and power $P_{\rm S}^j$ as a function of the coupling strength $g$. The superscripts l, g, t, refer to the local, global, and transmission approach respectively. (b) Lasso diagramm obtained by varying $\Omega_{\rm h}$ in the window where power is positive ($[\Omega_{\rm c},T_{\rm h}\Omega_{\rm c}/T_{\rm c}]$, for the local approach \cite{hofer:2016prb}). Along the arrow, $\Omega_{\rm h}$ is increased. The vertical line denotes the Carnot efficiency: $1-T_{\rm c}/T_{\rm h}$. For these plots, the PERLind approach is visibly indistinguishable from the transmission approach. Parameters: $\Omega_{\rm h} = 2\,\Omega_c$, $\kappa\equiv\kappa_{\rm c}=\kappa_{\rm h}=0.05\,\Omega_{\rm c}$, $k_{\rm B}T_{\rm h} = 2.5\,\Omega_{\rm c}$, $k_{\rm B}T_{\rm c} = 0.5\,\Omega_{\rm c}$ (b) $g=2\,\kappa$.}
\end{figure}

In Fig.~\ref{fig:heatenginebos}, power, heat current, and efficiency of the bosonic heat engine for a resonant drive are illustrated. The local and global approaches for this scenario were already compared in Ref.~\cite{hofer:2017njp}, where exact numerics for finite reservoirs was used as a benchmark. In agreement with these results, we find that the local and global approaches together reproduce exact solutions over the full range of parameters where the Born-Markov approximation is justified. This implies that our master equation, which reduces either to the local or to the global approach, describes this scenario very well. We note that the PERLind approach reproduces the transmission approach extremely well. The disadvantage of the PERLind approach is that it is not thermodynamically consistent. The disadvantage of our master equation is that it exhibits a discontinuity when the frequency grouping is adapted (i.e., when changing from the local to the global approach).

\begin{figure}
	\includegraphics[width=\textwidth]{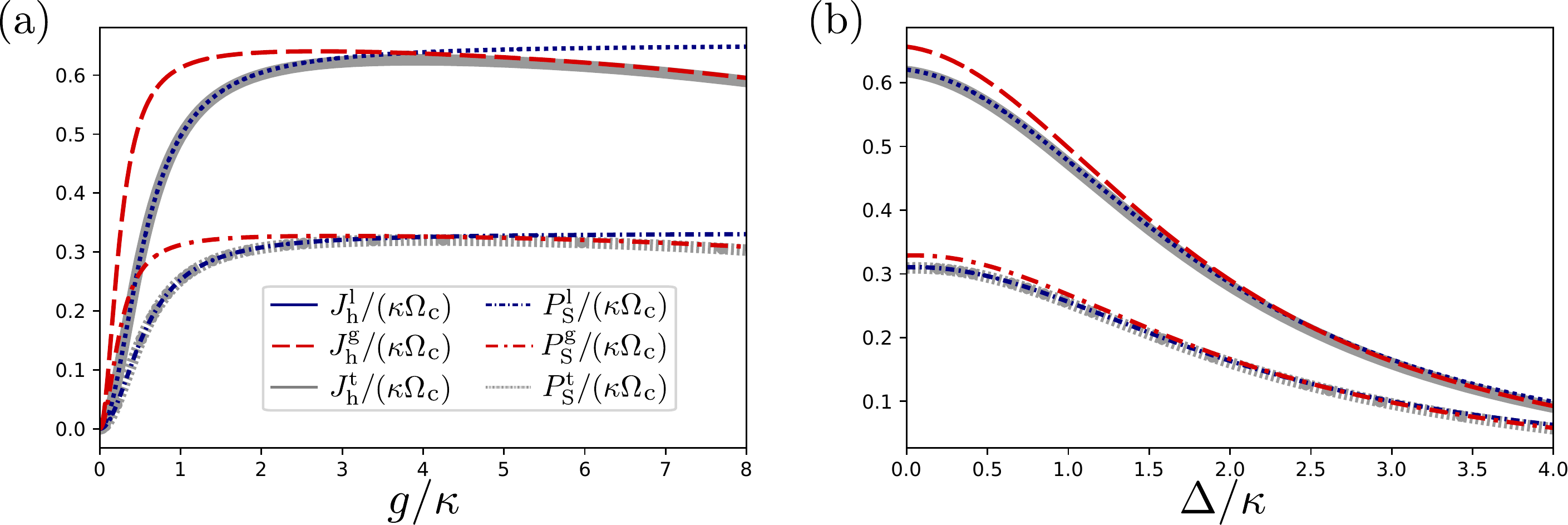}
	\caption{\label{fig:heatenginebosdelta} Bosonic heat engine for a a finite detuning $\Delta$. Steady state heat current from the hot bath $J^j_{\rm h}$ and power $P_{\rm S}^j$ as a function of (a) the coupling strength $g$, and (b) the detuning $\Delta$. The superscripts l, g, t, refer to the local, global, and transmission approach respectively. For these plots, the PERLind approach is visibly indistinguishable from the transmission approach. Parameters: $\Omega_{\rm h} = 2\,\Omega_c$, $\kappa\equiv\kappa_{\rm c}=\kappa_{\rm h}=0.05\,\Omega_{\rm c}$, $k_{\rm B}T_{\rm h} = 2.5\,\Omega_{\rm c}$, $k_{\rm B}T_{\rm c} = 0.5\,\Omega_{\rm c}$ (a) $\Delta = \kappa/4$ (b) $g=2\,\kappa$.}
\end{figure}

In Fig.~\ref{fig:heatenginebosdelta}, the performance of the heat engine at finite detuning $\Delta$ is investigated. For small detuning, we find that the global approach correctly reproduces the $g\rightarrow 0$ limit, as an exact degeneracy is no longer reached in this limit [panel (a)]. As $\Delta$ increases, the low-$g$ behavior is increasingly well captured by the global approach, because the secular approximation becomes better. For a fixed coupling strength $g$, we find a decrease in the heat engine performance as $\Delta$ is increased [panel (b)]. This behavior is expected, because energy transfer between the drive and the system works best on resonance.

\section{Interacting double quantum dot - beyond local and global approaches}
\label{sec:beyond}
\subsection{System}
In the previous examples, the derived master equation reduces to the well known global or local approaches. In this last example, we consider a system that has more transition frequencies, such that the master equation may differ from both the local as well as the global approach. To this end, we consider a spinless double quantum dot with Coulomb interactions described by the Hamiltonian
\begin{equation}
\label{eq:hamfermint}
\begin{aligned}
\hat{H}_{\rm S} &= \Omega\hat{d}_{\rm L}^\dagger\hat{d}_{\rm L} + \Omega\hat{d}_{\rm R}^\dagger\hat{d}_{\rm R}
+ g(\hat{d}_{\rm L}^\dagger\hat{d}_{\rm R}+\hat{d}_{\rm R}^\dagger\hat{d}_{\rm L})+U\hat{d}_{\rm L}^\dagger\hat{d}_{\rm L}\hat{d}_{\rm R}^\dagger\hat{d}_{\rm R}\\&=\Omega_-\hat{d}_-^\dagger\hat{d}_- + \Omega_+\hat{d}_+^\dagger\hat{d}_++U\hat{d}_-^\dagger\hat{d}_-\hat{d}_+^\dagger\hat{d}_+.
\end{aligned}
\end{equation}
Apart from the interaction term, the system as well as the bath properties are identical to Sec.~\ref{sec:fermions} (with $\Omega\equiv\Omega_{\rm L}=\Omega_{\rm R}$ and $\Omega_\pm=\Omega\pm g$) where all notation used in this section is defined. From Eq.~\eqref{eq:hamfermint}, we see that the Hamiltonian is diagonal in the occupation basis of the $\pm$ modes. The transition frequencies and jump operators are obtained from the equations
\begin{equation}
\label{eq:fermintfour}
\begin{aligned}
e^{i\hat{H}_{\rm S}t}\hat{d}_{\rm R}e^{-i\hat{H}_{\rm S}t} =& \frac{e^{-i\Omega_- t}}{\sqrt{2}}(1-\hat{d}_+^\dagger\hat{d}_+)\hat{d}_-+\frac{e^{-i(\Omega_-+U) t}}{\sqrt{2}}\hat{d}_+^\dagger\hat{d}_+\hat{d}_-\\& +\frac{e^{-i\Omega_+ t}}{\sqrt{2}}(1-\hat{d}_-^\dagger\hat{d}_-)\hat{d}_++\frac{e^{-i(\Omega_++U) t}}{\sqrt{2}}\hat{d}_-^\dagger\hat{d}_-\hat{d}_+,\\e^{i\hat{H}_{\rm S}t}\hat{d}_{\rm L}e^{-i\hat{H}_{\rm S}t} = &-\frac{e^{-i\Omega_- t}}{\sqrt{2}}(1-\hat{d}_+^\dagger\hat{d}_+)\hat{d}_--\frac{e^{-i(\Omega_-+U) t}}{\sqrt{2}}\hat{d}_+^\dagger\hat{d}_+\hat{d}_- \\&+\frac{e^{-i\Omega_+ t}}{\sqrt{2}}(1-\hat{d}_-^\dagger\hat{d}_-)\hat{d}_++\frac{e^{-i(\Omega_++U) t}}{\sqrt{2}}\hat{d}_-^\dagger\hat{d}_-\hat{d}_+.
\end{aligned}
\end{equation}
Considering only transition frequencies that correspond to particles leaving the system, we find
\begin{equation}
\label{eq:omegajfermint}
\{\omega_j\} = \{\Omega_-,\,\Omega_+,\,\Omega_-+U,\,\Omega_++U\},
\end{equation}
and the jump operators
\begin{equation}
\label{eq:jumpopsfermint}
\begin{aligned}
&\{\hat{S}_{{\rm R},-1;j}\} = \left\{\frac{1-\hat{d}_+^\dagger\hat{d}_+}{\sqrt{2}}\hat{d}_-,\,\frac{1-\hat{d}_-^\dagger\hat{d}_-}{\sqrt{2}}\hat{d}_+,\,\frac{\hat{d}_+^\dagger\hat{d}_+}{\sqrt{2}}\hat{d}_-,\,\frac{\hat{d}_-^\dagger\hat{d}_-}{\sqrt{2}}\hat{d}_+\right\},\\&\{\hat{S}_{{\rm L},-1;j}\} = \left\{-\frac{1-\hat{d}_+^\dagger\hat{d}_+}{\sqrt{2}}\hat{d}_-,\,\frac{1-\hat{d}_-^\dagger\hat{d}_-}{\sqrt{2}}\hat{d}_+,\,-\frac{\hat{d}_+^\dagger\hat{d}_+}{\sqrt{2}}\hat{d}_-,\,\frac{\hat{d}_-^\dagger\hat{d}_-}{\sqrt{2}}\hat{d}_+\right\}.
\end{aligned}
\end{equation}

\subsection{Global approach}
In the global approach, we choose a different set for each transition frequency. This results in the dissipator 
\begin{equation}
\label{eq:globaldissfermint}
\begin{aligned}
\mathcal{L}_\alpha^{\rm g} = \sum_{\sigma=\pm}\frac{\kappa_{\alpha}}{2}\big\{&n_{\rm F}^\alpha(\Omega_\sigma)\mathcal{D}[(1-\hat{d}^\dagger_{\bar{\sigma}}\hat{d}_{\bar{\sigma}})\hat{d}_\sigma^\dagger]+[1-n_{\rm F}^\alpha(\Omega_\sigma)]\mathcal{D}[(1-\hat{d}^\dagger_{\bar{\sigma}}\hat{d}_{\bar{\sigma}})\hat{d}_\sigma]\\&n_{\rm F}^\alpha(\Omega_\sigma+U)\mathcal{D}[\hat{d}^\dagger_{\bar{\sigma}}\hat{d}_{\bar{\sigma}}\hat{d}_\sigma^\dagger]+[1-n_{\rm F}^\alpha(\Omega_\sigma+U)]\mathcal{D}[\hat{d}^\dagger_{\bar{\sigma}}\hat{d}_{\bar{\sigma}}\hat{d}_\sigma]\big\},
\end{aligned}
\end{equation}
where $\bar{\sigma}\neq\sigma$. As usual, the thermodynamic Hamiltonian in the global approach is equal to $\hat{H}_{\rm S}$.
This frequency grouping is justified when
\begin{equation}
\label{eq:justglfermint}
|\omega_j-\omega_{j'}|\gg \kappa_{\alpha},
\end{equation}
for all $j$, $j'$, and $\alpha$.

\subsection{Local approach}
In the local approach, we group all transition frequencies into a single set. A natural set frequency is then the average frequency $\omega_q=\Omega+U/2$. The jump operators are given by summing the jump operators in Eq.~\eqref{eq:jumpopsfermint} over $j$, which results in the local dissipator
\begin{equation}
\label{eq:localdissfermint}
\mathcal{L}^{\rm l}_\alpha =  \kappa_{\alpha}\big\{n_{\rm F}^\alpha(\Omega+U/2)\mathcal{D}[\hat{d}^\dagger_\alpha]+[1-n_{\rm F}^\alpha(\Omega+U/2)]\mathcal{D}[\hat{d}_\alpha]\}.
\end{equation}
The thermodynamic Hamiltonian, obtained by rescaling all transitions to $\Omega+U/2$, is given by
\begin{equation}
\label{eq:hamtdlfint}
\hat{H}_{\rm TD}=(\Omega+U/2)(\hat{d}_{\rm L}^\dagger\hat{d}_{\rm L}+\hat{d}_{\rm R}^\dagger\hat{d}_{\rm R})=(\Omega+U/2)(\hat{d}_+^\dagger\hat{d}_++\hat{d}_-^\dagger\hat{d}_-).
\end{equation}
We note that apart from the choice of the transition frequency, the dissipator and thermodynamic Hamiltonian are identical to the one obtained for $U=0$, c.f., Eqs.~\eqref{eq:disslocalferm} and \eqref{eq:hamtdferm}.
The frequency grouping for the local approach is justified as long as
\begin{equation}
\label{eq:justlfermint}
|n_{\rm F}^\alpha(\omega_j)-n_{\rm F}^\alpha(\omega_{j'})|\ll 1,
\end{equation}
for all $j$, $j'$, and $\alpha$.

\subsection{Semi-local approach}
In addition to the local and global approaches, we consider the low-$g$ frequency grouping, where $\Omega_\pm$ can be replaced by $\Omega$. This results in the set frequencies
\begin{equation}
\label{eq:omegaqsl}
\{\omega_q\} = \{\Omega,\,\Omega+U\},
\end{equation}
with the corresponding jump operators
\begin{equation}
\label{eq:jumpopssl}
\begin{aligned}
&\{\hat{S}_{{\rm R},-1;q}\}=\left\{(1-\hat{d}^\dagger_{\rm L}\hat{d}_{\rm L})\hat{d}_{\rm R},\,\hat{d}^\dagger_{\rm L}\hat{d}_{\rm L}\hat{d}_{\rm R}\right\},\\&\{\hat{S}_{{\rm L},-1;q}\}=\left\{(1-\hat{d}^\dagger_{\rm R}\hat{d}_{\rm R})\hat{d}_{\rm L},\,\hat{d}^\dagger_{\rm R}\hat{d}_{\rm R}\hat{d}_{\rm L}\right\}.
\end{aligned}
\end{equation}
Note that the jump operators still locally change the number of electrons. However, the jumps are now dependent on the occupancy of the other dot due to the Coulomb interaction. Inserting these quantities into Eq.~\eqref{eq:bathops} results in the dissipator
\begin{equation}
\label{eq:disssl}
\begin{aligned}
\mathcal{L}_\alpha^{\rm sl} = \kappa_{\alpha}\{&n_{\rm F}^\alpha(\Omega)\mathcal{D}[(1-\hat{d}^\dagger_{\bar{\alpha}}\hat{d}_{\bar{\alpha}})\hat{d}^\dagger_\alpha]+[1-n_{\rm F}^\alpha(\Omega)]\mathcal{D}[(1-\hat{d}^\dagger_{\bar{\alpha}}\hat{d}_{\bar{\alpha}})\hat{d}_\alpha]\\&+n_{\rm F}^\alpha(\Omega+U)\mathcal{D}[\hat{d}^\dagger_{\bar{\alpha}}\hat{d}_{\bar{\alpha}}\hat{d}^\dagger_\alpha]+[1-n_{\rm F}^\alpha(\Omega+U)]\mathcal{D}[\hat{d}^\dagger_{\bar{\alpha}}\hat{d}_{\bar{\alpha}}\hat{d}_\alpha]\},
\end{aligned}
\end{equation}
where $\bar{\alpha}\neq\alpha$. The thermodynamic Hamiltonian corresponding to this frequency grouping reads
\begin{equation}
\label{eq:hamtdsl}
\hat{H}_{\rm TD} = \Omega(\hat{d}_{\rm L}^\dagger\hat{d}_{\rm L}+\hat{d}_{\rm R}^\dagger\hat{d}_{\rm R})+U\hat{d}_{\rm L}^\dagger\hat{d}_{\rm L}\hat{d}_{\rm R}^\dagger\hat{d}_{\rm R}.
\end{equation}

This frequency grouping is justified for
\begin{equation}
\label{eq:justsl}
|n_{\rm F}^\alpha(\Omega_+)-n_{\rm F}^\alpha(\Omega_-)|\ll 1,\hspace{.5cm}\text{and}\hspace{.5cm}|n_{\rm F}^\alpha(\Omega_++U)-n_{\rm F}^\alpha(\Omega_-+U)|\ll 1.
\end{equation}
To ensure that the frequencies in different sets obey $|\omega_j-\omega_{j'}|\gg \kappa_{\alpha}$, one may expect the additional condition $U\gg g,\kappa_{\alpha}$. However, because no coherences can build up between states with a different total number of electrons in the system, this additional condition is not required. The semi-local approach thus enjoys a strictly larger regime of validity than the local approach [c.f.~Eq.~\eqref{eq:justlfermint}], remaining valid even for a vanishing interaction strength $U=0$.

\subsection{Results}
\begin{figure}
	\includegraphics[width=\textwidth]{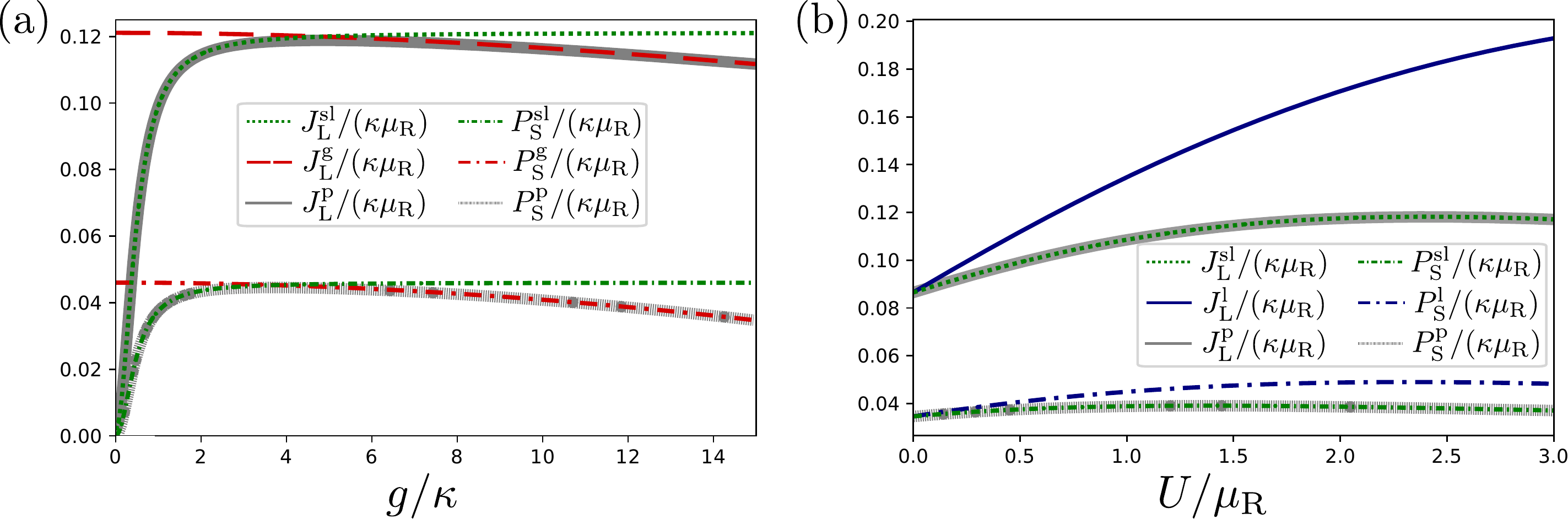}
	\caption{\label{fig:fermint} Steady state heat current from the hot bath $J_{\rm L}^j$ and output power $P_{\rm S}^j$ as a function of (a) the coupling constant $g$, and (b) the interaction energy $U$. the superscripts sl, g, p, refer to the semi-local, global, and PERLind approach respectively. (a) For finite interaction strength, the semi-local and the global approach together capture the behavior at all values of $g$. The local approach fails for all values of $g$ due to the finite value of $U=0.5\,\mu_{\rm R}$ (not shown). (b) For small coupling $g=\kappa$, the semi-local approach agrees with the PERLind approach for all interaction strengths while the local approach differs for finite $U$. The global approach fails for all values $U$ due to the smallness of $g$ (not shown). Parameters: $\Omega=2.5\,\mu_{\rm R}$, $\mu_{\rm L}=0$, $\kappa\equiv\kappa_{\rm c}=\kappa_{\rm h}=0.05\,\mu_{\rm R}$, $k_{\rm B}T_{\rm L} = 2.5\,\mu_{\rm R}$, $k_{\rm B}T_{\rm R} = \mu_{\rm R}$ (a) $U=0.5\,\mu_{\rm R}$ (b) $g=\kappa$.}
\end{figure}

Due to the interaction term, the transmission approach no longer applies. As a benchmark, we instead use the PERLind approach, see \ref{app:perlfermint}. As mentioned above, this approach is expected to provide accurate results whenever the Born-Markov approximations are justified, an expectation that was confirmed in the examples we considered above.

Figure \ref{fig:fermint} illustrates heat current and power in the interacting double quantum dot. We find that at finite interaction strengths, the semi-local approach should be employed instead of the local approach for small coupling $g$. In particular, comparing Fig.~\ref{fig:fermint}\,(a) with Fig.~\ref{fig:heatengineferm}\,(a) we find a very similar interplay between the semi-local and global approach at $U\neq0$ as we observed between the local and global approaches for $U=0$. We furthermore find that the semi-local approach agrees with the PERLind approach for any value of the interaction strength, as long as the coupling $g$ is sufficiently small to respect Eq.~\eqref{eq:justsl}, see Fig.~\ref{fig:fermint}\,(b).
These results highlight the importance to go beyond the local and global approaches for systems that have more than two competing transition frequencies.

\section{Conclusions}
\label{sec:conclusions}
Markovian master equations in GKLS form are approximate descriptions for the reduced system state. The approximations involved in deriving these equations may not preserve the laws of thermodynamics (as is the case, e.g., in the PERLind approach). A comparison to the transmission approach for non-interacting particles illustrates that the source of any thermodynamic inconsistency is an inconsistent assignment of heat to the different jump operators appearing in the master equation. To shed light on this issue, we performed a microscopic derivation of the probability distribution for heat, employing the same approximations that are used for deriving master equations in GKLS form. We found that, when employing master equations, the resolution in heat is limited by the approximations that are performed. Exploiting this limited resolution, we derived a thermodynamically consistent master equation. To this end, we adapted the thermodynamic bookkeeping to enforce a consistent assignment of heat to the jump operators. The changes in heat induced by this procedure remain smaller than the resolution that the master equation allows for and are thus consistent with the performed approximations. This refined thermodynamic bookkeeping is captured by a thermodynamic Hamiltonian, which may differ from the Hamiltonian that determines the dynamics of the quantum state.

We illustrated our master equation with three different examples, including a time-dependent model and a model that includes interactions. Our master equation may reduce to the well-known global approach, or to a thermodynamically consistent version of the local approach in their respective regime of validity. For systems where neither of these approaches is adequate, it provides a novel description.

Our master equation provides a thermodynamically consistent description for quantum systems that are amenable to a Markovian description. This allows future investigations on the thermodynamics of quantum devices to fully rely on the conclusions that follow from basic thermodynamics.

\section*{Acknowledgements}
We thank Anton Trushechkin for bringing his related work \cite{trushechkin:2021} to our attention and for stimulating discussions. We further thank Kacper Prech for proofreading parts of the manuscript.
We thank the Knut and Alice Wallenberg Foundation (project 2016.0089) and Nano\-Lund for financial support.
P.P.P. acknowledges funding from the European Union's Horizon 2020 research and innovation programme under the Marie Sk\l{}odowska-Curie Grant Agreement No. 796700, from the Swedish Research Council (Starting Grant 2020-03362), and from the Swiss National Science Foundation (Eccellenza Professorial Fellowship PCEFP2\_194268).

\appendix

\section{Thermodynamic consistency for a time-dependent Hamiltonian}
\label{app:tdtime}
Here we show how thermodynamic consistency is obtained for a time-dependent system Hamiltonian. We note that the 0th law only applies to the time-independent scenario, as no equilibration is expected in the presence of an external drive.

Let us first consider slow driving, where $\hat{H}_{\rm S}(t)$ changes on a time-scale that is much larger than the bath correlation time $\tau_{\rm B}$. In this case, a master equation may be derived by considering the Hamiltonian to be \textit{frozen}, treating its time argument as any other parameter \cite{alicki:1979}. We may still use Eq.~\eqref{eq:mastereqncf} but the transition frequencies $\omega_j$ and $\omega_q$, as well as the operators $\hat{S}_{\alpha,k;j}$ become time-dependent. The thermodynamic Hamiltonian, obtained as in the time-independent case outlined in Sec.~\ref{sec:hamtd}, then also becomes time-dependent and the first law reads
\begin{equation}
\label{eq:firstlawslow}
\partial_t U = {\rm Tr}\{(\partial_t\hat{H}_{\rm TD})\hat{\rho}\}+\sum_\alpha(P_\alpha+J_\alpha),
\end{equation} 
where the internal energy as well as the heat-current and power from reservoir $\alpha$ are still defined by Eqs.~\eqref{eq:inten} and \eqref{eq:heatpowbathtd} respectively. The first term on the right-hand side of Eq.~\eqref{eq:firstlawslow} denotes the power provided by the external drive. The second law still holds as discussed in Sec.~\ref{sec:hamtd}.

Next, we consider a time-periodic system Hamiltonian with period $t_{\rm p}=2\pi/\varpi$. In this case, the frequencies $\omega_j$ are not directly transition frequencies but rather fulfill
\begin{equation}
\label{eq:freqtdep}
\omega_{j,l}=\nu_j+l\varpi,
\end{equation}
where $l$ is an integer and $\nu_j$ are the transition frequencies of the averaged Hamiltonian $\hat{H}_{\rm av}$ defined in Eq.~\eqref{eq:avham}. For clarity, we extended the index of $\omega_{j,l}$ to explicitly include $l$, such that $j$ uniquely defines the transition frequency of the averaged Hamiltonian. We stress that in this scenario, Eqs.~\eqref{eq:mastereq} and \eqref{eq:mastereqncfin} hold. Throughout this appendix, we remain in the interaction picture, because going to the Schrödinger picture is non-trivial for time-dependent Hamiltonians [cf.~the comment below Eq.~\eqref{eq:bathdiag}]. Since any time-dependence may be thought of as a single-period of a periodic process, our master equation is applicable for any time-dependence. However, for a thermodynamically consistent description, we require the additional assumption 
\begin{equation}
\label{eq:freqtaus}
\varpi\gg 1/\tau_{\rm S},
\end{equation}
which is complementary to the slow driving regime discussed above. As in the time-independent case, we may group the frequencies into sets $x_q$, such that
\begin{equation}
\label{eq:setsindapp}
\begin{aligned}
&|\omega_{j,l}-\omega_{j',l'}|\ll1/\tau_{\rm B}\hspace{.5cm}{\rm for}\,\, \omega_{j,l}\in x_q,\, \omega_{j',l'}\in x_{q'} \,\,{\rm with}\,\, q=q',\\
&|\omega_{j,l}-\omega_{j',l'}|\gg1/\tau_{\rm S}\hspace{.5cm}{\rm for}\,\,\omega_{j,l}\in x_q,\, \omega_{j',l'}\in x_{q'} \,\,{\rm with}\,\, q\neq q'.
\end{aligned}
\end{equation}
We note that due to Eq.~\eqref{eq:freqtaus}, we may choose the sets $x_q$ such that $\omega_{j,l}$ and $\omega_{j,l'}$ are not in the same set for $l\neq l'$. In that case, all frequencies in a given set have different values for $j$, i.e., correspond to different transition frequencies $\nu_j$ of the averaged Hamiltonian. For a given set $x_q$, a thermodynamic Hamiltonian $\hat{H}_{\rm TD}^q$ may be obtained by rescaling $\nu_j\rightarrow \omega_q$ for all $j$ that occur in the set $x_q$. In contrast to the time-independent case, the same transition frequency $\nu_j$ may feature in different sets (as it may feature in different $\omega_{j,l}$). The rescaling therefore has to be done individually for each set. We note that this rescaling is even required in the secular approximation since $\nu_j$ and $\omega_{j,l}$ differ by $l\varpi$ \cite{levy:2012pre}. The heat current from bath $\alpha$, obtained from Eqs.~\eqref{eq:mastereq} and \eqref{eq:heatcurrpowerbath} can then be cast into
\begin{equation}
\label{eq:heatbathtd}
J_\alpha = \sum_{\{q|\omega_q>0\}}{\rm Tr}\{(\hat{H}_{\rm TD}^q-\mu_\alpha\hat{N}_{\rm S})\tilde{\mathcal{L}}_{\alpha;q} \hat{\rho}_{\rm S}\},
\end{equation}
with $\tilde{\mathcal{L}}_{\alpha;q}$ given in Eq.~\eqref{eq:bathdiagin}.

To arrive at the master equation in Eq.~\eqref{eq:mastereqncf}, we dropped terms that oscillate with frequencies much larger than $1/\tau_{\rm S}$. Unfortunately, these terms may be important when evaluating the standard expression for external power given in Eq.~\eqref{eq:powersys}. Loosely speaking, the oscillations of the density matrix cancel oscillations of $\partial_t\hat{H}_{\rm S}(t)$, thereby contributing to the average power. In general, one may no longer disentangle the internal energy from the external power \cite{alicki:2015}. We may then still consider the first law upon taking a long-time average, ensuring that changes in the internal energy become negligible. Similarly, one could average over a single period when the system has reached a limit cycle. The first law then reduces to
\begin{equation}
\label{eq:firstlawav}
0 = \bar{P}_{\rm ext}+\sum_{\alpha}(\bar{P}_\alpha+\bar{J}_\alpha),
\end{equation}
where the bar denotes any kind of average that ensures a vanishing change in internal energy.

Interestingly, one may still access the internal energy (and thus the time-dependent power) when the master equation takes on a local structure. To illustrate this, let us consider a coupling Hamiltonian of the form
	\begin{equation}
	\label{eq:couplloc}
	\hat{V}=\sum_{\alpha}\left(\hat{S}_\alpha\hat{B}^\dagger_\alpha+\hat{B}_\alpha\hat{S}^\dagger_\alpha\right).
	\end{equation}
A local master equation is obtained when we group the transition frequencies such that each set frequency corresponds to a single reservoir, i.e., $\{\omega_q\}=\{\omega_\alpha\}$. In this case, the jump operators (in the interaction picture and dropping the index $q$) are of the form $\hat{S}_{\alpha}(t)=\hat{U}_{\rm S}^\dagger(t)\hat{S}_\alpha\hat{U}_{\rm S}(t)$ [cf.~Eq.~\eqref{eq:jumps}]. For each jump operator, we then introduce a corresponding number operator $\hat{n}_\alpha$, such that $[\hat{S}_\alpha,\hat{n}_\alpha]=\hat{S}_\alpha$. In this case, we may define a single, albeit time-dependent, thermodynamic Hamiltonian (in the interaction picture)
\begin{equation}
\label{eq:hamtdsingle}
\hat{H}_{\rm TD} = \sum_{\alpha}\omega_{\alpha}\hat{U}_{\rm S}^\dagger(t)\hat{n}_\alpha\hat{U}_{\rm S}(t),
\end{equation}
which fulfills the defining property $[\hat{S}_{\alpha}(t),\hat{H}_{\rm TD}]=\omega_{\alpha}\hat{S}_{\alpha}(t)$. One may then still use Eq.~\eqref{eq:inten} to define the internal energy and the first law reads
\begin{equation}
\label{eq:firstlawtimeinq}
\partial_t U = -i{\rm Tr}\{[\hat{H}_{\rm TD},\hat{H}_{\rm S}(t)]\hat{\rho}_{\rm S}\}+\sum_{\alpha}(P_\alpha+ J_\alpha),
\end{equation}
which is valid both in the interaction, as well as in the Schrödinger picture where both the jump operators as well as the thermodynamic Hamiltonian become time-independent. The first term on the right-hand side of Eq.~\eqref{eq:firstlawtimeinq} can be identified with the power provided by the external drive. 

To verify the second law, we write the entropy production rate as
\begin{equation}
\label{eq:entprodrateapp}
\dot{\Sigma}=-k_{\rm B}\sum_{\alpha}\sum_{\{q|\omega_q>0\}}{\rm Tr}\{(\tilde{\mathcal{L}}_{\alpha;q}\hat{\rho}_{\rm S})[\ln\hat{\rho}_{\rm S}-\ln\hat{\rho}_{\rm G}^q]\}\geq 0,
\end{equation}
where we introduced 
\begin{equation}
\label{eq:gibbsq}
\hat{\rho}_{\rm G}^q=\frac{e^{-\beta(\hat{H}^q_{\rm TD}-\mu\hat{N}_{\rm S})}}{{\rm Tr}\{e^{-\beta(\hat{H}^q_{\rm TD}-\mu\hat{N}_{\rm S})}\}}.
\end{equation}
Spohns inequality ensures the positivity of the entropy production rate since $\tilde{\mathcal{L}}_{\alpha;q}\hat{\rho}_{\rm G}^q=0$.

\section{Supplemental calculations for the fermionic heat engine}
\subsection{Bath correlation functions}
\label{app:bathcorrferm}
The bath operators given in Eq.~\eqref{eq:bathopferm} result in the bath correlation functions [cf.~Eq.~\eqref{eq:bathcorr}]
\begin{equation}
\label{eq:bathcorrferm}
\begin{aligned}
&C^\alpha_{1}=\frac{\kappa_\alpha}{2}e^{is\mu_\alpha}\left[\delta(s)-i\frac{k_{\rm B}T_\alpha}{\sinh(\pi sk_{\rm B}T_\alpha)}\right],\\&
C^\alpha_{-1}=\frac{\kappa_\alpha}{2}e^{-is\mu_\alpha}\left[\delta(s)-i\frac{k_{\rm B}T_\alpha}{\sinh(\pi sk_{\rm B}T_\alpha)}\right],
\end{aligned}
\end{equation}
where we introduced the bath spectral density
\begin{equation}
\label{eq:kappa}
\frac{\kappa_\alpha}{2\pi} = \sum_l t_{\alpha,l}^2\delta(\omega-\varepsilon_{\alpha,l}),
\end{equation}
quantifying the system-bath coupling and which we assume to be constant as a function of $\omega$. We note that $C^\alpha_k\equiv C^\alpha_{k,k}$ and $C^\alpha_{k,k'}\propto \delta_{k,k'}$. The bath correlation functions decay on the time-scale $\beta_\alpha$. The bath correlation time can thus be identified by $\tau_{\rm B} = \max\{\beta_{\rm L},\beta_{\rm R}\}$. The system (in the interaction picture) changes on the time-scale $\tau_{\rm S} = \min\{1/\kappa_{\rm L},1/\kappa_{\rm R}\}$. Thus, the Born-Markov approximations are expected to hold for temperatures that are much higher then the coupling between system and bath. Furthermore, from the Redfield equation in Eq.~\eqref{eq:redfield}, together with Eq.~\eqref{eq:bathcorrferm}, we find that the integral over $s$ involves factors $\exp[is (\omega_j\pm\mu_\alpha)]$. Whenever $|\omega_j\pm\mu_\alpha|\gg k_{\rm B}T_\alpha$, the rapid oscillations of these factors imply that only values of $s\lesssim 1/(\omega_j\pm\mu_\alpha)$ contribute to the integral. For larger values of $s$, the term $k_{\rm B}T_\alpha/\sinh(\pi sk_{\rm B}T_\alpha)$ varies on a time-scale much slower than the oscillations. This implies that the Born-Markov approximations are also justified in the regime $|\omega_j\pm\mu_\alpha|\gg k_{\rm B}T_\alpha$ and $\min \{|\omega_j\pm\mu_{\rm L}|,|\omega_j\pm\mu_{\rm R}|\}\gg 1/\tau_{\rm S}$. Now in case $1/\tau_{\rm S}\ll |\omega_j\pm\mu|$ (dropping the bath index for ease of notation), either we have $k_{\rm B}T\ll |\omega_j\pm\mu|$, such that the Born-Markov approximations are justified, or we have $k_{\rm B}T\gtrsim |\omega_j\pm\mu|$ which implies $1/\tau_{\rm S}\ll k_{\rm B}T$, also ensuring the Born-Markov approximation. In the first case, the bath-correlation functions oscillate to zero over times much smaller than $\tau_{\rm S}$, in the latter case the bath correlation functions decay much faster than $\tau_{\rm S}$. As a consequence, the Born-Markov apprixmations are fulfilled whenever $\kappa$ is much smaller than either $k_{\rm B}T$, or $|\omega_j\pm\mu|$ as expressed in Eq.~\eqref{eq:bornmarkvalferm}.

The Fourier transforms of the bath correlation functions are given by Eq.~\eqref{eq:gammasferm}.
We reach the same conclusions about the validity of the Born-Markov approximation by demanding that these functions are flat over the energy scale $\kappa$ (the inverse of $\tau_{\rm S})$. For $\kappa\ll k_{\rm B}T$, the Fermi-Dirac distribution is everywhere slowly varying compared to $\kappa$. For $\kappa\ll|\omega-\mu|$ and $k_{\rm B}T\ll|\omega-\mu|$, the Fermi function is either very close to zero or one, depending on the sign of $\omega-\mu$, and remains so over the energy scale of $\kappa$.

\subsection{Power and heat currents}
\label{app:currsferm}
For the double dot system with non-interacting fermions the Landauer-like formula provides the following steady state heat current (out of the left lead)
\begin{equation}
J_{\rm L}^\mathrm{t} = \frac{1}{2\pi}\int_{-\infty}^\infty d\omega\mathcal{T}(\omega)(\omega-\mu_{\rm L})[n_{\rm F}^{\rm L}(\omega)-n_{\rm F}^{\rm R}(\omega)],
\end{equation}
and power
\begin{equation}
\label{eq:powtferm}
P_{\rm S}^\mathrm{t} = \frac{(\mu_{\rm R}-\mu_{\rm L})}{2\pi}\int_{-\infty}^\infty d\omega\mathcal{T}(\omega)[n_{\rm F}^{\rm L}(\omega)-n_{\rm F}^{\rm R}(\omega)],
\end{equation}
with the transmission function
\begin{equation}
\label{eq:transmissionf}
\mathcal{T}(\omega) = \frac{g^2\kappa_{\rm L}\kappa_{\rm R}}{|(\omega-\Omega_{\rm L}+i\frac{\kappa_{\rm L}}{2})(\omega-\Omega_{\rm R}+i\frac{\kappa_{\rm R}}{2})-g^2|^2}.
\end{equation}
Note that the transmission function is the same as for the bosonic system, cf.~Eq.~\eqref{eq:transmission}.

For the global and local approaches, the definitions for the heat current and the power are given in Eq.~\eqref{eq:heatpowbathtd} and \eqref{eq:powerout}, with $\hat{N}_{\rm S}=\hat{d}^\dagger_{\rm L}\hat{d}_{\rm L}+\hat{d}^\dagger_{\rm R}\hat{d}_{\rm R}$. For the local approach, we find 
\begin{equation}
\label{eq:heatcurrrf}
J^{\rm l}_{\rm L} = \frac{(\bar{\Omega}-\mu_{\rm L})}{\kappa_{\rm L}+\kappa_{\rm R}}\frac{4g^2\kappa_{\rm L}\kappa_{\rm R}[n_{\rm F}^{\rm L}(\bar{\Omega})-n_{\rm F}^{\rm R}(\bar{\Omega})]}{4g^2+\kappa_{\rm L}\kappa_{\rm R}+16\Delta^2\kappa_{\rm L}\kappa_{\rm R}/(\kappa_{\rm L}+\kappa_{\rm R})^2},
\end{equation}
and
\begin{equation}
\label{eq:powerlf}
P_{\rm S}^{\rm l}\equiv -P^{\rm l}_{\rm L}- P^{\rm l}_{\rm R} = \frac{(\mu_{\rm R}-\mu_{\rm L})}{(\bar{\Omega}-\mu_{\rm L})}J^{\rm l}.
\end{equation}
This tight-coupling condition between heat current and power [which can be inferred from the form of $\hat{H}_{\rm TD}$, cf.~Eq.~\eqref{eq:hamtdferm}] is a direct result of approximating transmission to happen at a single energy.

For the global approach, we find
\begin{equation}
\label{eq:heatcurrgf}
J^{\rm g}_{\rm L} = \sum_{\sigma=\pm}(\Omega_\sigma-\mu_{\rm L})\frac{\kappa_{\rm L}^\sigma\kappa_{\rm R}^\sigma}{\kappa_{\rm L}^\sigma+\kappa_{\rm R}^\sigma}[n_{\rm F}^{\rm L}(\Omega_\sigma)-n_{\rm F}^{\rm R}(\Omega_\sigma)],
\end{equation}
and
\begin{equation}
\label{eq:powergf}
P_{\rm S}^{\rm g}\equiv -P^{\rm g}_{\rm L}- P^{\rm g}_{\rm R}= (\mu_{\rm R}-\mu_{\rm L})\sum_{\sigma=\pm}\frac{\kappa_{\rm L}^\sigma\kappa_{\rm R}^\sigma}{\kappa_{\rm L}^\sigma+\kappa_{\rm R}^\sigma}[n_{\rm F}^{\rm L}(\Omega_\sigma)-n_{\rm F}^{\rm R}(\Omega_\sigma)].
\end{equation}

Finally, in the PERLind approach the dissipators read
\begin{equation}
\label{eq:dissperlindf}
\mathcal{L}^{\rm p}_\alpha = \sum_{k=-1,1}\mathcal{D}[\hat{J}_{\alpha,k}],
\end{equation}
with the jump operators
\begin{equation}
\label{eq:jumperlindf}
\begin{aligned}
\hat{J}_{\rm R,-1} &= \sqrt{\kappa_{\rm R}^+[1-n_{\rm F}^{\rm R}(\Omega_+)]}\hat{d}_++ \sqrt{\kappa_{\rm R}^-[1-n_{\rm F}^{\rm R}(\Omega_-)]}\hat{d}_-,\\
\hat{J}_{\rm R,1} &= \sqrt{\kappa_{\rm R}^+n_{\rm F}^{\rm R}(\Omega_+)}\hat{d}_+^\dagger+ \sqrt{\kappa_{\rm R}^-n_{\rm F}^{\rm R}(\Omega_-)}\hat{d}_-^\dagger,\\
\hat{J}_{\rm L,-1} &= \sqrt{\kappa_{\rm L}^+[1-n_{\rm F}^{\rm L}(\Omega_+)]}\hat{d}_+- \sqrt{\kappa_{\rm L}^-[1-n_{\rm F}^{\rm L}(\Omega_-)]}\hat{d}_-,\\
\hat{J}_{\rm L,1} &= \sqrt{\kappa_{\rm L}^+n_{\rm F}^{\rm L}(\Omega_+)}\hat{d}_+^\dagger- \sqrt{\kappa_{\rm L}^-n_{\rm F}^{\rm L}(\Omega_-)}\hat{d}_-^\dagger.
\end{aligned}
\end{equation}
In this approach, heat current and power are defined as
\begin{equation}
\label{eq:heatcurrpf}
J^{\rm p}_{\rm L} = {\rm Tr}\left\{\left(\hat{H}_{\rm S}-\mu_{\rm L}\hat{N}_{\rm S}\right)\mathcal{L}^{\rm p}_{\rm L}\hat{\rho}\right\},
\end{equation}
and
\begin{equation}
\label{eqpowerpf}
P_{\rm S}^{\rm p} = -\mu_{\rm L}{\rm Tr}\left\{\hat{N}_{\rm S}\mathcal{L}^{\rm p}_{\rm L}\hat{\rho}\right\}-\mu_{\rm R}{\rm Tr}\left\{\hat{N}_{\rm S}\mathcal{L}^{\rm p}_{\rm R}\hat{\rho}\right\}.
\end{equation}

\section{Supplemental calculations for the bosonic heat engine}

\subsection{Bath correlations functions}
\label{app:bathcorrbos}
The bath operators in Eq.~\eqref{eq:bathopbos} result in the bath correlation functions [cf.~Eq.~\eqref{eq:bathcorr}]
\begin{equation}
\label{eq:bathcorrbos}
\begin{aligned}
&C^\alpha_{1}=\int_{0}^{\infty}\frac{d\omega}{2\pi}\kappa_{\alpha}e^{is\omega}n_{\rm B}^\alpha(\omega),\\&
C^\alpha_{-1}=\int_{0}^{\infty}\frac{d\omega}{2\pi}\kappa_{\alpha}e^{-is\omega}[n_{\rm B}^\alpha(\omega)+1],
\end{aligned}
\end{equation}
where $\kappa_{\alpha}$ is given by Eq.~\eqref{eq:kappa} and we have $C^\alpha_k\equiv C^\alpha_{k,k}$ and $C^\alpha_{k,k'}\propto \delta_{k,k'}$. We note that these integrals only converge if $\kappa_{\alpha}\rightarrow 0$ for $\omega\rightarrow 0$. Here we do not give a specific form of $\kappa_\alpha$ and directly proceed to investigating the Fourier transforms of the bath correlation functions. For a detailed discussion on the bath correlation functions for an Ohmic spectral density, we refer to Ref.~\cite{rivas:2010}. By Fourier transforming Eq.~\eqref{eq:bathcorrbos}, we find the rates given in Eq.~\eqref{eq:gammasbos}.
We note that even if $\kappa_{\alpha}$ needs to vanish as $\omega\rightarrow 0$, we can still assume it to be constant over all energies that the system probes (i.e., all energies where the transmission function is non-zero).

Similarly to the fermionic case, the Born-Markov approximations are justified when the quantities in Eq.~\eqref{eq:gammasbos} remain constant over the energy interval $\kappa_{\alpha}$ around the transition frequencies of the system. This is the case if $\kappa_{\alpha}\ll\omega_j$. We note that for very small temperatures ($k_{\rm B}T\ll\omega_j$), $n_{\rm B}(\omega_j)$ becomes exponentially small and the dynamics is dominated by the temperature-independent term in Eq.~\eqref{eq:gammasbos} which corresponds to spontaneous emission. We further note that $\kappa_{\alpha}\ll k_{\rm B}T_\alpha$ is not a sufficient condition for the Born-Markov approximation to be justified. When $\kappa_{\alpha}\rightarrow \omega_j$, then $n_{\rm B}(\omega_j-\kappa_{\alpha})$ diverges, no matter the temperature. The conclusions drawn here from the Fourier transform of the bath correlation functions are in complete agreement with the conclusions obtained in Ref.~\cite{rivas:2010} by considering the bath correlation functions for an Ohmic spectral density.

\subsection{The average Hamiltonian}
\label{app:avham}
To derive the average Hamiltonian for the system given in Eq.~\eqref{eq:hambos}, we first consider the rotating frame determined by the unitary transformation 
\begin{equation}
\label{eq:ur}
\hat{U}_{\rm r} = e^{it[(\Omega_{\rm c}+\Delta)\hat{a}^\dagger_{\rm c}\hat{a}_{\rm c}+(\Omega_{\rm h}-\Delta)\hat{a}^\dagger_{\rm h}\hat{a}_{\rm h}]}.
\end{equation}
In the rotating frame, we find the time-independent Hamiltonian
\begin{equation}
\label{eq:hamrot}
\tilde{H}_{\rm S} = \hat{U}_{\rm r}(t)\hat{H}_{\rm S}(t)\hat{U}^\dagger_{\rm r}(t)-i\hat{U}_{\rm r}(t)\partial_t\hat{U}_{\rm r}^\dagger(t) = \Delta(\hat{a}_{\rm h}^\dagger\hat{a}_{\rm h}-\hat{a}_{\rm c}^\dagger\hat{a}_{\rm c})+g(\hat{a}_{\rm h}^\dagger\hat{a}_{\rm c}+\hat{a}_{\rm c}^\dagger\hat{a}_{\rm h}).
\end{equation}
The time-evolution operator in the lab frame fulfills the relation
\begin{equation}
\label{eq:usysbos}
\hat{U}_{\rm S}(t) = \mathcal{T}e^{-i\int_{0}^{t}dt'\hat{H}_{\rm S}(t')} = \hat{U}_{\rm r}^\dagger(t)\tilde{U}_{\rm S}(t),
\end{equation}
where $\tilde{U}_{\rm S}(t)$ denotes the time-evolution operator for the system in the rotating frame.

To obtain the average Hamiltonian, we use the relations
\begin{equation}
\label{eq:reltp}
e^{it_{\rm p} \varpi \hat{n}_j} = 1,
\end{equation}
for $t_{\rm p} = 2\pi/\varpi$, where $\hat{n}_j$ is any operator that only has integer eigenvalues (such as a photon number operator). With the help of this relation, one can show
\begin{equation}
\label{eq:avhambosapp}
\hat{U}_{\rm S}(t_{\rm p}) = e^{-it_{\rm p}[\Omega_{\alpha}^+\hat{a}^\dagger_+\hat{a}_++\Omega_{\alpha}^-\hat{a}^\dagger_-\hat{a}_-]},
\end{equation}
which holds both for $\alpha= {\rm c,h}$ [the frequencies are given in Eq.~\eqref{eq:bosfreq}], resulting in the two choices for the average Hamiltonian discussed in the main text.

\subsection{Power and heat currents}
In the rotating frame discussed in the last subsection, the bosonic heat engine looks just like the system discussed in Sec.~\ref{sec:compare}, upon setting $\bar{\Omega}=0$ and shifting the energies of the reservoir modes by $\Omega_{\rm c}+\Delta$ and $\Omega_{\rm h}-\Delta$ respectively (see also Ref.~\cite{hofer:2017njp}). We then obtain the Landauer-like formula for the heat current
\begin{equation}
\label{eq:heattbos}
J_{\rm h}^\mathrm{t} = \frac{1}{2\pi}\int_{\varpi}^\infty d\omega\mathcal{T}(\omega)\omega[n_{\rm B}^{\rm h}(\omega)-n_{\rm B}^{\rm c}(\omega-\varpi)],
\end{equation}
and for power we find
\begin{equation}
\label{eq:powtbos}
P_{\rm S}^\mathrm{t} = \frac{1}{2\pi}\int_{-\Omega_{\rm c}}^\infty d\omega\mathcal{T}(\omega)\varpi[n_{\rm B}^{\rm h}(\omega)-n_{\rm B}^{\rm c}(\omega-\varpi)],
\end{equation}
with the transmission function
\begin{equation}
\label{eq:transmissionb}
\mathcal{T}(\omega) = \frac{g^2\kappa_{\rm c}\kappa_{\rm h}}{|(\omega-\Omega_{\rm h}+i\frac{\kappa_{\rm c}}{2})(\omega-\Omega_{\rm c}-\varpi+i\frac{\kappa_{\rm h}}{2})-g^2|^2}.
\end{equation}

For the local approach, we find the heat current
\begin{equation}
\label{eq:locbosheat}
J^{\rm l}_{\rm h} = \frac{\Omega_{\rm h}-\Delta}{\kappa_{\rm c}+\kappa_{\rm h}}\frac{4g^2\kappa_{\rm c}\kappa_{\rm h}[n_{\rm B}^{\rm h}(\Omega_{\rm h}-\Delta)-n_{\rm B}^{\rm c}(\Omega_{\rm c}+\Delta)]}{4g^2+\kappa_{\rm c}\kappa_{\rm h}+16\Delta^2\kappa_{\rm c}\kappa_{\rm h}/(\kappa_{\rm c}+\kappa_{\rm h})^2},
\end{equation}
and power 
\begin{equation}
\label{eq:powlbos}
P^{\rm l} = \left(1-\frac{\Omega_{\rm c}+\Delta}{\Omega_{\rm h}-\Delta}\right)J_{\rm h}^{\rm l}.
\end{equation}

The global approach results in the heat current
\begin{equation}
\label{eq:heatcurrgb}
J^{\rm g}_{\rm h} = \sum_{\sigma=\pm}\Omega_{\rm h}^\sigma\frac{\kappa_{\rm c}^\sigma\kappa_{\rm h}^\sigma}{\kappa_{\rm c}^\sigma+\kappa_{\rm h}^\sigma}[n_{\rm B}^{\rm h}(\Omega_{\rm h}^\sigma)-n_{\rm B}^{\rm c}(\Omega_{\rm c}^\sigma)],
\end{equation}
and power
\begin{equation}
\label{eq:powergb}
P_{\rm S}^{\rm g}\equiv J_{\rm h}+J_{\rm c}= \sum_{\sigma=\pm}\varpi\frac{\kappa_{\rm c}^\sigma\kappa_{\rm h}^\sigma}{\kappa_{\rm c}^\sigma+\kappa_{\rm h}^\sigma}[n_{\rm B}^{\rm h}(\Omega_{\rm h}^\sigma)-n_{\rm B}^{\rm c}(\Omega_{\rm c}^\sigma)].
\end{equation}

Finally, for the PERLind approach, the dissipators read (in the rotating frame)
\begin{equation}
\label{eq:dissperlindb}
\mathcal{L}^{\rm p}_\alpha = \sum_{k=-1,1}\mathcal{D}[\hat{J}_{\alpha,k}],
\end{equation}
with the jump operators
\begin{equation}
\label{eq:jumperlindb}
\begin{aligned}
\hat{J}_{\rm c,-1} &= \sqrt{\kappa_{\rm c}^+[n_{\rm B}^{\rm c}(\Omega_{\rm c}^+)+1]}\hat{a}_++ \sqrt{\kappa_{\rm c}^-[n_{\rm B}^{\rm c}(\Omega_{\rm c}^-)+1]}\hat{a}_-,\\
\hat{J}_{\rm c,1} &= \sqrt{\kappa_{\rm c}^+n_{\rm B}^{\rm c}(\Omega_{\rm c}^+)}\hat{a}_+^\dagger+ \sqrt{\kappa_{\rm c}^-n_{\rm B}^{\rm c}(\Omega_{\rm c}^-)}\hat{a}_-^\dagger,\\
\hat{J}_{\rm h,-1} &= \sqrt{\kappa_{\rm h}^+[n_{\rm B}^{\rm h}(\Omega_{\rm h}^+)+1]}\hat{a}_+- \sqrt{\kappa_{\rm h}^-[n_{\rm B}^{\rm h}(\Omega_{\rm h}^-)+1]}\hat{a}_-,\\
\hat{J}_{\rm h,1} &= \sqrt{\kappa_{\rm h}^+n_{\rm B}^{\rm h}(\Omega_{\rm h}^+)}\hat{a}_+^\dagger- \sqrt{\kappa_{\rm h}^-n_{\rm B}^{\rm h}(\Omega_{\rm g}^-)}\hat{a}_-^\dagger.
\end{aligned}
\end{equation}
In this approach, the heat currents can be obtained from Eq.~\eqref{eq:redfield} by applying the approximations discussed in the supplemental material of Ref.~\cite{ptaszynski:2019}. For the present system, we obtain
\begin{equation}
\label{eq:heatcurrpb}
J^{\rm p}_\alpha =\sum_{\sigma,\sigma' = \pm}\frac{\Omega_{\alpha}^\sigma+\Omega_{\alpha}^{\sigma'}}{2} {\rm Tr}\left\{\left(\hat{J}_{\alpha,1;\sigma}^\dagger\hat{J}_{\alpha,1;\sigma'}-\hat{J}_{\alpha,-1;\sigma}^\dagger\hat{J}_{\alpha,-1;\sigma'}\right)\tilde{\rho}\right\},
\end{equation}
where
\begin{equation}
\label{eq:jaks}
\hat{J}_{\alpha,k}=\hat{J}_{\alpha,k;+}+\hat{J}_{\alpha,k;-},
\end{equation}
and $\hat{J}_{\alpha,k;\sigma}$ can be inferred from Eq.~\eqref{eq:jumperlindb} as the term that is proportional to $\hat{a}_\sigma^{(\dagger)}$. The power is given by
\begin{equation}
\label{eqpowerpp}
P_{\rm S}^{\rm p} = J^{\rm p}_{\rm h}+J^{\rm p}_{\rm c}.
\end{equation}

\section{PERLind approach for the interacting double quantum dot}
\label{app:perlfermint}
For the interacting double quantum dot discussed in Sec.~\ref{sec:beyond}, the dissipators in the PERLind approach read
\begin{equation}
\label{eq:dissperlindfint}
\mathcal{L}^{\rm p}_\alpha = \sum_{k=-1,1}\mathcal{D}[\hat{J}_{\alpha,k}],
\end{equation}
with the jump operators
\begin{equation}
\label{eq:jumperlindfint}
\begin{aligned}
\hat{J}_{\rm R,-1} &= \sqrt{\frac{\kappa_{\rm R}}{2}[1-n_{\rm F}^{\rm R}(\Omega_+)]}(1-\hat{d}^\dagger_-\hat{d}_-)\hat{d}_++ \sqrt{\frac{\kappa_{\rm R}}{2}[1-n_{\rm F}^{\rm R}(\Omega_++U)]}\hat{d}^\dagger_-\hat{d}_-\hat{d}_++\\&+\sqrt{\frac{\kappa_{\rm R}}{2}[1-n_{\rm F}^{\rm R}(\Omega_-)]}(1-\hat{d}^\dagger_+\hat{d}_+)\hat{d}_-+ \sqrt{\frac{\kappa_{\rm R}}{2}[1-n_{\rm F}^{\rm R}(\Omega_-+U)]}\hat{d}^\dagger_+\hat{d}_+\hat{d}_-,\\
\hat{J}_{\rm R,1} &= \sqrt{\frac{\kappa_{\rm R}}{2}n_{\rm F}^{\rm R}(\Omega_+)}(1-\hat{d}^\dagger_-\hat{d}_-)\hat{d}^\dagger_++ \sqrt{\frac{\kappa_{\rm R}}{2}n_{\rm F}^{\rm R}(\Omega_++U)}\hat{d}^\dagger_-\hat{d}_-\hat{d}_+^\dagger\\&+\sqrt{\frac{\kappa_{\rm R}}{2}n_{\rm F}^{\rm R}(\Omega_-)}(1-\hat{d}^\dagger_+\hat{d}_+)\hat{d}_-^\dagger+ \sqrt{\frac{\kappa_{\rm R}}{2}n_{\rm F}^{\rm R}(\Omega_-+U)}\hat{d}^\dagger_+\hat{d}_+\hat{d}_-^\dagger,\\
\hat{J}_{\rm L,-1} &= \sqrt{\frac{\kappa_{\rm L}}{2}[1-n_{\rm F}^{\rm L}(\Omega_+)]}(1-\hat{d}^\dagger_-\hat{d}_-)\hat{d}_++ \sqrt{\frac{\kappa_{\rm L}}{2}[1-n_{\rm F}^{\rm L}(\Omega_++U)]}\hat{d}^\dagger_-\hat{d}_-\hat{d}_++\\&-\sqrt{\frac{\kappa_{\rm L}}{2}[1-n_{\rm F}^{\rm L}(\Omega_-)]}(1-\hat{d}^\dagger_+\hat{d}_+)\hat{d}_-- \sqrt{\frac{\kappa_{\rm L}}{2}[1-n_{\rm F}^{\rm L}(\Omega_-+U)]}\hat{d}^\dagger_+\hat{d}_+\hat{d}_-,\\
\hat{J}_{\rm L,1} &= \sqrt{\frac{\kappa_{\rm L}}{2}n_{\rm F}^{\rm L}(\Omega_+)}(1-\hat{d}^\dagger_-\hat{d}_-)\hat{d}_+^\dagger+ \sqrt{\frac{\kappa_{\rm L}}{2}n_{\rm F}^{\rm L}(\Omega_++U)}\hat{d}^\dagger_-\hat{d}_-\hat{d}_+^\dagger+\\&-\sqrt{\frac{\kappa_{\rm L}}{2}n_{\rm F}^{\rm L}(\Omega_-)}(1-\hat{d}^\dagger_+\hat{d}_+)\hat{d}_-^\dagger- \sqrt{\frac{\kappa_{\rm L}}{2}n_{\rm F}^{\rm L}(\Omega_-+U)}\hat{d}^\dagger_+\hat{d}_+\hat{d}_-^\dagger.
\end{aligned}
\end{equation}

\section*{References}
\bibliographystyle{quantum_ph}
\bibliography{biblio}

\begin{thebibliography}{10}

\bibitem{breuer:book}
H.-P. Breuer and F.~Petruccione.
\newblock
  \href{http://dx.doi.org/10.1093/acprof:oso/9780199213900.001.0001}{\emph{The
  theory of open quantum systems}},  (Oxford University Press 2002).

\bibitem{weiss:book}
U.~Weiss.
\newblock \href{http://dx.doi.org/10.1142/8334}{\emph{Quantum Dissipative
  Systems}},  (World Scientific 1993).

\bibitem{kamenev:book}
A.~Kamenev.
\newblock \href{http://dx.doi.org/10.1017/CBO9781139003667}{\emph{Field Theory
  of Non-Equilibrium Systems}},  (Cambridge University Press 2011).

\bibitem{schaller:book}
G.~Schaller.
\newblock \href{http://dx.doi.org/10.1007/978-3-319-03877-3}{\emph{Open Quantum
  Systems Far from Equilibrium}},  (Springer 2014).

\bibitem{gorini:1976}
V.~Gorini, A.~Kossakowski, and E.~C.~G. Sudarshan.
\newblock Completely positive dynamical semigroups of {N}-level systems.
\newblock \emph{\href{http://dx.doi.org/10.1063/1.522979}{J. Math. Phys.}}
  \href{http://dx.doi.org/10.1063/1.522979}{\textbf{17}, 821}
  \href{http://dx.doi.org/10.1063/1.522979}{ (1976)}.

\bibitem{lindblad:1976}
G.~Lindblad.
\newblock On the generators of quantum dynamical semigroups.
\newblock \emph{\href{http://dx.doi.org/10.1007/BF01608499}{Commun. Math.
  Phys.}} \href{http://dx.doi.org/10.1007/BF01608499}{\textbf{48}, 119}
  \href{http://dx.doi.org/10.1007/BF01608499}{ (1976)}.

\bibitem{dann:2018}
R.~Dann, A.~Levy, and R.~Kosloff.
\newblock Time-dependent markovian quantum master equation.
\newblock \emph{\href{http://dx.doi.org/10.1103/PhysRevA.98.052129}{Phys. Rev.
  A}} \href{http://dx.doi.org/10.1103/PhysRevA.98.052129}{\textbf{98}, 052129}
  \href{http://dx.doi.org/10.1103/PhysRevA.98.052129}{ (2018)}.

\bibitem{carmichael:book}
H.~Carmichael.
\newblock \href{http://dx.doi.org/10.1007/978-3-540-47620-7}{\emph{An Open
  Systems Approach to Quantum Optics}},  (Springer 1991).

\bibitem{restrepo:2014}
J.~Restrepo, C.~Ciuti, and I.~Favero.
\newblock Single-polariton optomechanics.
\newblock \emph{\href{http://dx.doi.org/10.1103/PhysRevLett.112.013601}{Phys.
  Rev. Lett.}}
  \href{http://dx.doi.org/10.1103/PhysRevLett.112.013601}{\textbf{112}, 013601}
  \href{http://dx.doi.org/10.1103/PhysRevLett.112.013601}{ (2014)}.

\bibitem{cresser:1992}
J.~Cresser.
\newblock Thermal equilibrium in the {J}aynes-{C}ummings model.
\newblock \emph{\href{http://dx.doi.org/10.1080/09500349214552211}{J. Mod.
  Opt.}} \href{http://dx.doi.org/10.1080/09500349214552211}{\textbf{39}, 2187}
  \href{http://dx.doi.org/10.1080/09500349214552211}{ (1992)}.

\bibitem{scala:2007}
M.~Scala, B.~Militello, A.~Messina, J.~Piilo, and S.~Maniscalco.
\newblock Microscopic derivation of the {J}aynes-{C}ummings model with cavity
  losses.
\newblock \emph{\href{http://dx.doi.org/10.1103/PhysRevA.75.013811}{Phys. Rev.
  A}} \href{http://dx.doi.org/10.1103/PhysRevA.75.013811}{\textbf{75}, 013811}
  \href{http://dx.doi.org/10.1103/PhysRevA.75.013811}{ (2007)}.

\bibitem{rivas:2010}
A.~Rivas, A.~D.~K. Plato, S.~F. Huelga, and M.~B. Plenio.
\newblock Markovian master equations: a critical study.
\newblock \emph{\href{http://dx.doi.org/10.1088/1367-2630/12/11/113032}{New J.
  Phys.}} \href{http://dx.doi.org/10.1088/1367-2630/12/11/113032}{\textbf{12},
  113032} \href{http://dx.doi.org/10.1088/1367-2630/12/11/113032}{ (2010)}.

\bibitem{hofer:2017njp}
P.~P. Hofer, M.~Perarnau-Llobet, L.~D.~M. Miranda, G.~Haack, R.~Silva, J.~B.
  Brask, and N.~Brunner.
\newblock Markovian master equations for quantum thermal machines: local versus
  global approach.
\newblock \emph{\href{http://dx.doi.org/10.1088/1367-2630/aa964f}{New J.
  Phys.}} \href{http://dx.doi.org/10.1088/1367-2630/aa964f}{\textbf{19},
  123037} \href{http://dx.doi.org/10.1088/1367-2630/aa964f}{ (2017)}.

\bibitem{gonzalez:2017}
J.~O. Gonz\'alez, L.~A. Correa, G.~Nocerino, J.~P. Palao, D.~Alonso, and
  G.~Adesso.
\newblock Testing the validity of the ‘local’ and ‘global’ {GKLS}
  master equations on an exactly solvable model.
\newblock \emph{\href{http://dx.doi.org/10.1142/S1230161217400108}{Open Syst.
  Inf. Dyn.}} \href{http://dx.doi.org/10.1142/S1230161217400108}{\textbf{24},
  1740010} \href{http://dx.doi.org/10.1142/S1230161217400108}{ (2017)}.

\bibitem{seah:2018}
S.~Seah, S.~Nimmrichter, and V.~Scarani.
\newblock Refrigeration beyond weak internal coupling.
\newblock \emph{\href{http://dx.doi.org/10.1103/PhysRevE.98.012131}{Phys. Rev.
  E}} \href{http://dx.doi.org/10.1103/PhysRevE.98.012131}{\textbf{98}, 012131}
  \href{http://dx.doi.org/10.1103/PhysRevE.98.012131}{ (2018)}.

\bibitem{naseem:2018}
M.~T. Naseem, A.~Xuereb, and O.~E. M\"ustecapl\ifmmode \imath \else \i
  \fi{}o\ifmmode~\breve{g}\else \u{g}\fi{}lu.
\newblock Thermodynamic consistency of the optomechanical master equation.
\newblock \emph{\href{http://dx.doi.org/10.1103/PhysRevA.98.052123}{Phys. Rev.
  A}} \href{http://dx.doi.org/10.1103/PhysRevA.98.052123}{\textbf{98}, 052123}
  \href{http://dx.doi.org/10.1103/PhysRevA.98.052123}{ (2018)}.

\bibitem{cattaneo:2019}
M.~Cattaneo, G.~L. Giorgi, S.~Maniscalco, and R.~Zambrini.
\newblock Local versus global master equation with common and separate baths:
  superiority of the global approach in partial secular approximation.
\newblock \emph{\href{http://dx.doi.org/10.1088/1367-2630/ab54ac}{New J.
  Phys.}} \href{http://dx.doi.org/10.1088/1367-2630/ab54ac}{\textbf{21},
  113045} \href{http://dx.doi.org/10.1088/1367-2630/ab54ac}{ (2019)}.

\bibitem{farina:2020}
D.~Farina, G.~De~Filippis, V.~Cataudella, M.~Polini, and V.~Giovannetti.
\newblock Going beyond local and global approaches for localized thermal
  dissipation.
\newblock \emph{\href{http://dx.doi.org/10.1103/PhysRevA.102.052208}{Phys. Rev.
  A}} \href{http://dx.doi.org/10.1103/PhysRevA.102.052208}{\textbf{102},
  052208} \href{http://dx.doi.org/10.1103/PhysRevA.102.052208}{ (2020)}.

\bibitem{elouard:2020}
C.~Elouard, D.~Herrera-Mart{\'{\i}}, M.~Esposito, and A.~Auff{\`{e}}ves.
\newblock Thermodynamics of optical bloch equations.
\newblock \emph{\href{http://dx.doi.org/10.1088/1367-2630/abbd6e}{New J.
  Phys.}} \href{http://dx.doi.org/10.1088/1367-2630/abbd6e}{\textbf{22},
  103039} \href{http://dx.doi.org/10.1088/1367-2630/abbd6e}{ (2020)}.

\bibitem{scali:2021}
S.~Scali, J.~Anders, and L.~A. Correa.
\newblock Local master equations bypass the secular approximation.
\newblock \emph{\href{http://dx.doi.org/10.22331/q-2021-05-01-451}{{Quantum}}}
  \href{http://dx.doi.org/10.22331/q-2021-05-01-451}{\textbf{5}, 451}
  \href{http://dx.doi.org/10.22331/q-2021-05-01-451}{ (2021)}.

\bibitem{brunner:2012}
N.~Brunner, N.~Linden, S.~Popescu, and P.~Skrzypczyk.
\newblock Virtual qubits, virtual temperatures, and the foundations of
  thermodynamics.
\newblock \emph{\href{http://dx.doi.org/10.1103/PhysRevE.85.051117}{Phys. Rev.
  E}} \href{http://dx.doi.org/10.1103/PhysRevE.85.051117}{\textbf{85}, 051117}
  \href{http://dx.doi.org/10.1103/PhysRevE.85.051117}{ (2012)}.

\bibitem{mitchison:2019}
M.~T. Mitchison.
\newblock Quantum thermal absorption machines: refrigerators, engines and
  clocks.
\newblock \emph{\href{http://dx.doi.org/10.1080/00107514.2019.1631555}{Contemp.
  Phys.}} \href{http://dx.doi.org/10.1080/00107514.2019.1631555}{\textbf{60},
  164} \href{http://dx.doi.org/10.1080/00107514.2019.1631555}{ (2019)}.

\bibitem{levy:2014}
A.~Levy and R.~Kosloff.
\newblock The local approach to quantum transport may violate the second law of
  thermodynamics.
\newblock \emph{\href{http://dx.doi.org/10.1209/0295-5075/107/20004}{EPL}}
  \href{http://dx.doi.org/10.1209/0295-5075/107/20004}{\textbf{107}, 20004}
  \href{http://dx.doi.org/10.1209/0295-5075/107/20004}{ (2014)}.

\bibitem{novotny:2002}
T.~Novotn\'y.
\newblock Investigation of apparent violation of the second law of
  thermodynamics in quantum transport studies.
\newblock \emph{\href{http://dx.doi.org/10.1209/epl/i2002-00174-3}{EPL}}
  \href{http://dx.doi.org/10.1209/epl/i2002-00174-3}{\textbf{59}, 648}
  \href{http://dx.doi.org/10.1209/epl/i2002-00174-3}{ (2002)}.

\bibitem{trushechkin:2016}
A.~S. Trushechkin and I.~V. Volovich.
\newblock Perturbative treatment of inter-site couplings in the local
  description of open quantum networks.
\newblock \emph{\href{http://dx.doi.org/10.1209/0295-5075/113/30005}{EPL}}
  \href{http://dx.doi.org/10.1209/0295-5075/113/30005}{\textbf{113}, 30005}
  \href{http://dx.doi.org/10.1209/0295-5075/113/30005}{ (2016)}.

\bibitem{dann:2020}
R.~Dann and R.~Kosloff.
\newblock Quantum thermo-dynamical construction for driven open quantum
  systems.
\newblock \href{http://arxiv.org/abs/2012.07979}{ArXiv:2012.07979}.

\bibitem{chiara:2018}
G.~D. Chiara, G.~Landi, A.~Hewgill, B.~Reid, A.~Ferraro, A.~J. Roncaglia, and
  M.~Antezza.
\newblock Reconciliation of quantum local master equations with thermodynamics.
\newblock \emph{\href{http://dx.doi.org/10.1088/1367-2630/aaecee}{New J.
  Phys.}} \href{http://dx.doi.org/10.1088/1367-2630/aaecee}{\textbf{20},
  113024} \href{http://dx.doi.org/10.1088/1367-2630/aaecee}{ (2018)}.

\bibitem{hegwill:2021}
A.~Hewgill, G.~De~Chiara, and A.~Imparato.
\newblock Quantum thermodynamically consistent local master equations.
\newblock \emph{\href{http://dx.doi.org/10.1103/PhysRevResearch.3.013165}{Phys.
  Rev. Research}}
  \href{http://dx.doi.org/10.1103/PhysRevResearch.3.013165}{\textbf{3}, 013165}
  \href{http://dx.doi.org/10.1103/PhysRevResearch.3.013165}{ (2021)}.

\bibitem{kirsanskas:2018}
G.~Kir\ifmmode~\check{s}\else \v{s}\fi{}anskas, M.~Francki\'e, and A.~Wacker.
\newblock Phenomenological position and energy resolving lindblad approach to
  quantum kinetics.
\newblock \emph{\href{http://dx.doi.org/10.1103/PhysRevB.97.035432}{Phys. Rev.
  B}} \href{http://dx.doi.org/10.1103/PhysRevB.97.035432}{\textbf{97}, 035432}
  \href{http://dx.doi.org/10.1103/PhysRevB.97.035432}{ (2018)}.

\bibitem{ptaszynski:2019}
K.~Ptaszy\ifmmode~\acute{n}\else \'{n}\fi{}ski and M.~Esposito.
\newblock Thermodynamics of quantum information flows.
\newblock \emph{\href{http://dx.doi.org/10.1103/PhysRevLett.122.150603}{Phys.
  Rev. Lett.}}
  \href{http://dx.doi.org/10.1103/PhysRevLett.122.150603}{\textbf{122}, 150603}
  \href{http://dx.doi.org/10.1103/PhysRevLett.122.150603}{ (2019)}.

\bibitem{kleinherbers:2020}
E.~Kleinherbers, N.~Szpak, J.~K\"onig, and R.~Sch\"utzhold.
\newblock Relaxation dynamics in a hubbard dimer coupled to fermionic baths:
  Phenomenological description and its microscopic foundation.
\newblock \emph{\href{http://dx.doi.org/10.1103/PhysRevB.101.125131}{Phys. Rev.
  B}} \href{http://dx.doi.org/10.1103/PhysRevB.101.125131}{\textbf{101},
  125131} \href{http://dx.doi.org/10.1103/PhysRevB.101.125131}{ (2020)}.

\bibitem{nathan:2020}
F.~Nathan and M.~S. Rudner.
\newblock Universal {L}indblad equation for open quantum systems.
\newblock \emph{\href{http://dx.doi.org/10.1103/PhysRevB.102.115109}{Phys. Rev.
  B}} \href{http://dx.doi.org/10.1103/PhysRevB.102.115109}{\textbf{102},
  115109} \href{http://dx.doi.org/10.1103/PhysRevB.102.115109}{ (2020)}.

\bibitem{davidovic:2020}
D.~Davidovi{\'{c}}.
\newblock Completely positive, simple, and possibly highly accurate
  approximation of the redfield equation.
\newblock \emph{\href{http://dx.doi.org/10.22331/q-2020-09-21-326}{{Quantum}}}
  \href{http://dx.doi.org/10.22331/q-2020-09-21-326}{\textbf{4}, 326}
  \href{http://dx.doi.org/10.22331/q-2020-09-21-326}{ (2020)}.

\bibitem{schaller:2008}
G.~Schaller and T.~Brandes.
\newblock Preservation of positivity by dynamical coarse graining.
\newblock \emph{\href{http://dx.doi.org/10.1103/PhysRevA.78.022106}{Phys. Rev.
  A}} \href{http://dx.doi.org/10.1103/PhysRevA.78.022106}{\textbf{78}, 022106}
  \href{http://dx.doi.org/10.1103/PhysRevA.78.022106}{ (2008)}.

\bibitem{majenz:2013}
C.~Majenz, T.~Albash, H.-P. Breuer, and D.~A. Lidar.
\newblock Coarse graining can beat the rotating-wave approximation in quantum
  {M}arkovian master equations.
\newblock \emph{\href{http://dx.doi.org/10.1103/PhysRevA.88.012103}{Phys. Rev.
  A}} \href{http://dx.doi.org/10.1103/PhysRevA.88.012103}{\textbf{88}, 012103}
  \href{http://dx.doi.org/10.1103/PhysRevA.88.012103}{ (2013)}.

\bibitem{cresser:2017}
J.~D. Cresser and C.~Facer.
\newblock Coarse-graining in the derivation of {M}arkovian master equations and
  its significance in quantum thermodynamics.
\newblock \href{http://arxiv.org/abs/1710.09939}{ArXiv:1710.09939}.

\bibitem{farina:2019}
D.~Farina and V.~Giovannetti.
\newblock Open-quantum-system dynamics: Recovering positivity of the {R}edfield
  equation via the partial secular approximation.
\newblock \emph{\href{http://dx.doi.org/10.1103/PhysRevA.100.012107}{Phys. Rev.
  A}} \href{http://dx.doi.org/10.1103/PhysRevA.100.012107}{\textbf{100},
  012107} \href{http://dx.doi.org/10.1103/PhysRevA.100.012107}{ (2019)}.

\bibitem{mozgunov:2020}
E.~Mozgunov and D.~Lidar.
\newblock Completely positive master equation for arbitrary driving and small
  level spacing.
\newblock \emph{\href{http://dx.doi.org/10.22331/q-2020-02-06-227}{{Quantum}}}
  \href{http://dx.doi.org/10.22331/q-2020-02-06-227}{\textbf{4}, 227}
  \href{http://dx.doi.org/10.22331/q-2020-02-06-227}{ (2020)}.

\bibitem{schaller:2020}
G.~Schaller and J.~Ablaßmayer.
\newblock Thermodynamics of the coarse-graining master equation.
\newblock \emph{\href{http://dx.doi.org/10.3390/e22050525}{Entropy}}
  \href{http://dx.doi.org/10.3390/e22050525}{\textbf{22}}
  \href{http://dx.doi.org/10.3390/e22050525}{ (2020)}.

\bibitem{becker:2021}
T.~Becker, L.-N. Wu, and A.~Eckardt.
\newblock Lindbladian approximation beyond ultraweak coupling.
\newblock \emph{\href{http://dx.doi.org/10.1103/PhysRevE.104.014110}{Phys. Rev.
  E}} \href{http://dx.doi.org/10.1103/PhysRevE.104.014110}{\textbf{104},
  014110} \href{http://dx.doi.org/10.1103/PhysRevE.104.014110}{ (2021)}.

\bibitem{trushechkin:2021}
A.~Trushechkin.
\newblock Unified {G}orini-{K}ossakowski-{L}indblad-{S}udarshan quantum master
  equation beyond the secular approximation.
\newblock \emph{\href{http://dx.doi.org/10.1103/PhysRevA.103.062226}{Phys. Rev.
  A}} \href{http://dx.doi.org/10.1103/PhysRevA.103.062226}{\textbf{103},
  062226} \href{http://dx.doi.org/10.1103/PhysRevA.103.062226}{ (2021)}.

\bibitem{wang:2014}
J.-S. Wang, B.~K. Agarwalla, H.~Li, and J.~Thingna.
\newblock Nonequilibrium {G}reen’s function method for quantum thermal
  transport.
\newblock \emph{\href{http://dx.doi.org/10.1007/s11467-013-0340-x}{Front.
  Phys.}} \href{http://dx.doi.org/10.1007/s11467-013-0340-x}{\textbf{9}, 673}
  \href{http://dx.doi.org/10.1007/s11467-013-0340-x}{ (2014)}.
\newblock Note the slightly different system-bath coupling.

\bibitem{meir:1992}
Y.~Meir and N.~S. Wingreen.
\newblock Landauer formula for the current through an interacting electron
  region.
\newblock \emph{\href{http://dx.doi.org/10.1103/PhysRevLett.68.2512}{Phys. Rev.
  Lett.}} \href{http://dx.doi.org/10.1103/PhysRevLett.68.2512}{\textbf{68},
  2512} \href{http://dx.doi.org/10.1103/PhysRevLett.68.2512}{ (1992)}.

\bibitem{zhang:2007}
W.~Zhang, T.~S. Fisher, and N.~Mingo.
\newblock The atomistic {G}reen's function method: An efficient simulation
  approach for nanoscale phonon transport.
\newblock \emph{\href{http://dx.doi.org/10.1080/10407790601144755}{Numer. Heat
  Transf. B: Fundam.}}
  \href{http://dx.doi.org/10.1080/10407790601144755}{\textbf{51}, 333}
  \href{http://dx.doi.org/10.1080/10407790601144755}{ (2007)}.

\bibitem{agarwalla:2018}
B.~K. Agarwalla and D.~Segal.
\newblock Assessing the validity of the thermodynamic uncertainty relation in
  quantum systems.
\newblock \emph{\href{http://dx.doi.org/10.1103/PhysRevB.98.155438}{Phys. Rev.
  B}} \href{http://dx.doi.org/10.1103/PhysRevB.98.155438}{\textbf{98}, 155438}
  \href{http://dx.doi.org/10.1103/PhysRevB.98.155438}{ (2018)}.

\bibitem{spohn:1978}
H.~Spohn and J.~L. Lebowitz.
\newblock Irreversible thermodynamics for quantum systems weakly coupled to
  thermal reservoirs.
\newblock \emph{\href{http://dx.doi.org/10.1002/9780470142578.ch2}{Adv. Chem.
  Phys}} \href{http://dx.doi.org/10.1002/9780470142578.ch2}{\textbf{38}, 109}
  \href{http://dx.doi.org/10.1002/9780470142578.ch2}{ (1978)}.

\bibitem{brandner:2016}
K.~Brandner and U.~Seifert.
\newblock Periodic thermodynamics of open quantum systems.
\newblock \emph{\href{http://dx.doi.org/10.1103/PhysRevE.93.062134}{Phys. Rev.
  E}} \href{http://dx.doi.org/10.1103/PhysRevE.93.062134}{\textbf{93}, 062134}
  \href{http://dx.doi.org/10.1103/PhysRevE.93.062134}{ (2016)}.

\bibitem{geva:2000}
E.~Geva, E.~Rosenman, and D.~Tannor.
\newblock On the second-order corrections to the quantum canonical equilibrium
  density matrix.
\newblock \emph{\href{http://dx.doi.org/10.1063/1.481928}{J. Chem. Phys.}}
  \href{http://dx.doi.org/10.1063/1.481928}{\textbf{113}, 1380}
  \href{http://dx.doi.org/10.1063/1.481928}{ (2000)}.

\bibitem{subasi:2012}
Y.~Suba\c{s}\i, C.~H. Fleming, J.~M. Taylor, and B.~L. Hu.
\newblock Equilibrium states of open quantum systems in the strong coupling
  regime.
\newblock \emph{\href{http://dx.doi.org/10.1103/PhysRevE.86.061132}{Phys. Rev.
  E}} \href{http://dx.doi.org/10.1103/PhysRevE.86.061132}{\textbf{86}, 061132}
  \href{http://dx.doi.org/10.1103/PhysRevE.86.061132}{ (2012)}.

\bibitem{esposito:2010njp}
M.~Esposito, K.~Lindenberg, and C.~V. den Broeck.
\newblock Entropy production as correlation between system and reservoir.
\newblock \emph{\href{http://dx.doi.org/10.1088/1367-2630/12/1/013013}{New J.
  Phys.}} \href{http://dx.doi.org/10.1088/1367-2630/12/1/013013}{\textbf{12},
  013013} \href{http://dx.doi.org/10.1088/1367-2630/12/1/013013}{ (2010)}.

\bibitem{ptaszynski:2019ent}
K.~Ptaszy\ifmmode~\acute{n}\else \'{n}\fi{}ski and M.~Esposito.
\newblock Entropy production in open systems: The predominant role of
  intraenvironment correlations.
\newblock \emph{\href{http://dx.doi.org/10.1103/PhysRevLett.123.200603}{Phys.
  Rev. Lett.}}
  \href{http://dx.doi.org/10.1103/PhysRevLett.123.200603}{\textbf{123}, 200603}
  \href{http://dx.doi.org/10.1103/PhysRevLett.123.200603}{ (2019)}.

\bibitem{strasberg:2019}
P.~Strasberg and M.~Esposito.
\newblock Non-{M}arkovianity and negative entropy production rates.
\newblock \emph{\href{http://dx.doi.org/10.1103/PhysRevE.99.012120}{Phys. Rev.
  E}} \href{http://dx.doi.org/10.1103/PhysRevE.99.012120}{\textbf{99}, 012120}
  \href{http://dx.doi.org/10.1103/PhysRevE.99.012120}{ (2019)}.

\bibitem{esposito:2009rmp}
M.~Esposito, U.~Harbola, and S.~Mukamel.
\newblock Nonequilibrium fluctuations, fluctuation theorems, and counting
  statistics in quantum systems.
\newblock \emph{\href{http://dx.doi.org/10.1103/RevModPhys.81.1665}{Rev. Mod.
  Phys.}} \href{http://dx.doi.org/10.1103/RevModPhys.81.1665}{\textbf{81},
  1665} \href{http://dx.doi.org/10.1103/RevModPhys.81.1665}{ (2009)}.

\bibitem{perarnau:2017}
M.~Perarnau-Llobet, E.~B\"aumer, K.~V. Hovhannisyan, M.~Huber, and A.~Acin.
\newblock No-go theorem for the characterization of work fluctuations in
  coherent quantum systems.
\newblock \emph{\href{http://dx.doi.org/10.1103/PhysRevLett.118.070601}{Phys.
  Rev. Lett.}}
  \href{http://dx.doi.org/10.1103/PhysRevLett.118.070601}{\textbf{118}, 070601}
  \href{http://dx.doi.org/10.1103/PhysRevLett.118.070601}{ (2017)}.

\bibitem{hofer:2017q}
P.~P. Hofer.
\newblock Quasi-probability distributions for observables in dynamic systems.
\newblock \emph{\href{http://dx.doi.org/10.22331/q-2017-10-12-32}{{Quantum}}}
  \href{http://dx.doi.org/10.22331/q-2017-10-12-32}{\textbf{1}, 32}
  \href{http://dx.doi.org/10.22331/q-2017-10-12-32}{ (2017)}.

\bibitem{kerremans:2021}
T.~Kerremans, P.~Samuelsson, and P.~P. Potts.
\newblock Probabilistically violating the first law of thermodynamics in a
  quantum heat engine.
\newblock \href{http://arxiv.org/abs/2102.01395}{ArXiv:2102.01395}.

\bibitem{gasparinetti:2014}
S.~Gasparinetti, P.~Solinas, A.~Braggio, and M.~Sassetti.
\newblock Heat-exchange statistics in driven open quantum systems.
\newblock \emph{\href{http://dx.doi.org/10.1088/1367-2630/16/11/115001}{New J.
  Phys.}} \href{http://dx.doi.org/10.1088/1367-2630/16/11/115001}{\textbf{16},
  115001} \href{http://dx.doi.org/10.1088/1367-2630/16/11/115001}{ (2014)}.

\bibitem{silaev:2014}
M.~Silaev, T.~T. Heikkil\"a, and P.~Virtanen.
\newblock Lindblad-equation approach for the full counting statistics of work
  and heat in driven quantum systems.
\newblock \emph{\href{http://dx.doi.org/10.1103/PhysRevE.90.022103}{Phys. Rev.
  E}} \href{http://dx.doi.org/10.1103/PhysRevE.90.022103}{\textbf{90}, 022103}
  \href{http://dx.doi.org/10.1103/PhysRevE.90.022103}{ (2014)}.

\bibitem{friedman:2018}
H.~M. Friedman, B.~K. Agarwalla, and D.~Segal.
\newblock Quantum energy exchange and refrigeration: a full-counting statistics
  approach.
\newblock \emph{\href{http://dx.doi.org/10.1088/1367-2630/aad5fc}{New J.
  Phys.}} \href{http://dx.doi.org/10.1088/1367-2630/aad5fc}{\textbf{20},
  083026} \href{http://dx.doi.org/10.1088/1367-2630/aad5fc}{ (2018)}.

\bibitem{kilgour:2019}
M.~Kilgour, B.~K. Agarwalla, and D.~Segal.
\newblock Path-integral methodology and simulations of quantum thermal
  transport: Full counting statistics approach.
\newblock \emph{\href{http://dx.doi.org/10.1063/1.5084949}{J. Chem. Phys.}}
  \href{http://dx.doi.org/10.1063/1.5084949}{\textbf{150}, 084111}
  \href{http://dx.doi.org/10.1063/1.5084949}{ (2019)}.

\bibitem{kohler:1997}
S.~Kohler, T.~Dittrich, and P.~H\"anggi.
\newblock Floquet-{M}arkovian description of the parametrically driven,
  dissipative harmonic quantum oscillator.
\newblock \emph{\href{http://dx.doi.org/10.1103/PhysRevE.55.300}{Phys. Rev. E}}
  \href{http://dx.doi.org/10.1103/PhysRevE.55.300}{\textbf{55}, 300}
  \href{http://dx.doi.org/10.1103/PhysRevE.55.300}{ (1997)}.

\bibitem{levy:2012pre}
A.~Levy, R.~Alicki, and R.~Kosloff.
\newblock Quantum refrigerators and the third law of thermodynamics.
\newblock \emph{\href{http://dx.doi.org/10.1103/PhysRevE.85.061126}{Phys. Rev.
  E}} \href{http://dx.doi.org/10.1103/PhysRevE.85.061126}{\textbf{85}, 061126}
  \href{http://dx.doi.org/10.1103/PhysRevE.85.061126}{ (2012)}.

\bibitem{folland:1997}
G.~B. Folland and A.~Sitaram.
\newblock The uncertainty principle: A mathematical survey.
\newblock \emph{\href{http://dx.doi.org/10.1007/BF02649110}{J. Fourier Anal.
  Appl.}} \href{http://dx.doi.org/10.1007/BF02649110}{\textbf{3}, 207}
  \href{http://dx.doi.org/10.1007/BF02649110}{ (1997)}.

\bibitem{tscherbul:2015}
T.~V. Tscherbul and P.~Brumer.
\newblock Partial secular {B}loch-{R}edfield master equation for incoherent
  excitation of multilevel quantum systems.
\newblock \emph{\href{http://dx.doi.org/10.1063/1.4908130}{J. Chem. Phys.}}
  \href{http://dx.doi.org/10.1063/1.4908130}{\textbf{142}, 104107}
  \href{http://dx.doi.org/10.1063/1.4908130}{ (2015)}.

\bibitem{jeske:2015}
J.~Jeske, D.~J. Ing, M.~B. Plenio, S.~F. Huelga, and J.~H. Cole.
\newblock {B}loch-{R}edfield equations for modeling light-harvesting complexes.
\newblock \emph{\href{http://dx.doi.org/10.1063/1.4907370}{J. Chem. Phys.}}
  \href{http://dx.doi.org/10.1063/1.4907370}{\textbf{142}, 064104}
  \href{http://dx.doi.org/10.1063/1.4907370}{ (2015)}.

\bibitem{spohn:1978b}
H.~Spohn.
\newblock Entropy production for quantum dynamical semigroups.
\newblock \emph{\href{http://dx.doi.org/10.1063/1.523789}{J. Math. Phys.}}
  \href{http://dx.doi.org/10.1063/1.523789}{\textbf{19}, 1227}
  \href{http://dx.doi.org/10.1063/1.523789}{ (1978)}.

\bibitem{kosloff:1984}
R.~Kosloff.
\newblock A quantum mechanical open system as a model of a heat engine.
\newblock \emph{\href{http://dx.doi.org/10.1063/1.446862}{J. Chem. Phys.}}
  \href{http://dx.doi.org/10.1063/1.446862}{\textbf{80}, 1625}
  \href{http://dx.doi.org/10.1063/1.446862}{ (1984)}.

\bibitem{hofer:2016prb}
P.~P. Hofer, J.-R. Souquet, and A.~A. Clerk.
\newblock Quantum heat engine based on photon-assisted {C}ooper pair tunneling.
\newblock \emph{\href{http://dx.doi.org/10.1103/PhysRevB.93.041418}{Phys. Rev.
  B}} \href{http://dx.doi.org/10.1103/PhysRevB.93.041418}{\textbf{93},
  041418(R)} \href{http://dx.doi.org/10.1103/PhysRevB.93.041418}{ (2016)}.

\bibitem{alicki:1979}
R.~Alicki.
\newblock The quantum open system as a model of the heat engine.
\newblock \emph{\href{http://dx.doi.org/10.1088/0305-4470/12/5/007}{J. Phys A:
  Math. Gen.}} \href{http://dx.doi.org/10.1088/0305-4470/12/5/007}{\textbf{12},
  L103} \href{http://dx.doi.org/10.1088/0305-4470/12/5/007}{ (1979)}.

\bibitem{alicki:2015}
R.~Alicki and D.~Gelbwaser-Klimovsky.
\newblock Non-equilibrium quantum heat machines.
\newblock \emph{\href{http://dx.doi.org/10.1088/1367-2630/17/11/115012}{New J.
  Phys.}} \href{http://dx.doi.org/10.1088/1367-2630/17/11/115012}{\textbf{17},
  115012} \href{http://dx.doi.org/10.1088/1367-2630/17/11/115012}{ (2015)}.

\end{thebibliography}

\end{document}